%% file: heptesiII.tex
\documentclass[12pt,dvips,twoside]{report}
\def\dir{}

\def\figcond{1}

\def\figsty{0}

\def\authname{}

\newcount\figsubcountss \figsubcountss=0
\ifnum\figcond>0 
\input epsf 
  \ifnum\figsty>0 
  \fi
\fi
\def\figinsert#1#2#3#4{
\ifnum\figcond>0 
  \ifnum\figsty>0 
    \ifnum\figsubcountss=0
      \immediate\write9{\noexpand\catcode`\noexpand\@=11}
      \immediate\write9{\noexpand\newpage}
      \immediate\write9{\noexpand\pagestyle {empty}}
      \immediate\write9{\noexpand\section* {Figure Captions}}
      \immediate\write9{\noexpand\begin {enumerate}}
      \immediate\write9{\noexpand\renewcommand {\noexpand\theenumi}{}}
      \immediate\write9{\noexpand\begin {enumerate}}
      \immediate\write9{\noexpand\renewcommand 
        {\noexpand\theenumii}{\noexpand\arabic {enumii}}}
      \immediate\write9{\noexpand\renewcommand {\noexpand\labelenumii}
        {Fig. \noexpand\arabic {enumii}:}}
      \immediate\write10{\noexpand\newpage}
      \ifnum\figsty>1
        \immediate\write10{\noexpand\pagestyle {headings}}
        \immediate\write10{\noexpand\setcounter {page}{1}}
        \immediate\write10{\noexpand\renewcommand {\noexpand\thepage}
            {Figure \noexpand\arabic{page} -- \noexpand\authname}}
      \else
        \immediate\write10{\noexpand\pagestyle {empty}}
      \fi
    \fi
    \global\advance\figsubcountss by 1
    \immediate\write10{\noexpand\begin {figure}[p]}
    \immediate\write10{\noexpand\begin {center}}
    \immediate\write10{\noexpand\ }
    \immediate\write10{\noexpand\epsfbox {\dir #1}}
    \immediate\write10{\noexpand\ \noexpand\\}
    \ifnum\figsty<2
      \immediate\write10{\noexpand\vspace {1cm}}
      \immediate\write10{Figure \noexpand\ref {#3}}
    \fi
    \immediate\write10{\noexpand\end {center}}
    \immediate\write10{\noexpand\end {figure}}
    \immediate\write10{\noexpand\newpage}
   \let\save=\ref \let\ref=0 \let\savec=\cite \let\cite=0
    \immediate\write9{\noexpand\item #2}
   \let\ref=\save \let\cite=\savec
    \immediate\write9{\noexpand\label {#3}}
    \begin {figure}[htbp]
    \begin{center}
    \fbox{Fig. \ref{#3}}
    \end{center}
    \end {figure}
  \else
    \begin{figure}[#4]
    \begin{center}
    \ \epsfbox{\dir #1}
    \caption {#2 \label {#3}}
    \end{center}
    \end{figure}
  \fi
\else
  \begin {figure}[htbp]
  \begin{center}
  \fbox{Fig. \ref{#3}}
  \caption {#2 \label {#3}}
  \end{center}
  \end {figure}
\fi
}
\def\Closeout#1{%
   \immediate\closeout#1}
\def\figepsfout{
\Closeout10
  \immediate\write9{\noexpand\end {enumerate}}
  \immediate\write9{\noexpand\end {enumerate}}
  \immediate\write9{\noexpand\catcode`\noexpand\@=12}
\Closeout9
\input \jobname.cap
\input \jobname.fis}
\catcode`\@=11
\newbox\tempboxa
\newdimen\captionboxsubcount 
\def\capsize#1{\captionboxsubcount=#1pt}
\newdimen\captionboxsub
\captionboxsub=\hsize \advance\captionboxsub by -\captionboxsubcount
\advance\captionboxsub by -\captionboxsubcount
\long\def\@makecaption#1#2{
 \setbox\@tempboxa\hbox{#1: #2}
 \ifdim \wd\@tempboxa >\captionboxsub 
\rightskip=\captionboxsubcount \leftskip=\captionboxsubcount #1: #2 
\else \hbox to\hsize{\hfil\box\@tempboxa\hfil} 
 \fi}
\def\enddocument{
\ifnum\figcond>0
  \ifnum \figsty>0 \figepsfout
\fi\fi
\@checkend{document}\clearpage\begingroup  
\if@filesw \immediate\closeout\@mainaux 
\def\global\@namedef##1##2{}\def\newlabel{\@testdef r}%
\def\bibcite{\@testdef b}\@tempswafalse \makeatletter\input \jobname.aux
\if@tempswa \@warning{Label(s) may have changed.  Rerun to get
cross-references right}\fi\fi\endgroup\deadcycles\z@\@@end}
\catcode`\@=12
\capsize{30}

\newcommand{\ba}{\begin{array}}  
\newcommand{\ea}{\end{array}}  
\newcommand{\bea}{\begin{eqnarray}}  
\newcommand{\eea}{\end{eqnarray}}  
\newcommand{\be}{\begin{equation}}  
\newcommand{\ee}{\end{equation}}  
\newcommand{\gapproxeq}{\lower .7ex\hbox{$\;\stackrel{\textstyle
>}{\sim}\;$}}  
\newcommand{\lapproxeq}{\lower .7ex\hbox{$\;\stackrel{\textstyle
<}{\sim}\;$}}  

\def\1N{\displaystyle{1/N_c}}

\def\pp{\pi\pi}  
  
\def\mpp{m_{\pi}}

\def\il{^I_l}

\def\Ds{{D\hskip -2.80mm/}\, }

\usepackage{epsfig}
\usepackage{cite}
\usepackage{latexsym}

\begin{document}
%
\pagenumbering{roman}
\pagestyle{plain}
\include{hephead}

\addcontentsline{toc}{chapter}{Introduction}
  \begin{center}
     {\bf INTRODUCTION\\[2em]}
  \end{center}
\begin{quote}
\input{INTRODUCTION}

\end{quote}
\newpage
\addcontentsline{toc}{chapter}{Acknowledgments}
\input{acknowledgments}

\pagenumbering{arabic}
\pagestyle{headings}
\part{Chiral Resonance Model}
\include{chapterl1}

\include{chapterl2}

\include{chapterl3}

\part{Heavy Systems}
\include{chapterh1}

\include{chapterh2}
\include{chapterh3}

\addcontentsline{toc}{part}{Appendices}
\appendix

\input appendixtot

%
\flushleft
\addcontentsline{toc}{part}{Bibliography}

\include{bibliography}
\end{document}

%% file: hephead.tex
%
%
\newpage
\thispagestyle{empty}
\begin{flushright}
\begin{minipage}{5cm}
\begin{flushleft}
\small
\baselineskip = 13pt
YCTP--P19--97\\
hep-ph/yymmnn \\
\end{flushleft}
\end{minipage}
\end{flushright}
	
\begin{center}
{\Large \bf EFFECTIVE LAGRANGIAN MODELS FOR GAUGE THEORIES OF FUNDAMENTAL 
INTERACTIONS}
\end{center}
\begin{center}
\large
Francesco {\sc Sannino}
\footnote{\tt sannino@zen.physics.yale.edu}
\end{center}
\begin{center}
Dept. of Physics, Yale University, New Haven, CT 06520-8120
\end{center}
\begin{center}
\bf Abstract
\end{center}
\begin{quote}
\small
\baselineskip = 14pt
In this thesis we show that the effective Lagrangian models, encoding 
the relevant symmetries of the underlying fundamental gauge theory for 
strong interactions (QCD), 
provide a reasonable understanding of the interactions among light mesons 
at intermediate energies as well as of the properties of Heavy baryons. 
In the first part of the thesis we show  
that it is possible 
to build a chiral and crossing symmetric 
model for light meson interactions at intermediate 
energies, which 
we have called the Chiral Resonance Model (ChRM). 
Using the previous scheme we can understand  
 $\pi\pi-$scattering up to the $1~GeV$ region by considering the resonance 
exchange together with the contact term contributions. 
We also observe that to fully describe 
low energy
$\pi\pi-$scattering in the Chiral Resonance framework a 
broad scalar $\sigma(550)$ particle is needed.  
In the second part of the thesis we investigate 
the heavy baryon spectra in the bound state picture. 
In this picture the heavy baryon is treated as a heavy spin multiplet 
of mesons ($Q\bar{q}$) bound in the background field of the nucleon ($qqq$), 
which in turn arises as a soliton configuration of light meson fields. 
We show that a relativistic 
model with light vectors gives a very satisfactory 
account of the $\Sigma_c^* - \Sigma_c$ hyperfine splitting in contrast 
to the model without light vectors. 
In the last chapter we also present a generalization 
of the bound state model. Indeed by binding heavy spin excited multiplets 
to the background Skyrmion field we can 
describe the spectrum of excited heavy baryons of arbitrary spin.
\end{quote}
\newpage

\thispagestyle{empty}
\begin{center}
\begin{quote}
\it 
Patet omnibus veritas, nondum est occupata.\\ 
Multum ex illa etiam futuris relictum est. \\ 
Vale.\\
\small\
\rm
Lucius Annaeus Seneca, ``Ad Lucilium Epistulae Morales''. 
\end{quote}
\end{center}
\vskip 1cm
\begin{center}
\begin{quote}
\it 
Truth lies open for all; it has not yet been monopolized.\\
 And there 
is plenty of it left 
even for posterity 
to discover. \\
Farewell. \\
\small
\rm
Lucius Annaeus Seneca, ``Letters addressed to Lucilius''.
\end{quote}
\end{center}
\vskip 1cm
\begin{center}
\begin{quote}
\it 
La verit\`a \`e accessibile a tutti,  
non \`e dominio riservato di nessuno,\\ 
e il campo che essa lascia ai posteri 
\`e ancora vasto. \\
Addio.\\
\small
\rm
Lucius Annaeus Seneca, ``Lettere a Lucilio''.
\end{quote}
\end{center}

\newpage

\thispagestyle{empty}
\begin{center}
\Large
To
{\it Lucia}
and 
{\it my parents}
\end{center}
\newpage

\tableofcontents

\listoffigures
\listoftables

%% file: INTRODUCTION.tex
The non abelian gauge theory which describes, in the perturbative regime, 
the strong interactions is 
Quantum Chromodynamics (QCD). Quarks and gluons 
are the fundamental degrees of freedom of the theory. 
A key feature of the 
theory (due to quantum corrections) is asymptotic freedom, i.e. the 
strong coupling constant increases as the energy scale of interest 
decreases. The perturbative approach becomes unreliable below a 
characteristic scale of the theory ($\Lambda$). Quarks and gluons 
confine themselves into colorless particles called hadrons 
(pions, protons,..). 
The latter are the true physical
 states of the theory.  

We need to investigate alternative ways to describe strong interactions, 
and in general any asymptotically free theory, in the non perturbative regime. 
This is the fundamental motivation of the present thesis. Although the 
underlying gauge theory cannot be easily treated in the non perturbative 
regime we can still use its global symmetries as a guide to build Effective 
Lagrangian Models. These models will be written directly in terms of the 
colorless physical states of the theory, i.e. hadrons. 
Two relevant, approximate symmetries are of extreme theoretical as well 
as phenomenological interest: Chiral Symmetry and Heavy Spin Symmetry. 
 Chiral Symmetry is the approximate symmetry 
associated with the light quarks $u$, $d$ and $s$, when their 
masses $m_q$ 
are considered to be negligible compared to the invariant scale of the theory 
 $\Lambda$. Heavy Spin Symmetry on the other hand is associated 
with the heavy quarks $b$, $c$, and $t$, whose masses $M_Q$ are large 
with respect to $\Lambda$. 
These symmetries greatly reduce the number of unknown parameters and 
provide a systematic expansion of the effective Lagrangians in $m_q$ 
and $1/M$. 
Another useful guide for the construction of effective Lagrangians 
will be the expansion for large number of colors $N_c$ 
of QCD. 
In this thesis we will show that the effective Lagrangian models, encoding 
the relevant symmetries of the underlying fundamental gauge theory, 
provide a reasonable understanding of the interactions among light mesons 
at intermediate energies as well as of the properties of Heavy baryons. 

 In part I of the thesis we will 
focus our attention on the physics related to the   
three light quark flavors $u$, $d$ and $s$.  In part II we will consider 
the physics associated with the heavy quarks.   

In the first part of the thesis we will show  
that it is possible 
to build a reasonable chiral and crossing symmetric 
model for light meson interactions at intermediate 
energies, which 
we have called the Chiral Resonance Model (ChRM). Using the previous scheme 
we will demonstrate that we can understand  
 $\pi\pi-$scattering up to the $1~GeV$ region by considering the resonance 
exchange together with the contact term contributions and by 
employing a suitable
regularization procedure. We will also observe that to fully describe 
low energy
$\pi\pi-$scattering in the Chiral Resonance framework a 
broad scalar $\sigma(550)$ particle is needed. At the time when the 
research work was done this state was not present in the Particle data 
Group review (PDG). In the latest PDG this state is finally present. Although 
the parameters associated with the $\sigma$ are still not well 
known the quoted range of parameters (mass and width) is consistent 
with the one determined using the Chiral Resonance Model.  
We will also see that the resonance $f_0(980)$ has to be introduced 
with a mechanism \`a la Ramsauer-Townsend.
It seems likely that any crossing symmetric
approximation will have a similar form. We will regard the ChRM as 
a leading order $\1N$ {\it mean field} approximation for 
Quantum Chromodynamics. 

In the second part of the thesis we will investigate 
the heavy baryon spectra in the bound state picture. 
In this picture the heavy baryon is treated as a heavy spin multiplet 
of mesons ($Q\bar{q}$) bound in the background field of the nucleon ($qqq$), 
which in turn arises as a soliton configuration of light meson fields. 
A nice feature of this approach is that it permits, in principle, 
an exact expansion of the heavy baryon properties in simultaneous 
powers of $1/M$ and $1/N_c$, where $M$ is the heavy quark mass. 
We will show that a relativistic 
model with light vectors gives a very satisfactory 
account of the $\Sigma_c^* - \Sigma_c$ hyperfine splitting in contrast 
to the model without light vectors. We will also show that the 
source of hyperfine splitting is hidden in non manifest heavy spin 
breaking terms. In the last chapter we will present a generalization 
of the bound state model. Indeed by binding heavy spin excited multiplets 
to the background Skyrmion field we will see that we can 
describe the spectrum of excited heavy baryons of arbitrary spin.

%% file: acknowledgments.tex
\vskip 2cm
\begin{center}
\textbf{Acknowledgments}
\end{center}
\begin{quote}
I am deeply grateful to my mentor, 
Prof.~Joseph~Schechter for his excellent academic guidance,  
invaluable advice, and for strongly 
supporting and encouraging 
 me during my research work. 
I greatly benefited from the long afternoon discussions about 
the enchanting world of physics. 

I am very happy to thank Dr.~Masayasu Harada and Dr.~Herbert Weigel, 
for being very valuable   
collaborators, for their suggestions and for helpful discussions. 
I would like to thank Asif Qamar for his collaboration. 

I am indebted to Deirdre Black for careful reading of the thesis. 

I would like to express my gratitude to 
the Theoretical High Energy group at Syracuse 
University for being always very supportive and for providing  
a very fertile research environment.
 
I would like to thank the Dipartimento di Fisica Teorica of Napoli University 
(Italy) for always supporting me and for 
allowing me to do my research work at Syracuse University,  
the Italian National Institution for Nuclear Physics (INFN) and 
the Italian Doctoral Program for partial financial support.

\end{quote}
\newpage

%% file: chapterl1.tex
\chapter{Introduction to the Chiral Resonance Model}
\label{chi}

\section{Brief review of QCD, Chiral Symmetry and $1/N_c$}

The non abelian theory which describes, in the perturbative regime,
 the strong interactions is Quantum Chromdynamics (QCD). Quarks 
and gluons are the associated
 degrees of freedom\cite{Fritzsch-GellMann,Marciano-Pagels}.
SU(3) is the non abelian gauge group of QCD.  
We will denote by $q^{\alpha}_c$ 
the fermionic matter fields associated with {\it quarks}. 
 $\alpha=u,~d,~s,...$  is the flavor index, while $c=1,2,3$
is the color index. The gauge bosons ({\it gluons}) belong to the 
adjoint representation of the gauge group ($G^a_{\mu},~a=1,\ldots,8$). 
The classical QCD Lagrangian density (i.e. which does not include 
quantum corrections) is  
\be
L_{color}=-\frac{1}{4} G^{\mu \nu}_{a} G_{\mu \nu}^{a} + 
\sum_{\alpha}\bar{q}^{\alpha}_c\left(i \Ds^{cd} -
m^{\alpha}\delta^{cd} \right) q^{\alpha}_d \ ,
\label{cromodinamica}
\ee
where the color indices are summed via the Einstein convention. 
The field strength tensor is
\be
G^{a}_{\mu \nu}=\partial_{\mu} G^{a}_{\nu} -\partial_{\nu} G^{a}_{\mu}
- g_s f^{abc}G_{b{\mu}} G_{c{\nu}} \ ,
\ee
where $g_s$ is the coupling constant and $f^{abc}$ are the group 
structure constants.
The covariant derivative acting on a single quark is 
\be
D_{\mu}q=(\partial_{\mu} + i g_s G^{a}_{\mu} T^{a})q \ ,
\ee
where  $T^{a}$ are the SU(3) generators.
A key feature of the theory, due to quantum corrections, is  
{\it asymptotic freedom}, i.e. the strong coupling constant $g_s$, 
decreases as the renormalization scale  ($\mu$) increases
\cite{Gross-Wilczek-a,Gross-Wilczek-b,Politzer}.
Indeed at first order in perturbation theory quantum corrections provide 
\be
\alpha_s(\mu)=\frac{6 \pi}{(33 - 2
N_f)\ln\left(\frac{\mu}{\Lambda}\right)} + \cdots \ ,
\label{running}
\ee where $N_f$ is the number of flavors whose masses are less than $\mu$ and  
$\Lambda$ is a renormalization invariant scale while,  
$\displaystyle{g^2_s=4\pi \alpha_s}$.  
It is clear that the perturbative approach is not reliable when 
 $\mu \rightarrow \Lambda$. Hence Eq.~(\ref{running}) itself 
cannot be trusted in this regime. However,  
a commonly accepted working hypothesis is that the strong coupling constant  
keeps increasing at low energies, leading to the phenomenon of 
quark and gluon {\it confinement}. 
In Eq.~(\ref{running}) $\alpha_s$ is a function of $\Lambda$. So the 
latter can actually be substituted for the dimension free quantity  
$\alpha_s$ in any result  deduced using QCD. This phenomenon is called 
{\it dimensional trasmutation}.

We need to investigate alternative ways to describe strong interactions, 
and in general any {\it asymptotically free} theory, in the non perturbative 
regime. This is the fundamental motivation of the present thesis. 

It is, however, possible to deduce more information from QCD, by 
studying it for some limiting values of its parameters. The parameters
 which we can tune are $\alpha_s$ (i.e.
$\Lambda$), the flavor number, the quark masses and the color number 
\footnote{Among these parameters (for completness) we must mention 
the   
$\Theta-vacuum$ parameter, associated with the strong CP 
(Charge times Parity)  violation. 
The experimental determination of the neutron 
dipole moment provides a very stringent upper bound, i.e. 
 $\Theta \leq 2 \times 10 ^{-10}$.  
Hence, in the following discussion we will not consider it.}. 

For the purposes of the research work presented in part I we will consider 
only the lightest three flavors $u,~d$ and $s$ ($N_f=3$). In part II 
we will consider heavy quark physics. 
The QCD Lagrangian density in the limit of zero mass for the light 
quarks gains the classical symmetry $U_L(3)\otimes U_R(3)$. 
The quark fields will transform as 
(for convenience we omit the color indices)  
\be
q_L^{\alpha}\rightarrow {g_L}^{\alpha}_{\beta}q_L^{\beta}\ ,
\quad 
q_R^{\alpha}\rightarrow {g_R}^{\alpha}_{\beta}q_R^{\beta} \ ,
\ee
where $g_{L,R} \in U_{L,R}(3)$ acts, respectively, 
on the $Left$ and  $Right$ components of the Dirac spinors.

The classical symmetry $U_L(3)\otimes U_R(3)$ is {\it explicitly} broken 
via quantum corrections, i.e.
\be
U_L(3)\otimes U_R(3) \longrightarrow U_V(1)\otimes SU_L(3)\otimes SU_R(3)\ ,
\ee
This phenomenon is called the chiral anomaly.
$U_V(1)$ is associated with baryon number conservation. 
Invariance 
under the non abelian symmetry group $SU_L(3)\otimes SU_R(3)$ predicts 
that all the physical states must be classified according to 
irreducible representations of the group. 
Nature does not realize such a physical 
spectrum and it is assumed that the symmetry is spontaneously broken via
\be
 SU_L(3)\otimes SU_R(3)\stackrel{SSB}{\longrightarrow} SU_V(3) \ .
\ee
The Nambu-Goldstone theorem states that the lost symmetry must be compensated 
by the presence of massless bosons (Goldstone bosons). In QCD 
such Goldstone bosons, associated 
with the spontaneously broken chiral symmetry, can be identified with the 
octet of pseudoscalar mesons ${\pi, K, \bar{K},\eta }$.
All of the other particles can be classified according to the irreducible 
representations of the diagonal subgroup $SU_V(3)$.

Now we will consider the large $N_c$  limit of Quantum Chromodynamics 
\cite{1n}. 
The expansion in $\1N$ will allow us to deduce much new information 
whose key feature is its validity in the non perturbative as well as 
perturbative regime of QCD. 
In this paragraph we will review only some of the $\1N$ predictions, 
the reader will find more results in \cite{1n}. In particular 
here we will consider only the key features relevant for part I of the 
thesis. 

For  $N_c\rightarrow\infty$, keeping constant the following product 
$\alpha_s N_c$ one can prove the following meson properties. 

\begin{itemize}
\item [a] 
At the leading order in $\1N$ the two point Green's functions 
  built out of local quark bilinear operators 
($J$= $\bar{q} q,~\bar{q}\gamma_{\mu} q$, etc.) are saturated by the 
exchange of an infinite number of non exotic meson resonances, 
i.e. of the $\bar{q} q $ type.
\be
<J(k)J(-k)>=\sum_{n}^{\infty}\frac{a_n^2}{k^2 - m^2_n}\ .
\label{2-point}
\ee
 $m_n$ is $n^{\rm th}$ meson mass,  and  $a_n=<0|J|n>$ is the 
 $J$ matrix element which creates the $n^{\rm th}$ meson from the vacuum. 


\item[b] 
The meson-meson scattering amplitude described by a 4-point Green's function, 
schematically represented in Fig.\ref{figura1n}(a), is 
$O(\1N)$. So we can deduce that the interactions among the mesons 
are subleading in the $\1N$ expansion.

\item[c] Non exotic mesons are stable. 
The 3-point Green function associated with the decay amplitude is 
represented in Fig.\ref{figura1n}(b) and is   
 $O(1/ \sqrt{N_c})$. Hence the decay itself is  $O(\1N)$. 

\end{itemize} 
\figinsert{figura1n.eps}{(a) schematically represents the 4-point Green 
function. (b) represents the 3-point Green function.
 In (a) and (b) the external lines represent a quark {\it loop}, 
while ($\bullet$) are the insertions of the non exotic meson fields.}
{figura1n}{hptb}

The chiral limit together with the $\1N$ expansion are the two 
 ingredients motivating the chiral resonance model which 
we will describe in the following paragraph.

\section{Chiral Resonance Model}       

Asymptotic freedom strongly reduces the predictive power of QCD, 
in particular for the description of strong interactions among 
mesons at low energies. Chiral perturbation theory ($\chi$PT) \cite{chp}  
allowes a systematic investigation of the strong interactions 
of the octect of Goldstone bosons ($\pi, K, \overline{K}, \eta$). 
$\chi$PT, essentially, relies on the expansion in energies of the 
scattering amplitudes, and it improves the Born terms of the 
Chiral Lagrangian by including quantum corrections via $loops$ and 
necessary counterterms.     
The energy expansion converges very fast for energies less than 
 $400-500~MeV$. It is a very hard task to extend the 
$\chi$PT scheme 
at higher energies, since the physical singularities associated with 
the resonances which exist in this region cannot be easily reproduced 
by a truncated (in energy) power expansion.

In order to describe the interactions among light mesons up to 
an energy (in the center of mass) of $1-1.5$ GeV is clear 
that one is forced to include the effects of these resonances in this
region.  

The question is: How to include such resonances ? 
As noted in the previous paragraph the meson-meson scattering 
amplitudes in the leading order in $\1N$ can be obtained by 
summing up all possible tree diagrams obtained via an effective 
Lagrangian density, which contains an {\it infinite} number of 
bosonic resonances per each possible spin. An {\it infinite} number of 
contact terms is also allowed \cite{1n}. 

The  $1/N_c$ analysis is very interesting, since it supports the 
Lagrangian effective models as reasonable models, but at  the moment 
is so general as to appear useless.  
We will see that this is not the case. 

Using the $\1N$ analysis as a guide we will try to build a {\it new} effective model 
\cite{Sannino-Schechter,Sanninomrst,Harada-Sannino-Schechter}
to describe the meson scattering  up to about $1$ GeV.

A scattering amplitude built using an effective Lagrangian, automatically 
satisfies {\it crossing} symmetry. On the other hand just calculating 
the tree approximation to an effective Lagrangian will not guarantee that 
unitarity is satisfied. This is the handle we will use to try to investigate 
additional structure. Unitarity has of course the consequence that 
the amplitude must have some suitable imaginary term which in the usual 
field theory is provided  by loop diagrams. However the leading $\1N$ 
approximation will give a purely real amplitude away from the 
singularities at the direct $s-channel$ poles. 
We may consider the imaginary part of the leading $\1N$ amplitude to consist 
just of the sum of delta functions at each such singularity. Clearly, 
the real part has a much more interesting structure and we will mainly 
confine our attention to it. 
   
Unitarity has the further consequence that the real parts of the partial 
wave amplitudes must satisfy certain well known bounds. 
The crucial 
question is how these bounds are satisfied since, as will see, individual 
contributions tend to violate them badly. At first one might expect that 
all of the infinite number of resonances are really needed to obtain 
cancellations. However the success of chiral dynamics at very low 
energies where none of the resonances have been taken into account suggests 
that this might not be the case. 
It is clear that we need to postulate a new  principle which has 
been introduced in Ref.~\cite{Sannino-Schechter} and we will call it   
{\it local cancellation}. 
{\it Local cancellation} cannot be easily derived from QCD, but if 
true it would greatly simplify the task of extending the phenomenological 
description of scattering processes to higher energies. 

According to this simple principle the inclusion of the resonances, up to 
the energy one wants to describe, in the direct channel as well as in the 
cross channels, will allow us to saturate the unitarity bounds and also 
will allow us to describe the phenomenology involved. 
Now one can easily understand the relevant role played by  
{\it crossing} symmetry, already pointed out in Veneziano's famous paper 
\cite{string}.
Indeed, as we will see in the next chapter, local cancellations among 
the different contributions from the resonances will, non trivially,
 enforce unitarity.  
In $\pi\pi$ scattering, we will see that the introduction of the 
cross term due to $\rho$ meson exchange in a chirally invariant manner,
 in the $Isospin=0$ and 
zero orbital angular momentum case, 
substantially delays the onset of the severe unitarity 
violation which would be present in the simplest chiral Lagrangian of pions. 
  
The onset of the unitarity bound via the {\it local cancellation} 
principle will also require the existence of the scalar particle $\sigma(550)$
\cite{Sannino-Schechter,Sanninomrst,Harada-Sannino-Schechter} 
in order to succesfully describe $\pi\pi-$scattering up to about 1.2~GeV.

In order to start this investigation it seems reasonable to keep in the 
Lagrangian only local operators with the lowest number of derivatives. 
Operators with a higher number of derivatives destabilize the theory 
at high energy and will make it more difficult to accomplish the  required 
{\it local cancellation}. 
Further restrictions will be obtained by imposing chiral symmetry. 

In the last paragraph of the present chapter we will show how to include 
the scalar, vectorial and tensorial resonances in order to preserve 
chiral symmetry.

We will now summarize the chiral resonance model according to the 
following logical scheme.

\begin{itemize}
\item[a]{The effective Lagrangian will contain only terms with the 
lowest number of derivatives. These local 
operators will be chirally symmetric.}

\item[b]{The model will be suitable for the real part 
of the physical amplitude. Such an 
amplitude must obey {\it crossing} symmetry.}

\item[c]{We will include in the scheme only the resonances whose masses 
lie in the energy range of interest. Away from the poles in the direct 
channel unitarity is expected to be provided by the 
{\it local cancellation} principle.}

\item[d]{A suitable regularization method must be employed to 
avoid divergences at each direct channel pole for the physical amplitude.}

\end{itemize}

In the next paragraph we will explain how to regularize the 
$s-$channel poles. 
The other points will be investigated in the next two chapters. In 
particular we will study the  $\pi\pi \rightarrow \pi \pi$ and 
 $\pi\pi \rightarrow K \bar{K}$ scattering amplitudes. 
These two channels are a classically important check for any model 
whose goal is to describe meson interactions at intermediate energies. 

\section{Regularization of the Model}
In the large $N_c$ picture the leading 
amplitude (of order $\displaystyle{1/N_c}$) is a sum 
of polynomial contact terms and tree-type resonance exchanges. Furthermore 
the resonances should be of the simple $q\overline{q}$ type; 
glueball and multi-quark 
meson resonances are suppressed. In our phenomenological model there is 
no way of knowing {\it a priori} whether a given experimental state is 
actually of the $q\overline{q}$ type. 
For definiteness we will keep all relevant 
resonances even though the status of a low lying scalar resonances like 
the $f_0(980)$ has been considered especially controversial 
\cite{Tornqvist:95}. If such
resonances turn out 
in the future to be not of type $q\overline{q}$, their tree 
contributions would be 
of higher order than $\displaystyle{1/N_c}$. In this event the 
amplitude would still of course 
satisfy crossing symmetry.

The most problematic feature involved in comparing the leading 
$\displaystyle{1/N_c}$ amplitude 
with experiment is that it does not satisfy unitarity. In fact, resonance 
poles like 
\begin{equation}
\frac{1}{M^2-s}
\label{propagator}
\end{equation}
will yield a purely real amplitude, except at the singularity, where 
they will diverge and drastically violate the unitarity bound. 
Thus in order 
to compare the $\displaystyle{1/N_c}$ amplitude with experiment 
we must regularize the denominators in some way. The usual method, as 
employed in Ref.~\cite{Sannino-Schechter}, is to regularize 
the propagator so 
that the resulting partial wave amplitude has the
 locally unitary form
\begin{equation}
\frac{M\Gamma}{M^2-s-iM\Gamma}\ .
\label{Breit-Wigner}
\end{equation}
This is only valid for a narrow resonance in a region where the 
{\it background} is negligible. Note that the $-iM\Gamma$ is strictly 
speaking a higher order in $\displaystyle{1/N_c}$ effect. 

For a very broad resonance  
there is no guarantee that such a form is correct. Actually, in 
Ref.~\cite{Sannino-Schechter} it was found necessary 
to include a rather broad 
low lying scalar resonance (denoted $\sigma(550)$) to avoid violating the 
unitarity bound. A suitable form turned out to be of the type 
\begin{equation}
\frac{M G}{M^2-s-iMG^\prime}\ ,      
\label{sigma-propagator}
\end{equation}
where $G$ is not equal to the parameter $G^\prime$ which was
introduced to regularize the propagator. Here $G$ is the quantity
related to the squared coupling constant. 

Even if the resonance is narrow, the effect 
of the background may be rather important. 
Demanding local 
unitarity in this case yields a partial 
wave amplitude of the well known form \cite{Taylor}:
\begin{equation}
\frac{e^{2i\delta}M\Gamma}{M^2-s-iM\Gamma}+e^{i\delta}\sin \delta\ ,      
\label{rescattering}
\end{equation}
where $\delta$ is a background phase (assumed to be slowly varying). 
We will see in the next chapters that such a regularization method 
is needed to fully understand the $\pi\pi\rightarrow\pi\pi$ and 
 $\pi\pi\rightarrow K\bar{K}$ scattering in the $f_0(980)$ region. 
We will adopt a point of view in which this form is regarded as a kind of 
regularization of our model. Of course, non zero 
$\delta$ represents a rescattering  
effect which is of higher order in $\displaystyle{1/N_c}$. The quantity 
$\displaystyle{e^{2i\delta}}$, taking $\delta=constant$, can be  
incorporated, for example, into the squared 
coupling constant connecting the resonance to two 
pions. In this way, crossing symmetry can be preserved. {}From its origin, it 
is clear that the complex residue does not signify the existence of a 
{\it ghost} particle. The non-pole background term in 
Eq.~(\ref{rescattering}) and hence $\delta$ 
is to be predicted by the other pieces in the effective Lagrangian. 

Another point which must be 
addressed in comparing the leading $\displaystyle{1/N_c}$ amplitude 
with experiment is that it is purely real away from the singularities. The 
regularizations mentioned above do introduce some imaginary pieces but these 
are clearly very model dependent. Thus it seems reasonable to compare the 
real part of our predicted amplitude with the real part of the experimental 
amplitude. Note that the difficulties mentioned above arise only for 
the direct channel poles; the crossed channel poles and contact terms will 
give purely real finite contributions.

It should be noted that if we predict the real part of the amplitude, the 
imaginary part can always be recovered by assuming elastic unitarity.


\section{Chiral Lagrangian}
In the low energy physics of hadrons 
it is important to correctly introduce   
the spontaneous chiral symmetry breaking structure. In our approach 
we will introduce it via non linear realizations. It is known that 
this method reproduces the low energy theorems obtained via 
current algebra. 
We start here with the definition of the 3 $\times$ 3 matrix $U$,
\begin{equation}
 U = \xi^2 \ , \qquad \xi = e^{ i\phi/F_\pi} \ ,
 \end{equation}
 where $F_\pi$ is a pion decay constant. $U$ is parameterized by 
the pseudoscalar matrix $\phi$, which is identified 
with the pseudoscalar meson octect. Under the chiral group 
  $U(3)_{\rm L}\times U(3)_{\rm R}$, $U$ transforms linearly.  
 \begin{equation}
 U \rightarrow g_{\rm L} U g_{\rm R}^{\dag} \ ,
 \label{trans: U}
 \end{equation}
 where $g_{\rm L,R} \in \mbox{U(3)}_{\rm L,R}$.
 Under the chiral transformation Eq.~(\ref{trans: U}),
 $\xi$ transforms non-linearly:
 \begin{equation}
 \xi \rightarrow 
 g_{\rm L} \, \xi \, K^{\dag}(\phi,g_{\rm L},g_{\rm R}) = 
 K(\phi,g_{\rm L},g_{\rm R}) \, \xi \, g_{\rm R}^{\dag} \ .
\label{nonlinear} 
\end{equation}
The previous equation implicitly defines the matrix 
$K(\phi,g_{\rm L},g_{\rm R})$.
The vector meson nonet $\rho_\mu$ is introduced as a 
 {\it gauge field} \cite{Kaymakcalan-Schechter}
 which transforms as
 \begin{equation}
 \rho_\mu \rightarrow K \rho_\mu K^{\dag} - 
 \frac{i}{\widetilde{g}} K \partial_\mu K^{\dag} \ ,
 \end{equation}
 where $\widetilde{g}$ is a {\it gauge coupling constant}.
 (For an alternative approach see, for a review, 
 Ref.~\cite{Bando-Kugo-Yamawaki:PRep}.)
 It is convenient to define
 \begin{eqnarray}
 p_\mu &=& \frac{i}{2}
 \left( 
   \xi \partial_\mu \xi^{\dag} - \xi^{\dag} \partial_\mu \xi
 \right) \ ,
 \nonumber\\
 v_\mu &=& \frac{i}{2}
 \left( 
   \xi \partial_\mu \xi^{\dag} + \xi^{\dag} \partial_\mu \xi
 \right) \ ,
\label{maurer} 
\end{eqnarray}
 which transform as
 \begin{eqnarray}
 p_\mu &\rightarrow& K p_\mu K^{\dag} \ , \nonumber\\
 v_\mu &\rightarrow& K v_\mu K^{\dag} + i K \partial_\mu K^{\dag} \ ,
 \end{eqnarray}
and are the chiral group $Maurer-Cartan$ one form. 
 Using the above quantities
 we construct the chiral Lagrangian including both pseudoscalar and
 vector mesons:
 \begin{equation}
{\cal L} =
 +\frac{1}{2} m_v^2 \mbox{Tr} 
 \left[ \left( \widetilde{g}\rho_\mu + v_\mu \right)^2 \right]
 + \frac{F_\pi^2}{2} \mbox{Tr} 
 \left[ p_\mu p^\mu \right]
 -\frac{1}{4} \mbox{Tr} 
 \left[ F_{\mu\nu}(\rho) F^{\mu\nu}(\rho) \right],
 \label{Lag: sym}
 \end{equation}
 where
 $F_{\mu\nu} = \partial_\mu \rho_\nu - 
 \partial_\nu \rho_\mu + i \widetilde{g} 
 [ \rho_\mu , \rho_\nu ]$
 is a {\it gauge field strength} of vector mesons. 
  $\widetilde{g}$ is connected to the vector meson coupling to two pions 
 $g_{\rho \pi \pi}$ via 
\begin{equation}
g_{\rho \pi \pi}=\frac{m^2_{\rho}}{\widetilde{g}F^2_{\pi}} \ .
\end{equation}
 In the real world chiral symmetry is explicitly broken by the quark mass term
 $- \widehat{m} \overline{q} {\cal M} q$,
 where $\widehat{m} \equiv (m_u+m_d)/2$,
 and ${\cal M}$ is the dimensionless matrix:
 \begin{equation}
 {\cal M} = \left(
 \begin{array}{ccc}
 1+y & & \\ & 1-y & \\ & & x
 \end{array} \right) \ .
 \end{equation}
 Here $x$ and $y$ are the quark mass ratios:
 \begin{equation}
 x = \frac{m_s}{\widehat{m}} \ , \qquad
 y = \frac{1}{2}
 \left(\frac{m_d-m_u}{\widehat{m}}\right) \ .
 \end{equation}
 These quark masses lead to mass terms for pseudoscalar mesons.
 Moreover,
 in considering the processes related to the kaon,
 (in this thesis we will consider 
 $\pi\pi\rightarrow K \overline{K}$ scattering amplitude)
 we need to take account of the large mass gap between 
 the $s$ quark mass and 
 the $u$ and $d$ quark masses.
 These effects are included as 
 SU(3) symmetry breaking terms in the above Lagrangian,
 which are summarized, for example, in 
 Refs.~\cite{Schechter-Subbaraman-Weigel,Harada-Schechter}.
 Here we write the lowest order pseudoscalar mass term only:
 \begin{equation}
 {\cal L}_{\phi{\rm-mass}} = 
 \delta' \mbox{Tr} 
 \left[ {\cal M} U^{\dag} + {\cal M}^{\dag} U \right] \ ,
 \label{pi mass 1}
 \end{equation}
 where $\delta'$ is an arbitrary constant. 

 We next introduce higher resonances into our Lagrangian.
 {}Firstly, we write the interaction between the  
 scalar nonet field $S$ and pseudoscalar mesons.
 Under the chiral transformation,
 this $S$ transforms as 
 $S \rightarrow K S K^{\dag}$.
 A possible form which includes the minimum number of  derivatives is 
 proportional to 
\begin{equation}
\mbox{Tr}\left[ S p_\mu p^\mu \right] \ .
\end{equation}
 The coupling of a physical isosinglet field to two pions is then
 described by
 \begin{equation}
 {\cal L_{\sigma}}=+\frac{\gamma_0}{\sqrt{2}}\; \sigma \;
 \partial_{\mu}\vec{\pi}\cdot\partial^{\mu}\vec{\pi}\ . 
 \label{la:sigma}
 \end{equation}
 Here we should note that 
 chiral symmetry requires derivative-type interactions
 between scalar fields and pseudoscalar mesons.
 Secondly, we represent the tensor nonet field by
 $T_{\mu\nu}$
 (satisfying $T_{\mu\nu} = T_{\nu\mu}$, and $T_{\mu}^{\mu}= 0$),
 which transforms as
\begin{equation}
 T_{\mu\nu} \rightarrow K T_{\mu\nu} K^{\dag} \ .
\label{2-transf}
\end{equation} 
The interaction term is given by
 \begin{equation}
 {\cal L}_T = - \gamma_2 F_\pi^2
 \mbox{Tr}\left[ T_{\mu\nu} p^\mu p^\nu \right] \ .
\label{spin-2} 
\end{equation}
 The heavier vector resonances such as $\rho(1450)$ can be introduced
 in the same way as $\rho$ in Eq.~(\ref{Lag: sym}).

In general, for a complete description of the mesonic processes one 
should include the effects due to chiral anomalies. Since in this 
part of the thesis we will not deal with such processes we refer 
the reader to the following references 
\cite{Schechter-Subbaraman-Weigel,Harada-Schechter,Harada-Sannino-Schechter-Weigel}.

%% file: chapterl2.tex
\chapter{A fundamental process:  $\pi\pi$ scattering}
\section{Introduction to $\pi\pi$ scattering}

Historically, the analysis of $\pi\pi$ scattering has been considered an 
important test of our understanding of strong interaction physics 
(QCD, now) at low energies. It is commonly accepted that the key feature is  
the approximate spontaneous breaking of chiral symmetry. Of course, the {\it 
kinematical} requirements of unitarity and crossing symmetry  
should be respected. The chiral perturbation scheme ($\chi$PT) \cite{chp}, 
which improves the tree  
Lagrangian approach by including loop corrections and counterterms, can 
provide a description of the scattering up to the energy region slightly 
above threshold $(400-500~MeV)$. 
Hence $\chi$PT is very useful for describing 
kaon decays into $2\pi$ and $3\pi$ 
(for an application see Ref.~\cite{D'Ambrosio-Miragliuolo-Sannino}). 
In Ref.~\cite{Buccella-Pisanti-Sannino} the derivative expansion has been 
used to estimate some of the parameters associated with CP violation 
for the $3\pi$ decays of neutral kaons. 
In order to describe the scattering up to energies beyond this region (say 
to around $1~GeV$) it is clear that the effects of particles lying in this 
region must be included. The Chiral 
Resonance Model (ChRM) provides a simple prescription for including 
these resonances. We will see that by applying the ChRM recipe we 
will be able to understand the scattering processes schematically 
represented in Fig.\ref{figuraa}.
\figinsert{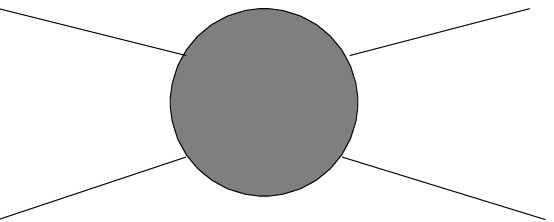}
{Schematic representation of a scattering process.}{figuraa}{hptb}       
The kinematics, the unregularized invariant scattering amplitudes and the 
generic scattering matrix $S$ are presented in Appendix A. 

In the previous chapter we stated that the ChRM predicts the real 
part of the scattering amplitude. 
The imaginary term can be recovered via the unitarity relations. 
 Specializing Eq.~(\ref{real-imaginary}) in Appendix A  to the 
 $\pi \pi$ channel we have, for the imaginary piece $I^I_l$ of the 
 $I$, $l$ partial wave amplitude 
\be
I^I_l=\frac{1}{2}\left[1\pm\sqrt{{\eta^I_l}^2 - 4 {R^I_l}^2} \right]\ ,
\label{imaginary}
\ee
where $\eta^I_l$ is the elasticity parameter. $I$ and $l$
are the isospin and the orbital angular momentum.   
Obviously, this formula is only meaningful if the real part obeys the bound 
\be
|R^I_l|\le \frac{\eta^I_l}{2} \ .
\label{limite}
\ee
The main difficulty one has to overcome in obtaining a unitary amplitude 
by the present method is the satisfaction of this bound Eq.~(\ref{limite}). 

We want to remark that making the regularizations of direct channel poles 
such as in Eqs.~(\ref{Breit-Wigner}) and (\ref{rescattering}) 
which provide unitarity 
in the immediate region of a narrow resonance, is not at all tantamount 
to unitarizing the model by hand. 

In this chapter we will present the first evidence of {\it
local cancellation}, obtained among the current algebra 
and the vector meson $\rho$ contributions.
We will then show that we need to introduce a broad scalar resonance  
(i.e. $\Gamma \gapproxeq M$) to fully understand the scattering 
process at low energies ($\le 800$ MeV). The effect of chirally invariant 
contact terms with a higher number of derivatives will also be investigated.  

\section{Current Algebra and $\rho$ vector--meson exchange}

In this section we will study the partial waves for 
$\pi\pi$ scattering computed
in a chiral Lagrangian model which contains both the pseudoscalar and vector
mesons, (i.e., the lowest lying s-wave quark--antiquark bound states).
Near threshold
$\eta\il=1$, $R\il$ is small and we should choose the minus sign in 
Eq.~({\ref{imaginary}}) so that
\be
{I\il}(s)\approx {[R\il]}^2  .
\label{approssimazione}
\ee
\noindent
In the large $N_c$ limit the amplitude near threshold is purely real 
and of the order
 $\1N$. This is consistent with Eq.~(\ref{approssimazione}) which
shows that $I^I_l(s)$ is of order $1/N_c^2 $ and hence
comes in at the second order. This agrees with the chiral perturbation theory
approach \cite{chp} in which $R^I_l(s)$ comes from the lowest order tree diagram while
$I^I_l$ arises from the next order loop diagram. On the other hand, when we
depart from the threshold region the $\1N$ approach treats the
contribution of the $\rho$-meson at first order while the chiral perturbation
theory approach treats it at second and higher orders.
A straightforward computation using the pion lagrangian 
(second term in Eq.~(\ref{Lag: sym})) 
together with the 
explicit chiral breaking term in Eq.~(\ref{pi mass 1}) 
yields the $\pi\pi$
scattering amplitude \cite{weinbergold} defined in 
Eq.~(\ref{eq:def}):  
\be
A_{CA}(s,t,u)=2\frac{s-m_{\pi}^2}{F_{\pi}^2} \ . 
\label{Aca}
\ee
This equation will be called the {\it current algebra result}. With
(\ref{eq:isospin}) and (\ref{eq:wave}) we obtain $R^0_0(s)=T^0_{11;0}(s)$ 
as illustrated in Fig.~{\ref{fig1bt}}.
The experimental Roy curves \cite{Roy} are also shown.
Up until about $0.5~GeV$ the
agreement is quite reasonable (and can be fine--tuned with second order
chiral perturbation terms) but beyond this point  $R^0_0$ keeps increasing
monotonically and badly violates the unitarity bound (\ref{limite}). 
\figinsert{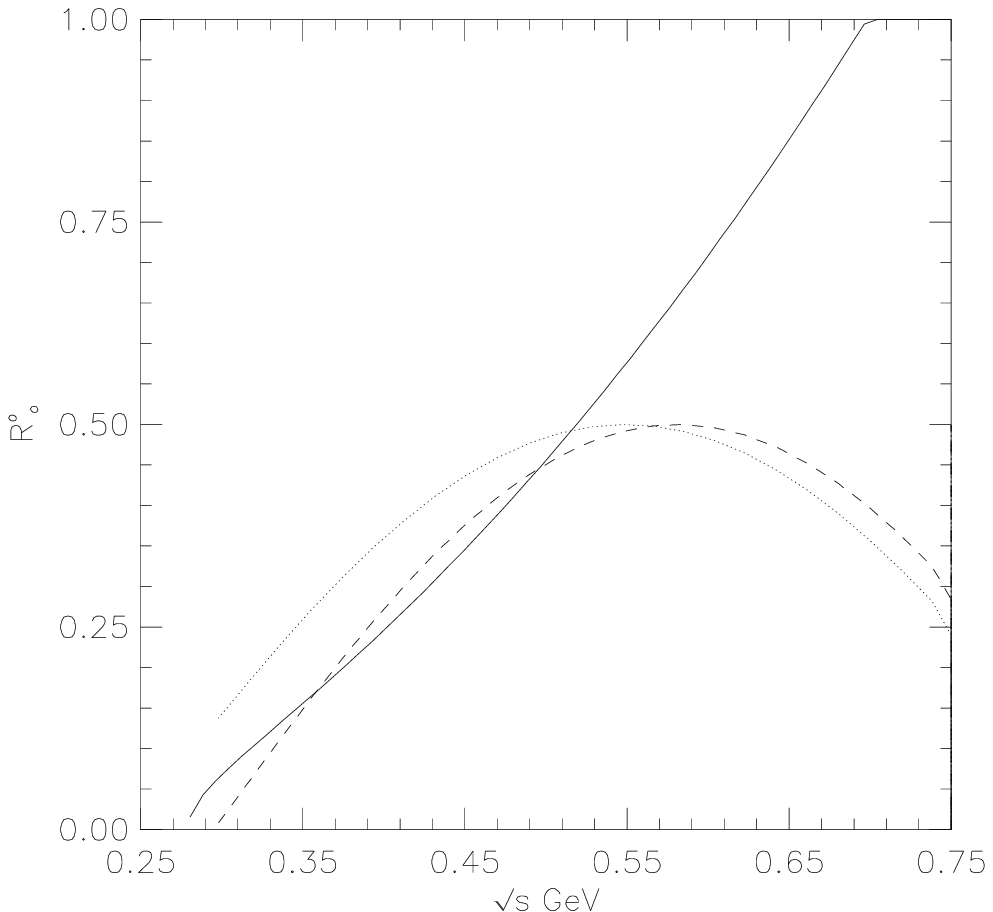}{The solid line is the current algebra 
result for $R^0_0$. The dotted and dot-dashed lines are the Roy curves for 
$R^0_0$.}{fig1bt}{hpbt}
We will
see now that the introduction of the $\rho$-meson greatly 
improves the situation.
The chiral Lagrangian for the vector--meson  pion system is displayed in   
Eq.~(\ref{Lag: sym}). 
The Lagrangian piece in (\ref{Lag: sym}) yields
both a pole-type contribution (from the $\rho_{\mu}v^{\mu}$ cross term)
 and
a contact term contribution (from the $v_{\mu}v^{\mu}$ term) to the
amplitude at tree level:
\be
A_{\rho}(s,t,u)=-\frac{g^2_{\rho \pi \pi}}{2} \left(\frac{u-s}{m_{\rho}^2 -
t} +\frac{t-s}{m_{\rho}^2-u} \right) + \frac{g^2_{\rho \pi \pi}}{2
m_{\rho}^2} \left[(t-s)+(u-s)\right] \ .
\label{Avect}
\ee 
where $m_{\rho}=0.769~GeV$ and $g_{\rho\pi\pi}=8.56$.
We notice that the entire first term in (\ref{Lag: sym}) 
 \be
\mbox{Tr}\left[\left(\tilde{g}\rho_{\mu} + v_{\mu}\right)^2 \right]\ ,
\ee
is chiral invariant
since $v_{\mu}$ and $-\widetilde{g} \rho_{\mu}$ transform identically. However
the $\mbox{Tr}(\rho_{\mu}v^{\mu})$ and 
$\mbox{Tr}(v_{\mu}v^{\mu})$ pieces are not
separately chiral invariant. This shows that the addition of the 
 $\rho$--meson in a chiral invariant manner necessarily introduces a contact term
in addition to the minimal pole term. Adding up the two terms in 
Eq.~(\ref{Avect})
and the term in Eq.~(\ref{Aca}) yields finally
\be
A_{CA+\rho}(s,t,u)=2\frac{s-m_{\pi}^2}{F_{\pi}^2} 
-\frac{g^2_{\rho \pi \pi}}{2 m_{\rho}^2} \left[\frac{t(u-s)}{m_{\rho}^2 -
t} +\frac{u(t-s)}{m_{\rho}^2-u} \right] \ .
\label{Aca-vect}
\ee
In this form we see that the threshold (current algebra) results
are unaffected since the second term drops out at $t=u=0$. An
alternative approach \cite{tensore-vettore} to obtaining Eq.~(\ref{Aca-vect}) involves
introducing a chiral invariant $\rho\pi\pi$ interaction with two more
derivatives.
 $A(s,t,u)$ has no singularities in the physical region. Reference to 
 Eq.~(\ref{eq:isospin})
shows that the isospin amplitudes $T^0_{11}$ and $T^2_{11}$ 
also have no singularities.
 However the $T^1_{11}$ amplitude has the expected singularity at $s=m^2_\rho$.
This may be cured in a conventional way, while still maintaining crossing
symmetry, by the replacements
\be
\frac{1}{m_{\rho}^2 - t,u} \rightarrow        
\frac{1}{m_{\rho}^2 - t,u - i m_{\rho}\Gamma_{\rho}\theta(t,u - 4
m^2_{\pi})} \ .
\label{propagatore-vettore}
\ee
A modification of this sort would enter automatically
if we were to carry the computation to order $\displaystyle{\frac{1}{N^2_c}}$.
However we shall regard (\ref{propagatore-vettore}) as a phenomenological 
regularization of
the leading amplitude.

Now let us look at the actual behaviour of the real parts of the partial wave 
amplitudes. $R^0_0$, as obtained from 
Eq.~(\ref{Aca-vect}), is graphed in  Fig.~\ref{fig2bt} for an extensive 
range of $\sqrt{s}$, together with the {\it current algebra} result.
\figinsert{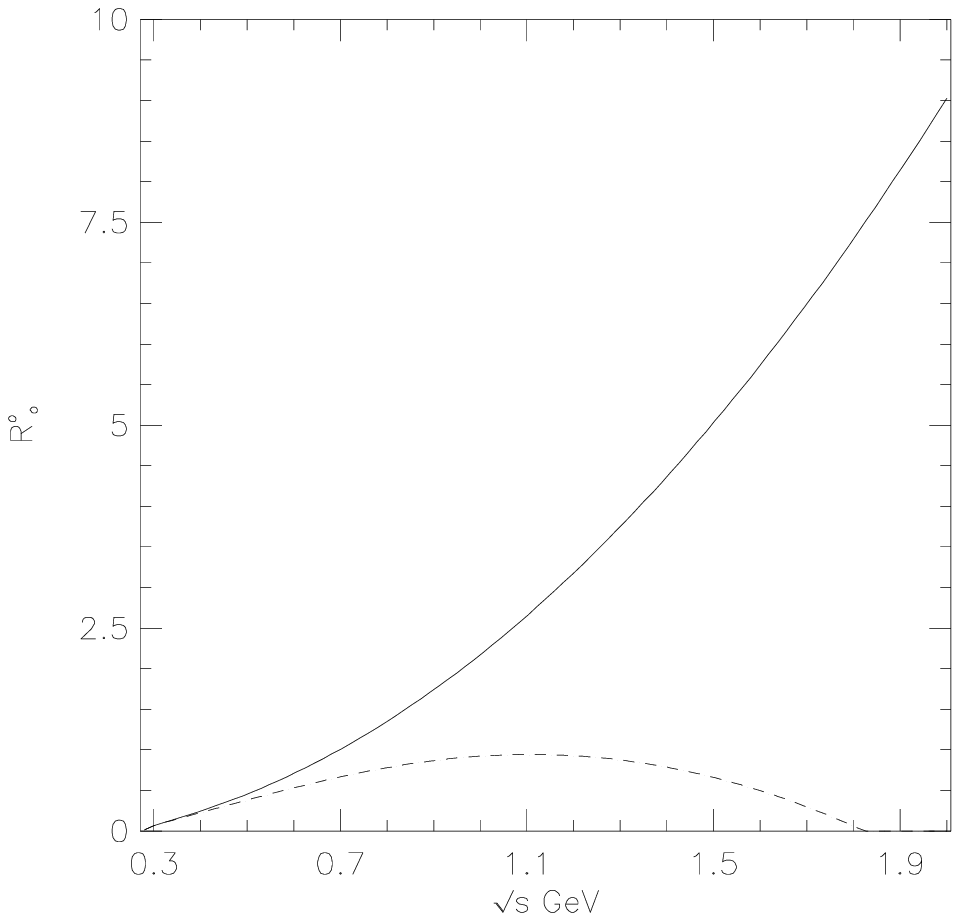}{The solid line  is the current algebra result 
for $R^0_0$. The dot-dashed line is the $\rho + \pi$ for 
$R^0_0$.}{fig2bt}{hpbt}

We immediately see that there is a remarkable improvement; the effect of
adding $\rho$ is to bend back the rising $R^0_0(s)$ so there is no longer a
drastic violation of the unitarity bound until after $\sqrt{s}=2~GeV$.
There is still a relatively small violation which we will discuss later.
Note that the modification in 
Eq.~(\ref{propagatore-vettore}) plays no role in the improvement
since it is only the non--singular $t$ and $u$ 
channel exchange diagrams which
contribute.

It is easy to see that the {\it delayed} drastic violation of the
unitarity bound $\displaystyle{\big|{R\il\big|}\leq\frac{1}{2}}$ is a property
of all partial waves. We have already learned from (\ref{Aca-vect}) that the
amplitude $A(s,t,u)$ starts out rising linearly with $s$. Now Eq.~(\ref{Avect})
 shows (for fixed scattering angle) that for large $s$ the $\rho$ 
exchange terms behave
as $s^0$. The leading large $s$ behavior will therefore come from the sum of
the original {\it current-algebra} term and the new {\it contact-term}:
\be
A_{CA+\rho}(s,t,u)\simeq \frac{2 s}{F^2_{\pi}}\left(1-3\frac{k}{4}\right),
\quad
k\equiv \frac{m_{\rho}^2}{\tilde{g}F^2_{\pi}} \ .
\label{asintoto}
\ee   
But $k$ is numerically around $2$ \cite{Kawa-Suzu}, so $A(s,t,u)$ 
eventually {\it decreases}
linearly with $s$.
This turn-around, which is due to the contact term that enforces chiral
symmetry, delays the onset of drastic unitarity violation until well
after the $\rho$ mass. It thus seems natural to speculate that, as we go up
in energy, the leading tree contributions from the resonances we encounter 
 (including both crossed channel as well as $s$-channel exchange) 
conspire to
keep the $R\il(s)$ within the unitarity bound
This is the first evidence of the {\it local cancellation} phenomenon.
We notice that the cancellation among resonance and contact term 
contributions for large $s$ allowed Weinberg in Ref.~\cite{Weinberg} to 
deduce some remarkable asymptotic relations about the resonance 
spectrum and the pion couplings.

In Figure~\ref{fig3bt} we show the partial waves $R^1_1$ and $I^1_1$
computed using Eq.~(\ref{Aca-vect}) and 
Eq.~(\ref{propagatore-vettore}). 
\figinsert{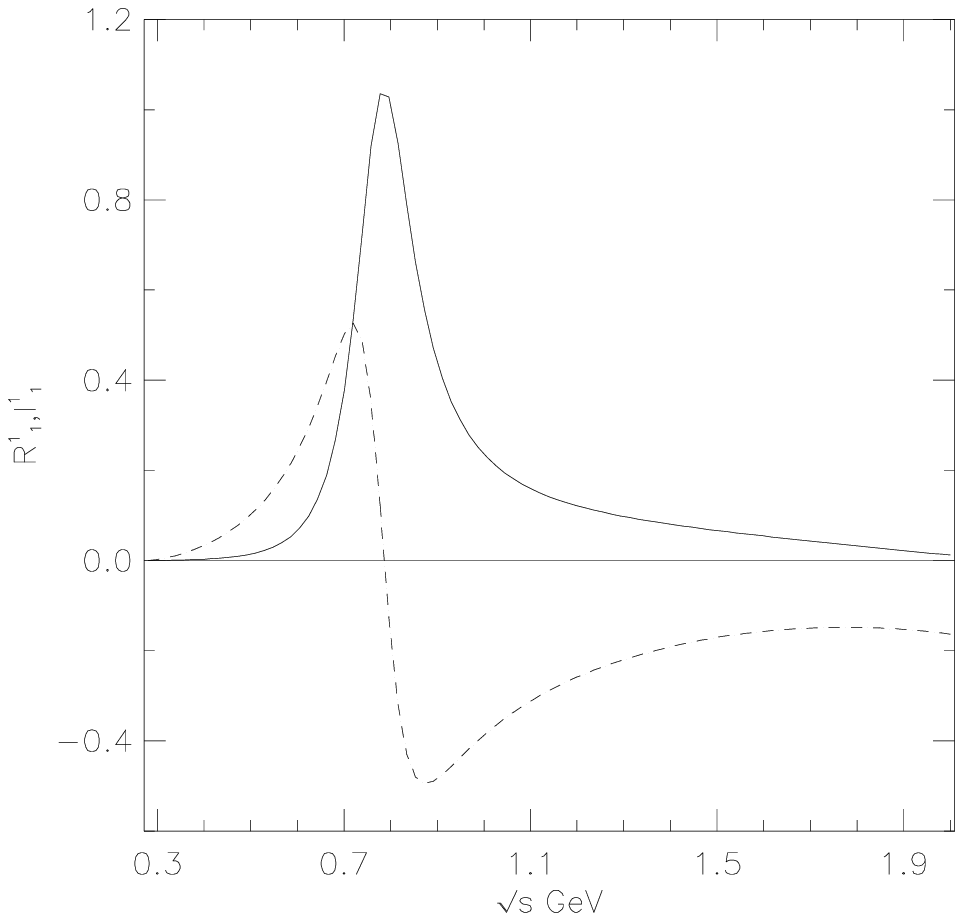}{The solid line represents the imaginary term $I^1_1$.
 The dot-dashed line is $R^1_1$.}{fig3bt}{hbtp}
Not surprisingly,
these display the standard resonant forms. The dominance of the 
vector--meson $\rho$ in the $I=l=1$ channel is used in $\chi$PT to estimate 
some of the counterterm coefficients \cite{donoghue}.
 For completeness we present the $R^2_0$ and $R^0_2$
amplitudes in Fig.~{\ref{fig4bt}}.
We notice that these amplitudes obey the unitarity limits up to 
about $2$ GeV. 

In this section we have shown that the inclusion of the $\rho$--meson 
dramatically reduces the onset of unitarity violation in the 
$R^0_0$ channel (see. Fig.~{\ref{fig2bt}}). 
We notice that this channel contains the exchange vector meson 
diagram which provides a decisive contribution. 
This result is a confirmation, not only, of the exchange symmetry 
but also of the relevance of the {\it crossing} symmetry.
 
The cancellation for large $s$ is due to the contact term 
 $v^{\mu}v_{\mu}$. The last term is fundamental in order to introduce the 
 $\rho$ in a chiral symmetric way. 
 In the following we will mainly study the $R^0_0$ channel, since 
this has the worst behaviour for large $s$.
\figinsert{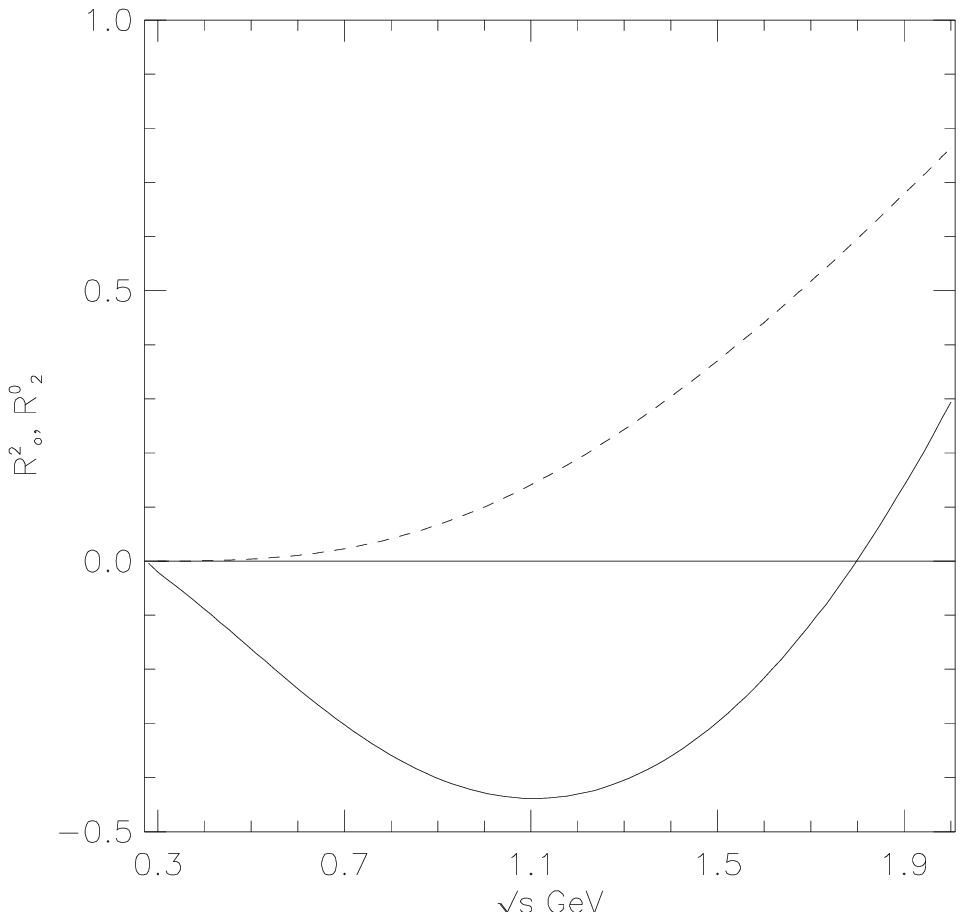}{The solid line is the  $\pi + \rho$ contribution for 
the real part of $I=2,~l=0$. The dot-dashed line is the  
 $\pi+\rho$ contribution for the  real part of $I=0,~l=2$.}{fig4bt}{hpbt}

\section{Is the $\sigma(550)$ alive ?}

As shown in Figure~\ref{figura1}, although the introduction of the 
$\rho$--meson
dramatically improves unitarity up to about 2~$GeV$, $R^0_0$ violates 
unitarity to a lesser extent starting at around 500~MeV. 
    What is needed to restore unitarity over the full range of interest and
to give better agreement with the experimental data for 
$\sqrt{s}\lapproxeq 900~MeV$?
\begin{itemize}
\item[{\it i.}]{Below  $450~MeV$, 
$R^0_0(s)$ actually lies a little below the Roy curves (see Fig.~\ref{fig1bt}).
Hence it would be nice to find a tree level mechanism which yields a small
positive addition in this region.} 
\item[{\it ii.}]{In the $600-1300~MeV$ range, an
increasingly negative contribution is clearly required to keep $R^0_0$
within the unitarity bound.} 
\end{itemize}

\noindent
In Reference~\cite{Sannino-Schechter} it has been shown that it 
is possible to satisfy both of these criteria
by introducing a broad scalar resonance (like the old $\sigma$) with
a mass around $550~MeV$. 
\figinsert{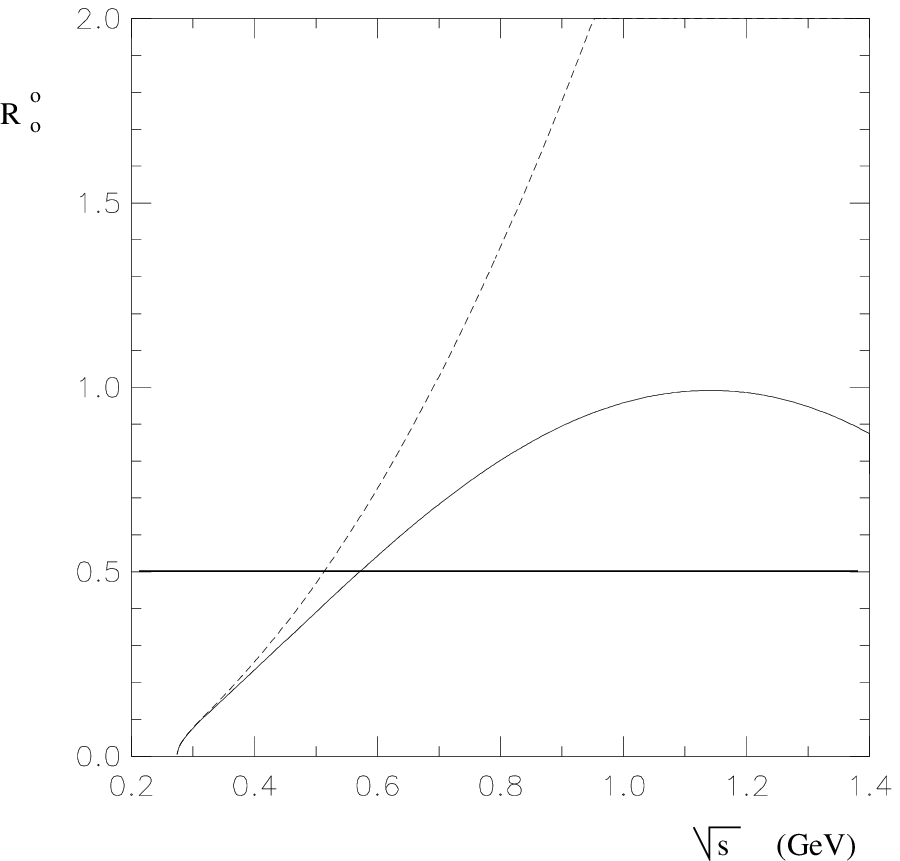}{Enlarged version of Fig.~\ref{fig2bt}. 
The solid line which shows the current algebra + $\rho$ result for $R^0_0$ 
is much closer to the unitarity bound of 0.5 than the dashed line 
which shows the current algebra result alone.}{figura1}{hptb}
In this model the $\sigma$ is not realized as a {\it chiral partner} but 
is introduced as a matter field with respect to chiral transformations. 
The $\sigma$ contribution to the invariant amplitude  $A(s,t,u)$ is 
\begin{equation}
Re A_{\sigma}(s,t,u)=Re
\frac{32\pi}{3H}
\frac{G}{M^3_\sigma}(s-2m_\pi^2)^2
\frac{(M_\sigma^2-s)+i M_{\sigma}G^{\prime}}{(s-M_\sigma^2)^2 +
M_\sigma^2{G^\prime}^2}\ ,
\label{eq:sigma}
\end{equation}
where 
 \be
H=\left(1-4\frac{m_{\pi}^2}{M^2_\sigma}\right)
 ^{\frac{1}{2}}
\left(1-2\frac{m_{\pi}^2}{M^2_\sigma}\right)^2\approx 1\ ,
\ee
and $G$ is related to the coupling constant $\gamma_0$ defined in 
Eq.~(\ref{la:sigma}) as 
\be
G={\gamma}^2_0\frac{3 H M_{\sigma}^3}{64\pi}\ .
\label{sigma-coupling} 
\ee

Note that the factor $(s-2m_{\pi}^2)^2$ is due to the derivative-type coupling 
required for chiral symmetry in Eq.~(\ref{la:sigma}). 
The total amplitude will be 
crossing symmetric since $A(s,t,u)$ and $A(u,t,s)$ in Eq.~(\ref{eq:def}) are obtained 
by performing the indicated permutations. $G^{\prime}$ is a parameter 
which we introduce to regularize the propagator. It can be called a width, 
but it turns out to be rather large so that, after the $\rho$ and $\pi$ 
contributions are taken into account, the partial wave amplitude 
$R^0_0$ does not clearly display the characteristic resonant behavior. In the 
most general situation one might imagine that $G$ could become complex 
as in Eq.~(\ref{rescattering}) due to higher order in $\displaystyle{1/N_c}$ corrections. It should be noted, 
however, that Eq.~(\ref{rescattering}) expresses nothing more than the assumption of 
unitarity for a {\it narrow} resonance and hence should not really be  
applied to the present broad case. 

A reasonable fit was found in Ref.~\cite{Sannino-Schechter} for $G$ purely real, but not equal to 
$G^{\prime}$. By the use of Eq.~(\ref{imaginary}), unitarity is in fact 
locally satisfied.   In \cite{Harada-Sannino-Schechter}
a best overall fit is obtained with the parameter 
choices $M_{\sigma}=559~MeV$, $G/G^{\prime}=0.29$ and $G^{\prime}=370~MeV$. 
These have 
been slightly fine-tuned from the values in Ref.~\cite{Sannino-Schechter}  
in order to obtain a better fit in the $1~GeV$ region. 
The contribution to $R^0_0$ due only to the presence of the low 
mass broad scalar $\sigma(550)$ is displayed in Fig.~{\ref{fig5tnew}}. 

\figinsert{fig5tnew.eps}{Contribution to  $R^0_0$ due to the 
$\sigma(550)$ particle.}{fig5tnew}{hptb}
 
The curve crosses zero at about $\sqrt{s}=560$~MeV. The small 
deviation is due to the effect of the cross diagrams. 
The result for the 
real part of $R^0_0$ due to the inclusion of the $\sigma$ contribution along 
with the $\pi$ and $\rho$ contributions is shown in Fig.~\ref{figura2}. 
\figinsert{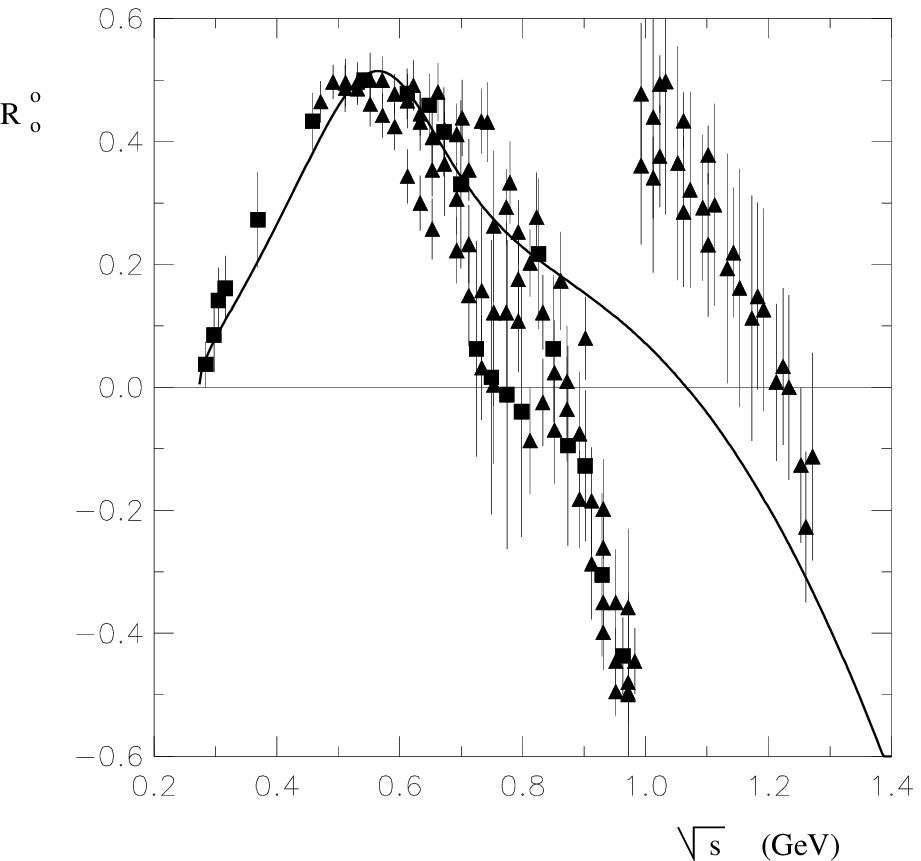}{The solid line is the {\it current algebra}
$+~\rho+\sigma$ result for $R^0_0$. The experimental points, in this
and suceeding figures, are
extracted from the phase shifts using Eq.~(\ref{real-imaginary}) and
actually correspond to $R^0_0/\eta^0_0$. ($\Box$) are extracted from
the data of Ref.~\cite{lowenergy} while ($\triangle$) are extracted from the data of Ref.~\cite{highenergy}. The predicted $R^0_0$ is small around the $1~GeV$ region.}{figura2}{htpb}
It is seen 
that the unitarity bound is satisfied and there is a reasonable agreement 
with the experimental points \cite{lowenergy,highenergy} up to about $800~MeV$. 
Beyond this point the effects of other resonances (mainly the $f_0(980)$) 
are required. {}From Eqs.~(\ref{eq:sigma}), 
(\ref{eq:isospin}) and (\ref{eq:wave}) we see that the contribution of
 $\sigma$ to $R^0_0$ becomes  
negative when $s>M^2_{\sigma}$. This is the mechanism which leads to 
satisfaction of the unitarity bound (c.f. Fig.~\ref{figura1}). 
For $s<M^2_{\sigma}$ 
one gets a positive contribution to $R^0_0$. This is helpful to push 
the predicted curve upwards and closer to the experimental results  
in this region, as shown in Fig.~\ref{figura3}. 
\figinsert{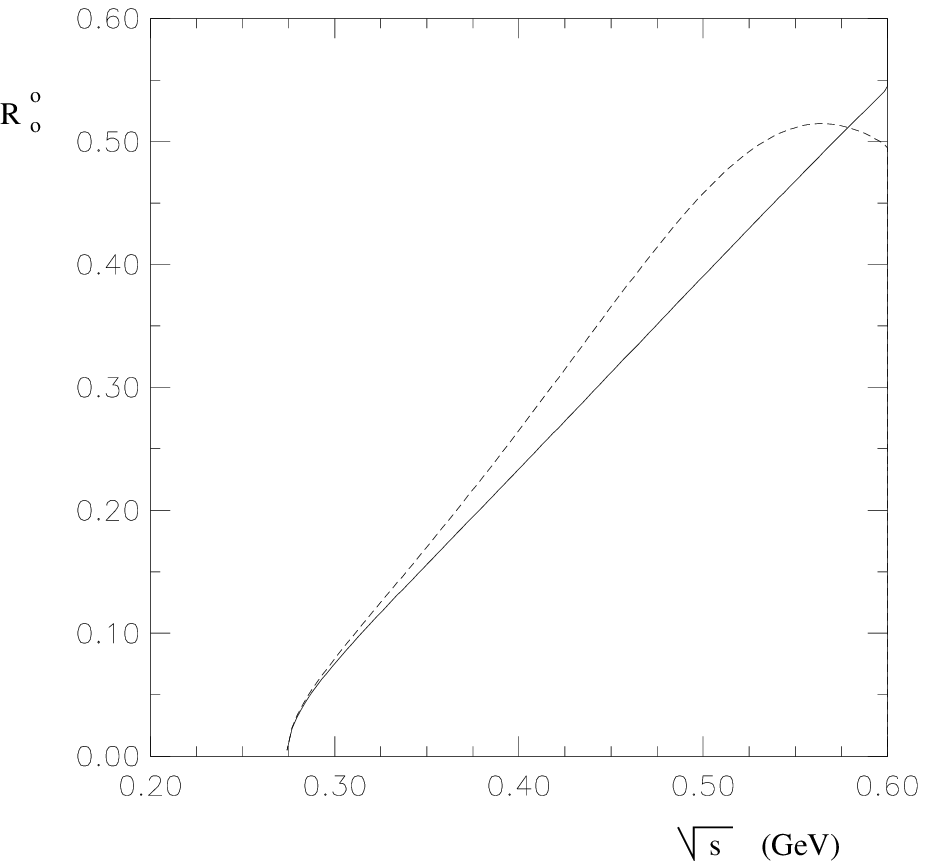}{A blowup of the low energy region. 
The solid line is the {\it current algebra}
$+~\rho$ contribution to $R^0_0$. The dashed line includes the
$\sigma$ particle and has the effect of turning the curve down to avoid
unitarity violation while boosting it at lower energies.}{figura3}{htbp}
The four-derivative contribution in 
the chiral perturbation theory approach performs the same function, 
however it does not change sign and hence does not satisfy  the unitarity 
bound above the $450~MeV$ region \cite{gasser-meissener}.

It is also interesting to notice that the main effect of the sigma particle
comes from its tail in Fig.~\ref{fig5tnew}. 
Near the pole region, its effect is hidden
by the dominant $\pi+\rho$ contribution. This provides a possible
explanation of why such a state may have escaped definitive
identification.
It is interesting to remark that a particle 
with mass and width very similar
to those given above for the $\sigma$ was predicted \cite{Jaffe} 
as part of a multiquark $qq\bar q\bar q$ nonet on the basis of the 
$MIT$ bag model. 
Hence, even though
they do not give rise to formally leading $\pi\pi$ amplitudes in the $\1N$
scheme, the picture has a good deal of plausibility from a polology 
point of view. It is not hard to
imagine that some $\1N$ subleading effects might be important
at low energies where the QCD coupling constant is strongest.
For completeness in Fig.~\ref{I2l0tutto} we also show the real 
part of the $I=2$ $l=0$ scattering amplitude ($R^2_0$).
\figinsert{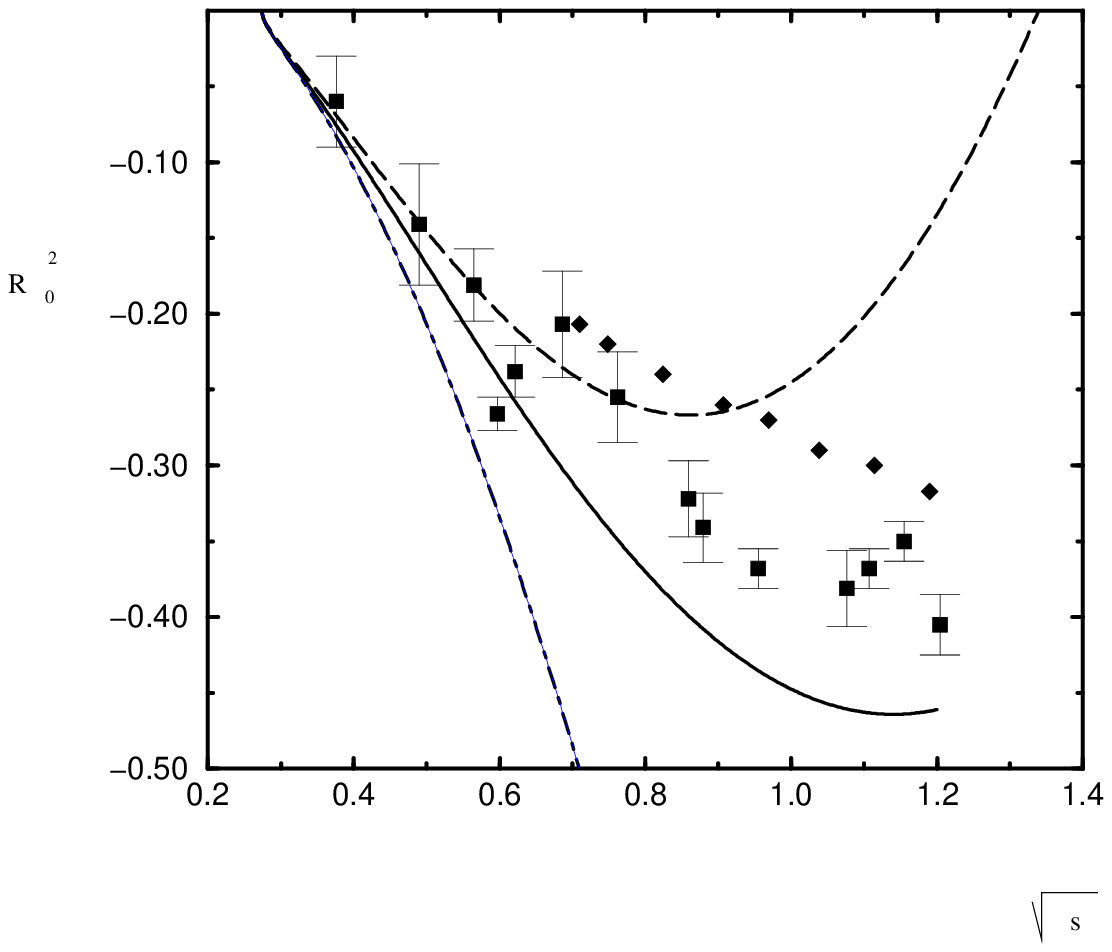}{The solid line is the {\it current algebra} 
 $+ \rho$ contribution for $R^2_0$. The dashed line is the     
{\it current algebra} $+ \rho + \sigma$ contribution for $R^2_0$. 
The dot-dashed line is the {\it current algebra} contribution 
for $R^2_0$.}{I2l0tutto}{hptb}  
We have used for the $\sigma$ parameters those presented in the first 
column of Table~\ref{tabella2}. It is clear that there is a fair 
agreement with the data up until about 1~GeV for the dashed line 
which represents the {\it current algebra} $+ \rho + \sigma$, while 
the {\it current algebra} prediction alone departs quite soon 
from the experimental results. The solid line ({\it current algebra} 
 $+\rho$) seems to better 
follow the experimental data up to about 1.2~GeV. However since 
we have not tried to fit this channel we will merely stress 
that the results are also consistent with the 
{\it local cancellation} principle. 
 {}For other authors \cite{Weinberg,Tornqvist:95,Tornqvist:96} the $\sigma$ 
is a $q\bar{q}$ state, while in \cite{Brown-Rho} is considered 
as an interpolating multi pionic state.
In the actual model it 
is not possible to uncover what makes the $\sigma$.   
This state is however essential according to the Chiral Resonance Model.
 $M_{\sigma},~G$ and $G^{\prime}$ are in practice the only 
unknown parameters in the present model.
Here we remark that the $\sigma$ has been introduced to understand 
low energy $\pi\pi$ scattering in the following two papers: 
{\it Exploring $\pi\pi$ Scattering in the $\1N$ Picture}
(Ref.~\cite{Sannino-Schechter}) and {\it Simple description of $\pi\pi$
scattering to 1 GeV} (Ref.~\cite{Harada-Sannino-Schechter}). At the 
time when these papers were written this state was not present in 
the {\it Particle Data Group} (PDG) \cite{pdg} review. In the latest 
PDG \cite{pdg-new} this state is finally present. 
Although the parameters associated with the $\sigma$ are still 
not well known the quoted range of parameters (mass and width) is 
consistent with the one determined in the Chiral Resonance framework.
As the conclusion to this paragraph we can say that the $\sigma$ is 
alive and well.   

\section{Comment on higher derivative contact terms}

In the previous paragraph we concluded that the $\sigma$ is crucial 
 in order 
to satisfy the unitarity bounds and to fit the low energy data in the $R^0_0$ 
channel. But {\it why do 
we need to introduce a new particle if we 
still have at our disposal higher derivative contact terms ?} Here 
we will analyze this possibility. {}For simplicity, here we will 
investigate four-derivative contact terms. 
 There are
two four-derivative chiral invariant contact interactions which are single
traces in flavor space:
\be
L_{4}=a \mbox{Tr} \left[ \partial_{\mu} U \partial_{\nu}U^{\dagger}  
\partial^{\mu} U \partial^{\nu}U^{\dagger}\right] + 
b \mbox{Tr} \left[ \partial_{\mu} U \partial^{\mu}U^{\dagger}
                   \partial_{\nu} U \partial^{\nu}U^{\dagger}\right] \ ,
\label{quattro}
\ee
where $a$ and $b$ are real constants. The single traces should be leading in
the $\1N$ expansion. Notice that the magnitudes of $a$ and $b$ will differ
from those in the chiral perturbation theory approach \cite{chp} since the
latter essentially also includes the effects of expanding the $\rho$
exchange amplitude up to order $s^2$. The four pion terms which result from
Eq.~(\ref{quattro}) are:
\be
L_{4}=\frac{8}{F^4_{\pi}}\left[ 2 a 
\left(\partial_{\mu}\vec{\pi}\cdot\partial_{\nu}\vec{\pi} \right)^2 
 + (b-a)\left(\partial_{\mu}\vec{\pi}\cdot\partial^{\mu}\vec{\pi} \right)^2 
\right] + \cdots    \ .
\ee
This leads to the following 
contribution to the $\pi\pi$ amplitude:
\be
A_{4}(s,t,u)=\frac{16}{F^4_{\pi}}\left\{
a\left[(t-2 m_{\pi}^2)^2 + (u-2 m_{\pi}^2)^2\right] + (b-a)(s-2
m^2_{\pi})^2 \right\} \ .
\label{A4}
\ee
It is reasonable to require that Eq.~(\ref{A4})
yields no correction at threshold, i.e. at $s=4\mpp^2$, $t=u=0$. This gives the
condition $b=-a$ and leaves the single parameter $a$ to play with. 
In Figure~\ref{figcontatto}
 we show $R^0_0$, as gotten by adding the piece obtained
from Eq.~(\ref{A4}) for several values of $a$ to the contribution of
 $\pi+\rho$. 
{}For $a=+1.0\times 10^{-3}$ the 
four-derivative contact term can pull the curve for $R^0_0$
down to avoid violation of the unitarity bound until around
$\sqrt{s}=1.0~GeV$. The price to be paid is that $R^0_0$ decreases very
rapidly beyond this point. We consider this to be an undesirable feature
since it would make a possible local cancellation scheme very unstable.
Another drawback of the four-derivative contact term scheme is that it lowers
$R^0_0(s)$ just above threshold, taking it further away from the Roy curves.
Hence in the following we will stick with the {\it deus ex machina} $\sigma$ 
particle and will keep only the minimal number of derivatives. 
\figinsert{figcontatto.eps}{Four-derivative contact term contribution for 
 $R^0_0$. 
The solid line corresponds to  $a=+1.0$. The dotted line corresponds to 
 $a=+0.7$. The dot-dashed line corresponds to 
 $a=0.5$. $a$ is represented in units of  $10^{-3}$. }{figcontatto}{hptb}

%% file: chapterl3.tex
\chapter{Exploring the 1 GeV region}
\section{What is happening in the 1 GeV region}

In the previous chapter we have shown \cite{Sannino-Schechter} that 
the Chiral Resonance Model is capable of describing $\pi\pi$ scattering 
up to about $800$~MeV in the $I=l=0$ channel (see Fig.\ref{figura2}). 
In this chapter we will investigate the $800-1200$~MeV energy range 
\cite{Harada-Sannino-Schechter}. We will 
 study in some detail the 
$f_0(980)$ properties and the problems connected with inelastic 
effects induced by the opening of the $K\bar{K}$ threshold.

  The neutral resonances which can contribute to the 
$I=l=0$ channel have the quantum numbers 
$J^{PC}=0^{++},~1^{--}$ and $ 2^{++}$. 
In the quark model these states are naturally interpreted as $s$ and 
 $p$ wave $q\bar{q}$ bound states.
We show in Table~\ref{tabella1} 
\noindent
\begin{table}
\begin{center}
\begin{tabular}{|c||c|| c c c|} \hline \hline
  &$I^G(J^{PC})$ & $M(MeV)$ & $\Gamma_{tot}(MeV)$ & $Br(2\pi)\%$\\ \hline \hline
$\sigma(550)$ & $0^+(0^{++})$  & 559     & 370      & $-$  \\
$\rho(770)$   & $1^+(1^{--})$  & 769.9   & 151.2    & 100  \\
$f_0(980)$   & $0^+(0^{++})$  & 980     & 40$-$400 & 78.1 \\
$f_2(1270)$  & $0^+(2^{++})$  & 1275    & 185      & 84.9 \\
$f_0(1300)$  & $0^+(0^{++})$  &1000-1500&150$-$400 & 93.6  \\
$\rho(1450)$  & $1^+(1^{--})$  &1465   & 310    & seen \\ \hline \hline
\end{tabular}
\end{center}
\caption{Resonances included in the $\pi \pi \rightarrow \pi\pi$ channel as 
listed in the PDG. Note that the $\sigma$ was not present in the 1994 
PDG and is not being described exactly as a {\it Breit-Wigner} shape;
we listed the fitted parameters shown in column 1 of Table~\ref{tabella2} 
where
$G^\prime$ is the analog of the {\it Breit-Wigner} width.}
\label{tabella1}
\end{table}
\noindent
the specific 
ones which are included, together with their masses and widths, where 
available from the Particle Data Group (PDG) \cite{pdg}  listings. 

Reference to Fig.~\ref{figura2}  shows that the experimental data for
 $R^0_0$ lie considerably below the $\pi+\rho+\sigma$ contribution between
$0.9$ and $1.0~GeV$ and then quickly reverse sign above this point.
We will now see that this distinctive shape is almost completely
explained by the inclusion of the relatively narrow scalar resonance $f_0(980)$
in a suitable manner. One can understand what is going on very simply
by starting from the real part of Eq.~(\ref{rescattering}):
\begin{equation}
M\Gamma 
\frac{(M^2-s)\cos (2\delta)-M\Gamma \sin (2\delta)}{(M^2-s)^2+M^2 {\Gamma}^2}
+ \frac{1}{2}\sin (2\delta)\ .
\label{980-mechanims}
\end{equation}
This expresses nothing more than the restriction of local unitarity in
the case of a narrow resonance 
in the presence of a background. We have seen that the difficulty
of comparing the tree level $\displaystyle{1/N_c}$ amplitude to experiment is enhanced
in the neighborhood of a direct channel pole. Hence it is probably
most reliable to identify the background term
$\displaystyle{\frac{1}{2} \sin(2\delta)}$ with our prediction for
$R^0_0$. In the region of interest, Fig.~\ref{figura2} shows that $R^0_0$ is very
small so that one expects $\delta$ to be roughly $90^\circ$ (assuming a monotonically increasing phase shift). Hence
the first pole term is approximately 
\begin{equation}
-\frac{(M^2-s)M\Gamma}{(M^2-s)^2+M^2 {\Gamma}^2}\ ,
\label{980-pole}
\end{equation}
which contains a crucial reversal of sign compared to the real part of
Eq.~(\ref{Breit-Wigner}). Thus, just below the resonance there is a sudden
{\it negative} contribution which jumps to a positive one above the
resonance. This is clearly exactly what is needed to bring experiment
and theory into agreement up until about $1.2~GeV$, as is shown
in Fig.~\ref{figura4}. 
\figinsert{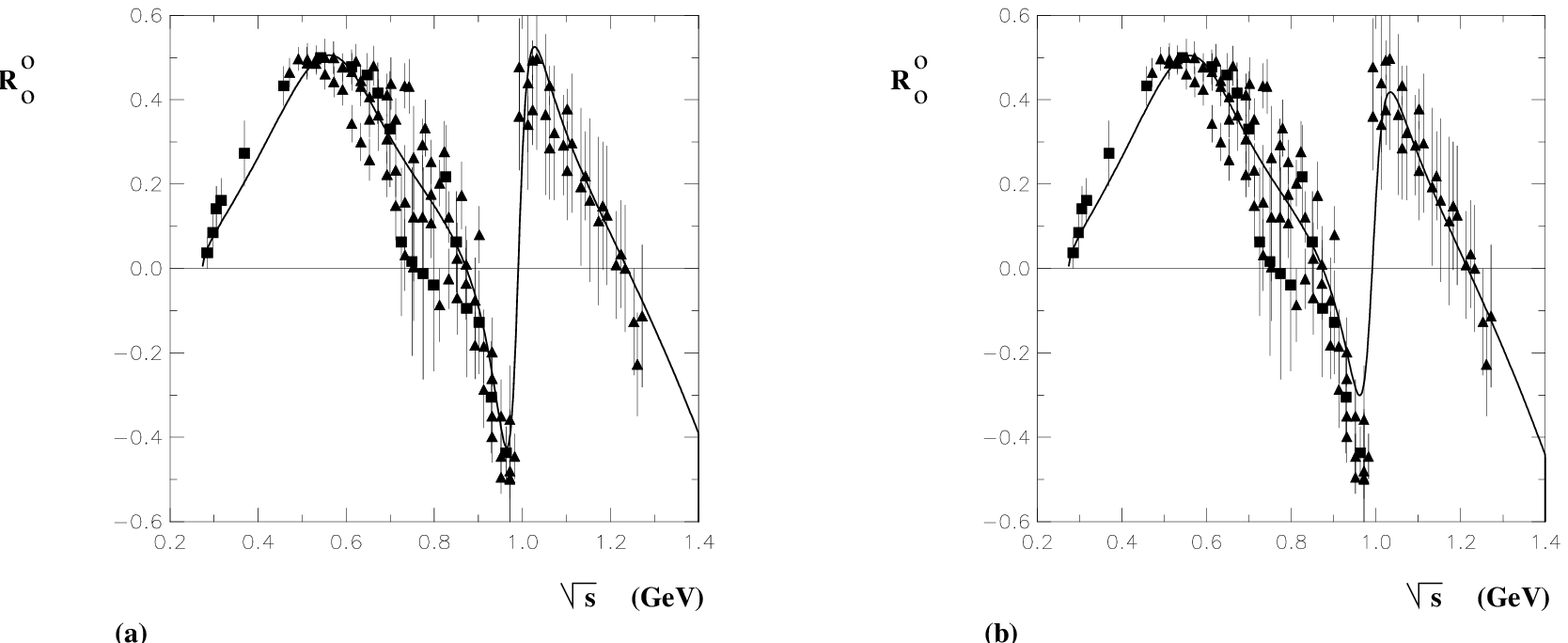}{(a): The solid line is the {\it current algebra}
$+~\rho~+~\sigma~+~f_0(980)$ result for $R^0_0$ obtained by assuming 
column 1 in Table~\ref{tabella2} for the $\sigma$ and $f_0(980)$ parameters
($Br(f_0(980)\rightarrow 2\pi)=100\%$).
(b): The solid line is the {\it current algebra}
$+~\rho~+~\sigma~+~f_0(980)$ result for $R^0_0$ obtained by assuming 
column 2 in Table~\ref{tabella2} ($Br(f_0(980)\rightarrow 2\pi)=78.1\%$) .}
{figura4}{htbp} 
The actual amplitude used for this calculation properly
contains the effects of the pions' derivative coupling to the
$f_0(980)$ as in Eq.~(\ref{eq:sigma}). 

The above mechanism, which leads to a sharp dip in the $I=J=0$ partial
wave contribution to the $\pi\pi$-scattering cross section, can be
identified with the very old {\it Ramsauer-Townsend} effect
\cite{shiff} which concerned the scattering of $0.7~eV$ electrons on
rare gas atoms. The dip occurs because the background phase of $\pi/2$
causes the phase shift to go through $\pi$ (rather than $\pi/2$) at
the resonance position. (Of course, the cross section is proportional
to $\sum_{I,J}^{~} (2J+1) \sin^2(\delta^J_I)$.) This simple mechanism seems
to be all that is required to understand the main feature of $\pi\pi$
scattering in the $1~GeV$ region.

\subsection{The {\it Ramsauer-Townsend} Effect}

Here we will compare the real part of the
$I=J=0$ partial wave amplitude which results from our crossing symmetric 
model with experimental data. 
{}Firstly we will consider the sum of the contributions of the
{\it current algebra}, $\rho$-{\it meson}, $\sigma$ and $f_0(980)$
pieces. Then we will add pieces corresponding to the {\it next group}
of resonances, namely the $f_2(1270)$, the $\rho (1450)$ and the
$f_0(1300)$. In this section we will continue to neglect the
$K\overline{K}$ channel.  

The current algebra plus $\rho$ contribution to the quantity $A(s,t,u)$ is 
defined in Eq.~(\ref{Aca-vect}). 
Note that for the $I=J=0$ channel the expression in Eq.~(\ref{Aca-vect}) 
will yield a purely real
contribution to the partial wave amplitude. The contribution of the 
low--lying $\sigma$ meson was given in Eq.~(\ref{eq:sigma}). For the
important $f_0(980)$ piece we have 
\begin{equation}
Re A_{f_0(980)}(s,t,u)=Re\left[ 
\frac{\gamma_{f_0\pi\pi}^2 e^{2i\delta} (s-2 m_{\pi}^2)^2}
{m_{f_0}^2 - s - i m_{f_0}\Gamma_{tot}(f_0)\theta ( s - 4 m_{\pi}^2)}
\right]\ ,
\label{f0(980)}
\end{equation}
where $\delta$ is a background phase parameter and the real coupling
constant $\gamma_{f_0\pi\pi}$ is related to the $f_0(980)\rightarrow \pi\pi$
width by 
\begin{equation}
\Gamma(f_0(980)\rightarrow
\pi\pi)=\frac{3}{32\pi}\frac{\gamma^2_{f_0\pi\pi}}
{m_{f_0}}\sqrt{1 - \frac{4 m_{\pi}^2}{m_{f_0}^2}}\ .
\end{equation}
We will not consider $\delta$ to be a new parameter but shall predict
it as being
\begin{equation}
\frac{1}{2}\sin (2\delta)\equiv \tilde{R}^0_0 (s=m_{f_0}^2)\ ,
\end{equation}
where $\tilde{R}^0_0$ is computed as the sum of the current algebra,
$\rho$, and sigma pieces.
Since the $K\overline{K}$ channel is being neglected, one might want
to set the {\it regularization parameter}  $\Gamma_{tot}(f_0)$ in the
denominator to $\Gamma(f_0(980)\rightarrow \pi\pi)$. We shall try both this
possibility as well as the experimental one 
$\displaystyle{\frac{\Gamma(f_0(980 )\rightarrow \pi\pi)}
{ \Gamma_{tot}(f_0)}\approx 78.1\%}$.

A best fit of our parameters to the experimental data results in the
curves shown in Fig.~\ref{figura4} 
for both choices of branching ratio. Only the
three parameters $G/G^{\prime}$, $G^{\prime}$ and $M_{\sigma}$ are
essentially free. The others are restricted by experiment.
Unfortunately the total width $\Gamma_{tot}(f_0)$ has a large
uncertainty; it is claimed by the PDG to lie in the $40-400~MeV$
range. Hence this is effectively a new parameter. In addition we have
considered the precise value of $m_{f_0}$ to be a parameter for fitting purposes. The parameter values
for each fit are given in Table~{\ref{tabella2}} 
together with the $\chi^2$ values.
It is clear that the fits are good and that the parameters are stable
against variation of the branching ratio. The predicted background
phase is seen to be close to $90^\circ$ in both cases. Note that the
fitted width of the $f_0(980)$ is near the low end of the experimental
range. The low--lying sigma has a mass of around $560~MeV$ and a width
of about $370~MeV$. As explained in section 3, we are not using
an exactly conventional {\it Breit-Wigner} type form for this very
broad resonance. The numbers characterizing it do however seem reasonably
consistent with other determinations thereof 
\cite{Tornqvist:95,Janssen-Pearce-Holinde-Speth,Morgan-Pennington:93}.

\begin{table}

\begin{center}
\begin{tabular}{ c||c c c c||c c c c||c } \hline \hline
\multicolumn{1}{ c||}{}&\multicolumn{4}{|c||}{}&
\multicolumn{4}{|c||}{With Next Group}& 
\multicolumn{1}{|c|}{ No $\rho(1450)$}\\

{\small $BR(f_0(980)\rightarrow 2\pi)\%$}            & 100   & 78.1  & 78.1 & 78.1  &
100 & 78.1 & 78.1 & 78.1 & 100 \\ 

$\eta^0_0 $         & 1     & 1   & 0.8     & 0.6  & 1   & 1    & 0.8
& 0.6 & 1
\\ \hline \hline 

$M_{f_0 (980)}~(MeV)$ & 987 & 989 & 990& 993  & 991 &992 &993 & 998 &992 \\  

$\Gamma_{tot}~(MeV)$  &64.6  &  77.1& 75.9& 76.8  &66.7 &77.2 &78.0&84.0&64.6 \\

$M_{\sigma}~(MeV) $   &559  & 557 & 557& 556  &537& 537 & 535 & 533 & 525 \\  

$G^\prime ~(MeV)$   &370  & 371 & 380& 395  &422& 412 & 426 & 451 &467 \\ 

$G/G^\prime$   & 0.290  & 0.294 & 0.294& 0.294  &0.270& 0.277
&0.275&0.270 & 0.263\\ \hline \hline

$\delta $~(deg.)      & 85.2  & 86.4   & 87.6 & 89.6 &   89.2  & 89.7  &
91.3&94.4  &90.4 \\

$\chi^2 $      & 2.0   & 2.8   &  2.7  & 3.1 &   2.4  & 3.2   & 3.2 &
3.4 & 2.5\\ \hline \hline
\end{tabular}

\end{center}
\caption{Fitted parameters for different cases of interest.}
\label{tabella2}
\end{table}

\section{Next group of Resonances}
The {\it local cancellation} principle together with {\it crossing} 
works up to 1~GeV in energy. What about above 1~GeV ? Does the inclusion 
of the resonances present in this range of energy modifies the results 
obtained at low energy ?  

In the $\1N$ approximation the $q\bar{q}$ mesons belong to 
ideally mixed nonets \cite{1n}. Following the $\1N$ prescriptions, 
at the first order, only excited $p-$wave and radially excited states 
in the $s-$wave must be included.  

The neutral members of the $p-$wave $q\bar{q}$ nonets have the quantum numbers
$J^{PC}=0^{++}$, $1^{++}$, $1^{+-}$ and $2^{++}$. Of course,
the neu\-tral members
of the radially excited $s-wave$ $q\bar{q}$ nonets have
$J^{PC}=0^{-+}$ and $1^{--}$. Only members of the
$0^{++}$, $1^{--}$ and $2^{++}$ nonets can couple to two pseudoscalars
\footnote{It is
possible to write down a two point mixing interaction between $0^{-+}$ and
radially excited $0^{-+}$ particles etc., but we shall neglect such effects
here.}. By $G$-parity conservation we finally note that it is the $I=0$
member of the $0^{++}$ and $2^{++}$ nonets and the $I=1$ member of the
$1^{--}$ nonet which can couple to two pions. Are there good experimental
candidates for these three particles?

The cleanest case is the lighter $I=0$ member of the $2^{++}$ nonet: the
$f_2(1270)$ has, according to the August 1994 Review of Particle Properties
(PDG)
\cite{pdg}, the right quantum numbers, a mass of $1275\pm 5~MeV$, a width of
$185\pm20~MeV$, a branching ratio of $85\%$ into two pions, and a branching
ratio of only $5\%$ into $K\overline{K}$. On the other hand the
$f^{\prime}_2(1525)$ has a $1\%$ branching ratio into $\pp$ and a $71\%$
branching ratio into $K\overline{K}$. It seems reasonable to approximate
the $2^{++}$ nonet as an ideally mixed one and to regard the $f_2(1270)$ as
its non-strange member.

The $\rho(1450)$ is the lightest listed \cite{pdg} particle which is a
candidate for a radial excitation of the usual $\rho(770)$. It has a less
than $1\%$ branching ratio into $K\overline{K}$ but the 
$\pi\pi$ branching ratio,
while presumably dominant, is not yet known. 
In the following analysis, for definitness, we will assume 
that the $\rho(1450)$ mainly decays into $\pi\pi$.  However, we notice 
that the $K^*(1410)$, which presumably belongs to the same $\rho(1450)$ 
 $SU(3)$ multiplet has a $K\pi$ branching ratio of only $7\%$. Hence 
we might also expect a small coupling to the two pions. We will 
also consider this effect by also excluding 
the $\rho(1450)$ from the analysis.
The $\rho(1700)$ is a little too high for our region of
interest.

An understanding of the $I=0$, $0^{++}$ channel has been elusive, despite
much work. 
We have already included the low mass scalar $\sigma(550)$ in order 
to understand $\pi\pi$ scattering at low energy. 
In the 1~GeV region we expect the narrow resonance $f_0(980)$ to play 
a key role.
The $f_0(980)$ has a $22\%$ branching ratio
into $K\overline{K}$ even though its central mass is below the
 $K\overline{K}$ threshold. 
The PDG also lists the scalar $f_0(1300)$ which has about a $93\%$
branching ratio into $\pp$ and a $7\%$ branching ratio into
$K\overline{K}$.
 We shall use the $f_0(1300)$ here. 
It is hard to
understand why, if the $f_0(980)$ is the $\overline{s}s$ member of a
conventional $0^{++}$ nonet, it is lighter than the $f_0(1300)$.
Most likely, the $f_0(980)$ is an exotic or a $K\overline{K}$
{\it molecule} \cite{Jaffe}. If that is the case, 
its coupling to two pions ought to be
suppressed in the $\1N$ picture.

Now we will give, in turn, the $\pi\pi$ scattering amplitudes due to the
exchange of the $f_0(1300)$, the $f_2(1270)$ and the $\rho(1450)$.

\subsection{The Tensor $f_2(1270)$.}
In the first chapter we already observed that we  
represent the $3\times 3$ matrix of tensor fields by a symmetric 
traceless matrix $T_{\mu\nu}$. $T_{\mu\nu}$ transforms 
covariantly under a generic chiral transformation 
(Eq.~(\ref{2-transf})). 
The 2--pion coupling can be deduced from the following chirally  
preserving term obtained from Eq.~(\ref{spin-2}) 
\be
-\frac{\gamma_2}{\sqrt{2}}(f_2)_{\mu\nu}
\left[\partial^{\mu}\vec{\pi}\cdot \partial^{\nu}\vec{\pi} \right]\ .
\ee
In this case we note that the chiral invariant interaction is just the same
as the minimal one we would have written down without using chiral symmetry.
The partial width is then
\be
\Gamma(f_2(1270)\rightarrow \pi\pi)=\frac{\gamma_2^2}{20 \pi}
\frac{p_{\pi}^5}{M^2_{f_2}} \ ,
\ee
where $p_\pi$ is the pion momentum in the $f_2$ rest frame. This leads to
$|\gamma_2|=13.1~GeV^{-1}$.

To calculate the $f_2$ exchange diagram we need the spin 2 propagator 
\cite{tensor} (see also Eq.~(\ref{eq:tensorpropag})).
\be
\frac{i}{q^2-m^2_{f_2}}\left[
\frac{1}{2}\left(
\theta_{\mu_1\nu_1} \theta_{\mu_2\nu_2}+
\theta_{\mu_1\nu_2}\theta_{\mu_2\nu_1}\right)-
\frac{1}{3}\theta_{\mu_1\mu_2}\theta_{\nu_1\nu_2}\right]\ ,
\ee
\noindent
where
\be
\theta_{\mu\nu}=-g_{\mu\nu}+\frac{q_\mu q_\nu}{m^2_{f_2}}\ .
\ee
\noindent
A straightforward computation then yields the $f_2$ contribution to the
 $\pi\pi$ scattering amplitude:
\bea
A_{f_2}(s,t,u)&=&\frac{\gamma^2_2}{2(m^2_{f_2}-s)}
\left(
-\frac{16}{3}\mpp^4
+\frac{10}{3}\mpp^2 s
-\frac{1}{3}s^2
+\frac{1}{2}(t^2+u^2)\right.\nonumber\\
&~&\left.-\frac{2}{3}\frac{\mpp^2s^2}{m^2_{f_2}}
-\frac{s^3}{6m^2_{f_2}}
+\frac{s^4}{6m^4_{f_2}}
\right)\ .
\eea
We notice that in the previous expression the behavior at high 
energy is dominated by the $s^3$ power. This is due to the fact 
that the present tensor is a massive particle not protected 
by any symmetry. This indicates that at higher energy we need 
to include a larger number of resonant states in order to 
bring the amplitudes inside the unitarity bounds. 
From this point of view our model is very close to the string 
mode \cite{string}, where an infinite number of resonances are 
identified with the string vibrational modes. 
The singularity in the propagator will be regulated as prescribed 
in the first chapter.

\subsection{The vector meson $\rho(1450)$ and the  $f_0(1300)$ scalar.}

The contribution of the $s-$wave radially excited $\rho(1450)$ meson 
to the scattering amplitude is 
\be
A_{\rho^{\prime}}(s,t,u)=-\frac{g^2_{\rho^{\prime} \pi \pi}}{2 m_{\rho^{\prime}}^2}
\left[\frac{t(u-s)}{m_{\rho^{\prime}}^2 - t} +\frac{u(t-s)}{m_{\rho^{\prime}}^2-u} \right] \ ,
\label{Arhop}
\ee
where $g_{\rho^{\prime}\pi\pi}$ is related to the 
 $\rho(1450)\rightarrow \pi\pi$ partial width by
\be
\Gamma(\rho(1450)\rightarrow \pi\pi)=\frac{g_{\rho^{\prime}\pi\pi}^2
p_{\pi}^3}{12 \pi m^2_{\rho^{\prime}}} \ .
\ee
With a branching ratio of $100 \%$ into two pions we get 
$|g_{\rho^{\prime}\pi\pi}|\simeq 7.9$. 
In order to get the invariant amplitude in 
Eq.~(\ref{Arhop}) there is no need to include the 
$\rho(1450)$ as a massive chiral gauge field. 
\cite{tensore-vettore}. 

By using Eq.~(\ref{la:sigma})) we can easily deduce the chiral 
coupling to two pions  
\begin{equation}
+\frac{\gamma_0}{\sqrt{2}}\; f_0 \;
\partial_{\mu}\vec{\pi}\cdot\partial^{\mu}\vec{\pi}\ . 
\end{equation}  
The partial decay width is 
\be
\Gamma(f_0(1300)\rightarrow\pi\pi)=\frac{3 \gamma_0^2}{64 \pi M_{f_0}}
\sqrt{ 1 - \frac{4 m^2_{\pi}}{M_{f_0}^2}} \times \left(M^2_{f_0} - 2
m^2_{\pi}\right)^2 \ .
\ee
Since these resonance parameters are not well defined, for definiteness 
we will assume the PDG central values \cite{pdg}, i.e. 
$\Gamma_{tot}(f_0(1300))=0.275$ GeV and $M_{f_0}=1.3$ GeV. 
Hence we deduce $|\gamma_0|\simeq 2.88$
GeV$^{-1}$. 
The unregularized invariant scattering amplitude is 
\be
A_{f_0}(s,t,u)=\frac{\gamma^2_0}{2}\frac{\left(s - 2
m^2_{\pi}\right)^2}{M^2_{f_0} - s} \ .
\ee
We will regularize this resonance propagator as for the $f_0(980)$

\subsection{$f_2(1270)+f_0(1300)+\rho(1450)$}

 Now we are in a position to appraise the contribution to $R^0_0$ of the next
group of resonances.
In order to better understand the {\it local cancellation} mechanism in 
this energy range we will not consider the background effect at the moment.
The contributions of each resonance are shown in 
Fig.~\ref{fig5bt}
\noindent
\figinsert{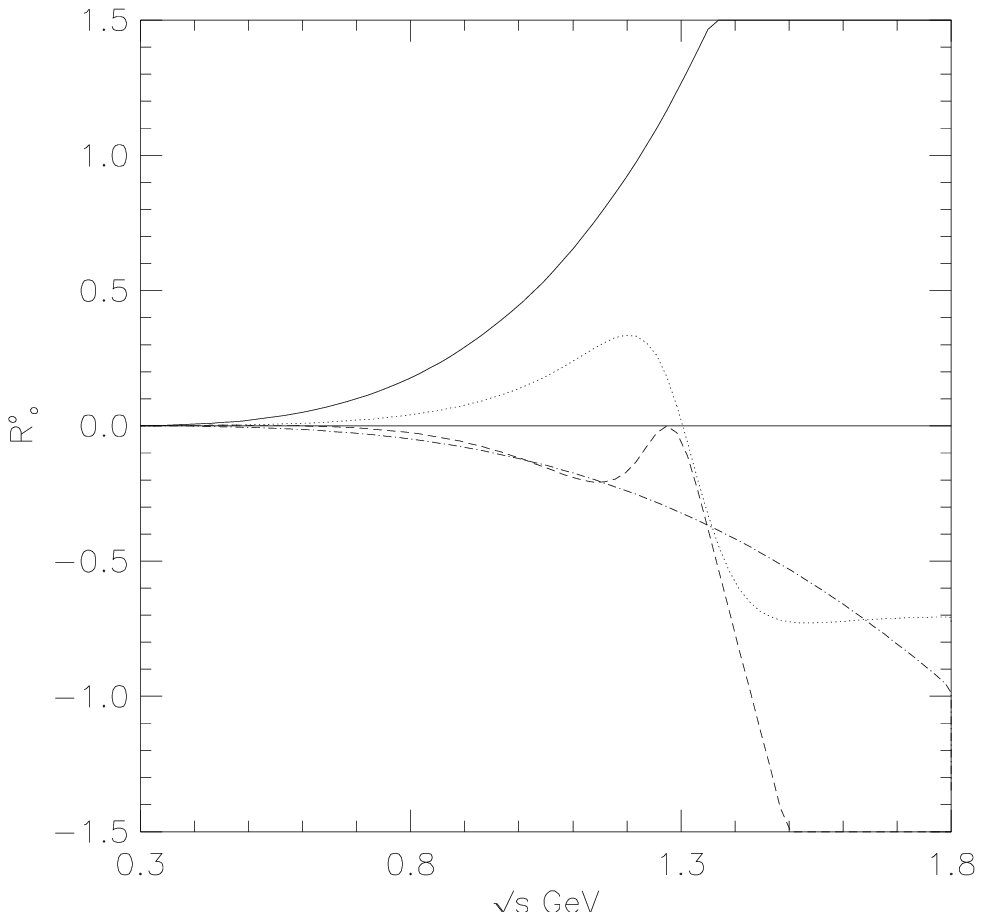}{Contributions for $R^0_0$. Solid line:
 $f_2(t+u)$. Dashed line: $f_2(s)$. Dotted line: 
 $f_0(1300)$. Dot--dashed line: $\rho(1450)$}{fig5bt}{hptb}
\noindent
Note that the $f_0(1300)$ piece is not the largest, as one might at first
expect. That honor goes to the $f_2$ contribution which is shown divided
into the $s$-channel pole piece and the $(t+u)$ pole piece. 
We observe that the $s$-channel pole piece, 
associated with the $f_2$, vanishes at 
$\sqrt{s}=M_{f_2}$. This happens because the numerator of the propagator in
(\ref{eq:tensorpropag}) is precisely a spin--2 projection operator at that
point. The $\rho(1450)$ contribution is solely due to the $t$ and $u-$channel
poles. It tends to cancel the $t$ and $u-$channel pole contributions of the
 $f_2(1270)$ but does not quite succeed. The $t$ and $u-$channel pole
contributions of the $f_0(1300)$ turn out to be negligible. Notice the
difference in characteristic shapes between the $s$ and $(t+u)$ 
exchange curves.
Fig.~\ref{fig6bt} shows the sum of all these individual contributions.
There does seem to be cancellation. At the high 
end, $R^0_0$ starts to run negative well
past the unitarity bound around $1.5~GeV$. But it is
reasonable to
expect resonances in the $1.5-2.0~GeV$ region to modify this. The maximum
positive value of $R^0_0$ is about $1$ at $\sqrt{s}=1.2~GeV$. 
This would be a problem if we did not have the low energy contributions 
({\it current algebra} $+ \rho + \sigma$). The background contribution 
is providentially negative (Fig.~\ref{figura2}), showing 
once again the need for the $\sigma$.
\figinsert{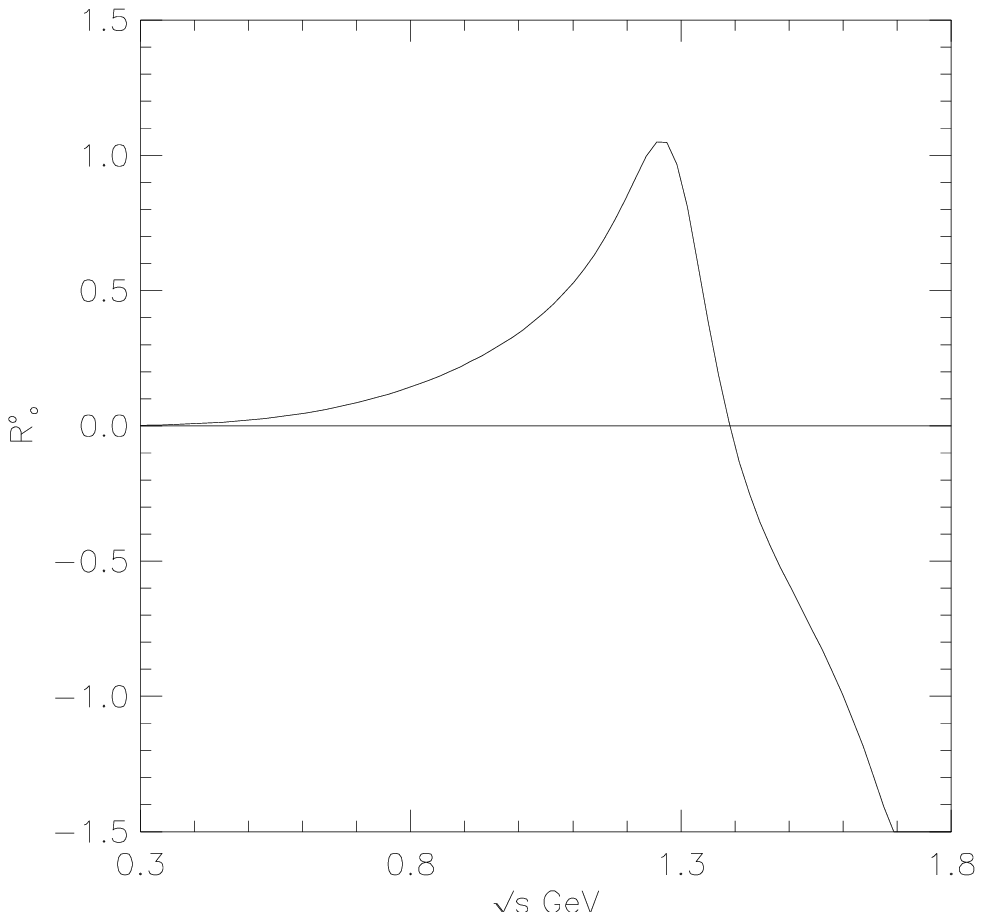}{Sum of the contributions in 
Fig.~\ref{fig5bt}.}{fig6bt}{hptb}
Let us  now check that the next group does not essentially modify the 
low energy results up to 1.2~GeV. 
 The somewhat positive net contribution of these resonances to $R^0_0$
is compensated for by readjustment of the parameters describing the low
lying sigma.

It may be interesting to include the
effect of the background phase for the $f_0 (1300)$ as we have just
seen that it was very important for a proper understanding of the
 $f_0 (980)$. To test this possibility we reversed the sign of the 
 $f_0 (1300)$ contribution and show the result as the solid curve in
Fig.~\ref{figura5}. This sign reversal is reasonable since our model suggests a
background phase of about $270^\circ$ in the vicinity of the 
 $f_0(1300)$. It can be seen that there is now a significantly greater
cancellation of the {\it next group} particles among themselves up to
about $1.2~GeV$. 
\noindent
\figinsert{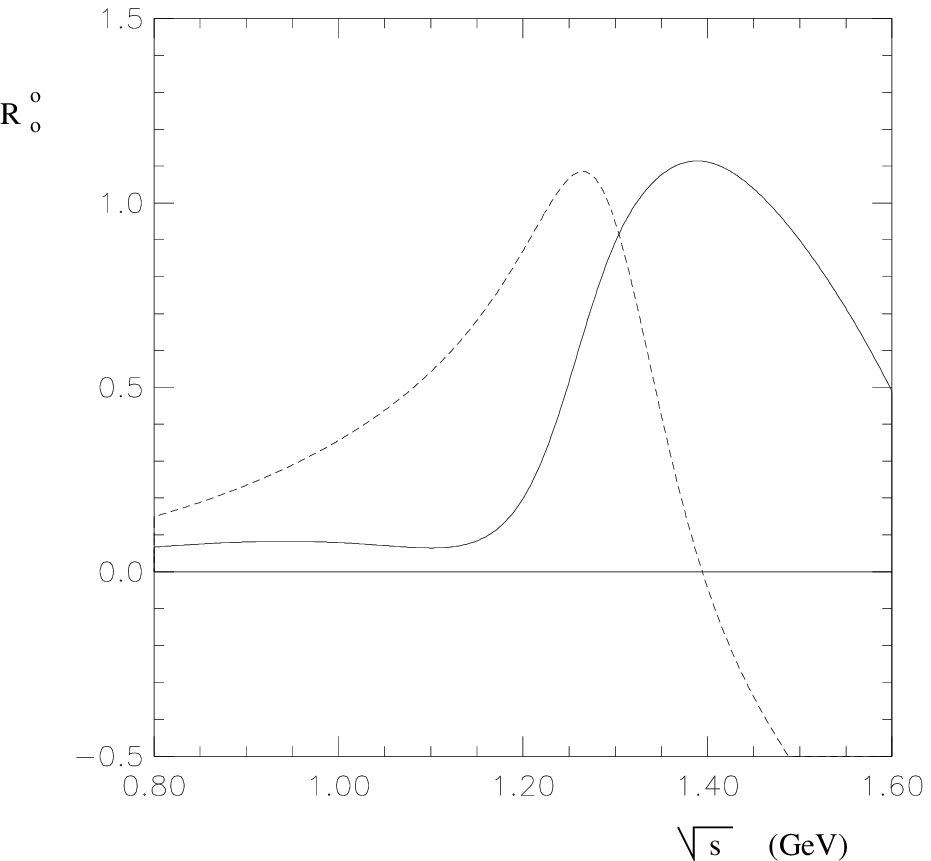}{Contribution from the {\it next 
group} of resonances; the solid line is obtained with the reverse sign 
of the $f_0(1300)$ piece; the dashed line is as in 
Fig.~\ref{fig6bt}.}{figura5}{hptb}   
\noindent
The resulting total fits are shown in Fig.~\ref{figura6} 
\figinsert{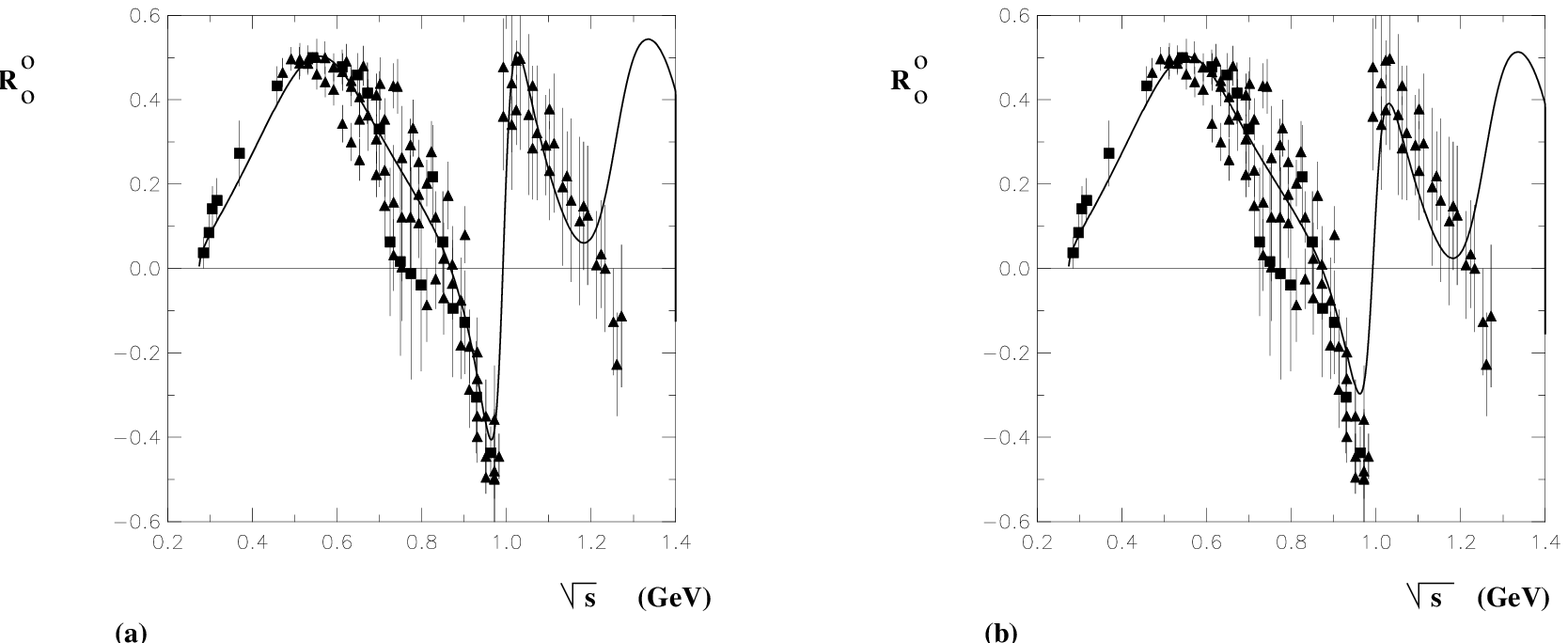}{Prediction for $R^0_0$ with the {\it next
group} of resonances. (a) assumes (column 5 in Table~\ref{tabella2})
 ($BR(f_0(980)\rightarrow 2\pi)=100\%$) while (b) assumes (column 6) ($BR(f_0(980)\rightarrow 2\pi)=78.1\%$).}{figura6}{htpb}
 for assumed $f_0 (980)\rightarrow \pi\pi$ branching ratios of 
both $100\%$ and $78.1\%$ 
and the parameters associated with the fits are shown in Table~\ref{tabella2}. It
is clear that the fitted parameters and results up to about $1.2~GeV$
are very similar to the cases when the {\it next group} was absent.
Above this region there is now, however, a positive bump in $R^0_0$
at around $1.3~GeV$. This could be pushed further up by choosing a
higher mass (within the allowable experimental range) for the $f_0
(1300)$. Resonances in the $1500~MeV$ region, which have {\it not}
been taken into account here, would presumably also have an important
effect in the region above $ 1.2~GeV$. Clearly there is not much
sense, at the present stage, in trying to produce a fit above
$1.2~GeV$.
In the last column of Table~\ref{tabella2} we have neglected 
the effect of the $\rho(1450)$. 
The resulting fit is shown in the last column of 
Table~\ref{tabella2}
and it is seen to leave the other parameters essentially unchanged. 

It thus seems that the results are consistent with the
hypothesis of {\it local cancellation}, wherein the physics up to a
certain energy $E$ is described by including only those resonances up
to slightly more than $E$ and it is furthermore hypothesized that the 
individual particles cancel in such a way that unitarity is maintained.

\section{Inelastic effects}

Up to now we have completely neglected the effects of coupled
inelastic channels. Of course the $4\pi$ channel opens at $540~MeV$, 
the $6\pi$ channel opens at $810~MeV$ and, probably most
significantly, the $K\overline{K}$ channel opens at $990~MeV$. We have
seen that a nice undestanding of the $\pi\pi$ elastic channel up to
about $1.2~GeV$ can be gotten with complete disregard for inelastic
effects. Nevertheless it is interesting to see how our results would
change if experimental data on the elasticity parameter $\eta^0_0$ are
folded into the analysis. Figure~\ref{figura7} 
\figinsert{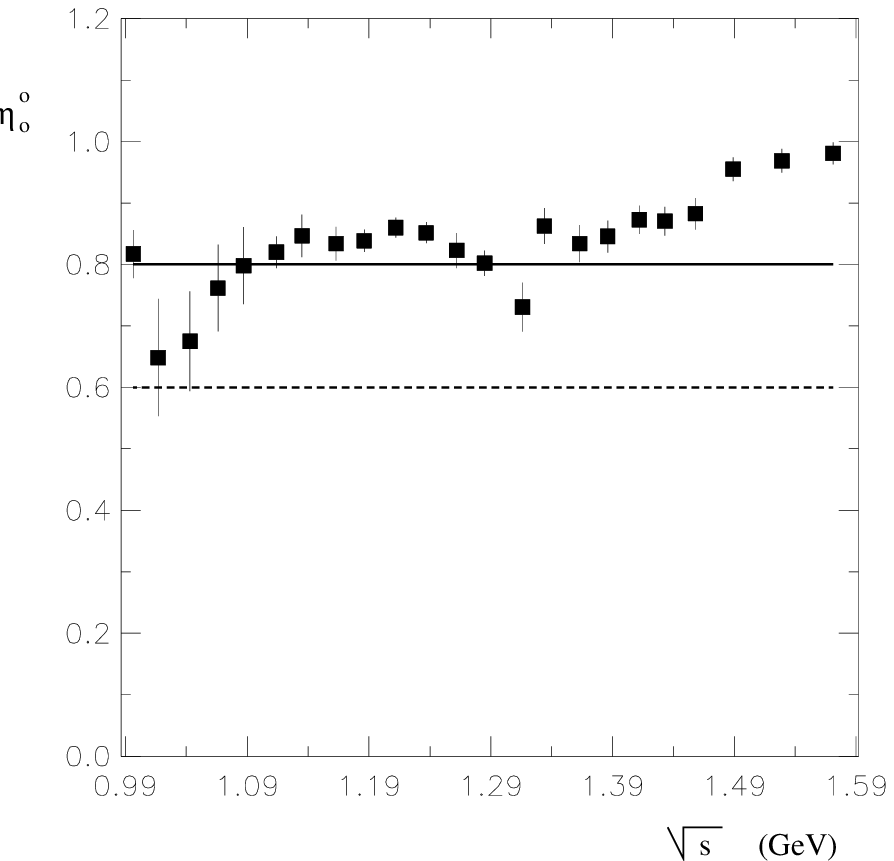}{An experimental determination of $\eta^0_0=
\sqrt{1 - 4 |T^0_{12,0}|^2}$ \cite{Cohen:80}.}{figura7}{htpb}
illustrates the results for
$\eta^0_0(s)$ obtained from an experimental analysis \cite{Cohen:80}
of $\pi\pi\rightarrow K\overline{K}$ scattering. For simplicity, we
approximated the data by a constant value $\eta^0_0 = 0.8$ above the
$K\overline{K}$ threshold. Figure~\ref{figura8}(a) 
\figinsert{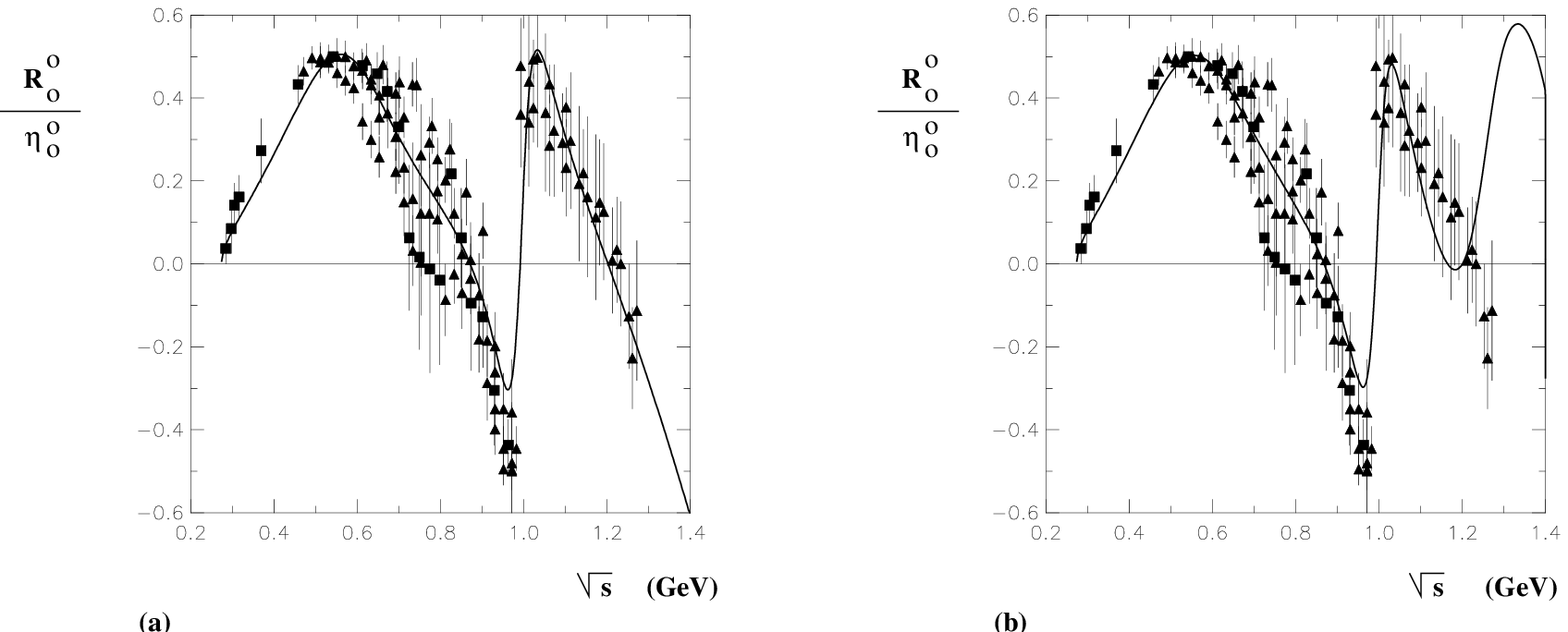}{Predictions with phenomenological treatment of
inelasticity ($\eta^0_0 = 0.8$) above $K\overline{K}$ threshold. (a): 
without {\it next group}. (b): with {\it next group}. }{figura8}{htbp}
shows the effect of this choice on
$R^0_0(s)$ computed without the inclusion of the {\it next group} of
resonances, while Fig.~\ref{figura8}(b) shows the effect when the {\it next group}
is included. Comparing with Fig.~\ref{figura4}(b) and \ref{figura6}(b), we see that setting $\eta^0_0 =
0.8$ has not made any substantial change. The parameters of the fit
are shown in Table~\ref{tabella2} 
as are the parameters for an alternative fit with 
$\eta^0_0 = 0.6$. The latter choice leads to a worse fit for $R^0_0$.

We conclude that inelastic effects are not very important for
understanding the main features of $\pi\pi$ scattering up to about
$1.2~GeV$. However, we will discuss the calculation of $\eta^0_0 (s) $
from our model in the $\pi\pi\rightarrow K\overline{K}$ paragraph.

\section{Phase Shifts}

Strictly speaking the Chiral Resonance Model only entitles us 
to compare the real part of the predicted amplitude with the real part of
the amplitude deduced from experiment. Since the predicted $R^0_0(s)$
up to $1.2~GeV$ satisfies the unitarity bound (within the fitting
error) we can calculate the imaginary part $I^0_0(s)$, and hence the
phase shift $\delta_0^0(s)$ on the assumption that full unitarity
holds. This is implemented by substituting $R^0_0(s)$ into
Eq.~(\ref{imaginary}) and resolving the discrete sign ambiguities by
demanding that $\delta^0_0(s)$ be continuous and monotonically
increasing (to agree with experiment). It is also necessary to know
$\eta^0_0(s)$ for this purpose; we will be content with the
approximations above which seem sufficient for understanding the main
features of $\pi\pi-$scattering up to $1.2~GeV$.

In this procedure there is a practical subtlety already discussed at the
end of section IV of Ref.~\cite{Sannino-Schechter}. In order for
$\delta^0_0(s)$ to increase monotonically it is necessary that the
sign in front of the square root in Eq.~(\ref{imaginary}) change. This can
lead to a discontinuity unless $2|R^0_0(s)|$ precisely reaches
$\eta^0_0(s)$. However the phase shift is rather sensitive to small
deviations from this exact matching. Since the fitting procedure does
not enforce that $|R^0_0(s)|$ go precisely to $\eta^0_0(s)/2 \approx
0.5$, this results in some small discontinuities. (These could be
avoided by trying to fit the phase shift directly.) 

Figure~\ref{figura9} 
\figinsert{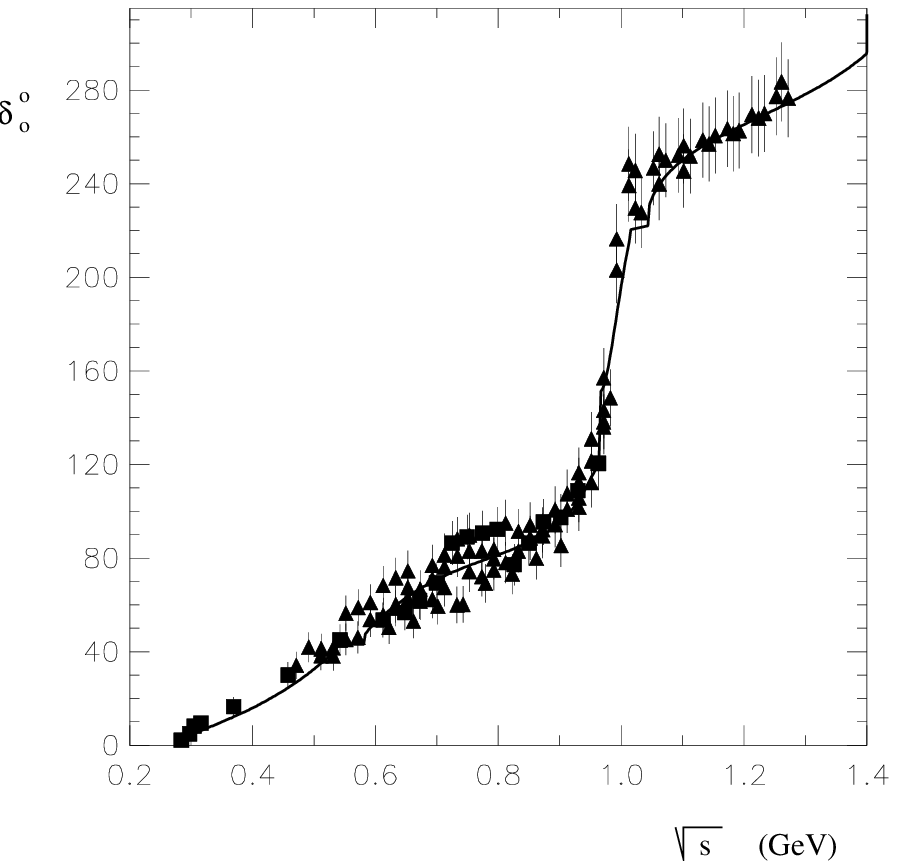}{Estimated phase shift using the
predicted real part and unitarity relation.}{figura9}{htpb}
shows the phase shift $\delta^0_0(s)$ estimated in this manner
for parameters in the first column of 
Table~\ref{tabella2}. As expected, the
agreement is reasonable. A very similar estimate is obtained when
(column 3 of 
Table~\ref{tabella2}) $\eta^0_0$ is taken to be $0.8$ while considering
the $\pi\pi$ branching ratio of $f_0(980)$ to be its experimental
value of $78.1\%$. It appears that these two parameter changes are
compensating for one other so that one may again conclude that the
turning on of the $K\overline{K}$ channel really does not have a major
effect. When the {\it next group} of resonances is included (column 7
of Table~\ref{tabella2}) 
the estimated $\delta^0_0(s)$ is very similar up to about
$1.2~GeV$. Beyond this point it is actually somewhat worse, as we
would expect by comparing Fig.~\ref{figura8}(b) with Fig.~\ref{figura8}(a).

\section{The inelastic channel $\pi\pi \rightarrow K \bar{K}$}
  
We have seen that $\pi\pi \rightarrow \pi \pi $ scattering can be understood
up to about $1.2~GeV$ without including this inelastic channel.  In
particular, a phenomenological description of the inelasticity did not
change the overall picture. However we would like to begin to explore
the predictions of the present model for this channel also. The whole
coupled channel problem is a very complicated one so we will be
satisfied here to check that the procedure followed for the $\pi\pi$
elastic channel can lead to an inelastic amplitude which also
satisfies the unitarity bounds. Specifically we will confine our
attention to the real part of the $I=J=0$  
 $\pi\pi \rightarrow K\overline{K}$ amplitude, 
 $R^0_{12;0}$ defined in Eq.~(\ref{eq:wave}).

In exact analogy to the $\pi\pi \rightarrow \pi\pi$ case we first consider
the contribution of the contact plus the $K^*(892)$ plus the
$\sigma(550)$ terms. It is necessary to know the coupling strength of
the $\sigma$ to $K\overline{K}$, defined by the effective Lagrangian
piece
\begin{equation}
+\frac{\gamma_{\sigma K \overline{K}}}{2}\sigma \partial_\mu
{\overline{K}}
\partial^\mu {K}\ .
\end{equation}
If the $\sigma$ is ideally mixed and there is no OZI rule--violating
piece we would have $\gamma_{\sigma K \overline{K}}=\gamma_0$ as 
defined in Eq.~(\ref{la:sigma}). For definiteness, we shall adopt this
standard mixing assumption. The appropriate amplitudes are listed in
Appendix~A. Figure~\ref{figura10} 
\figinsert{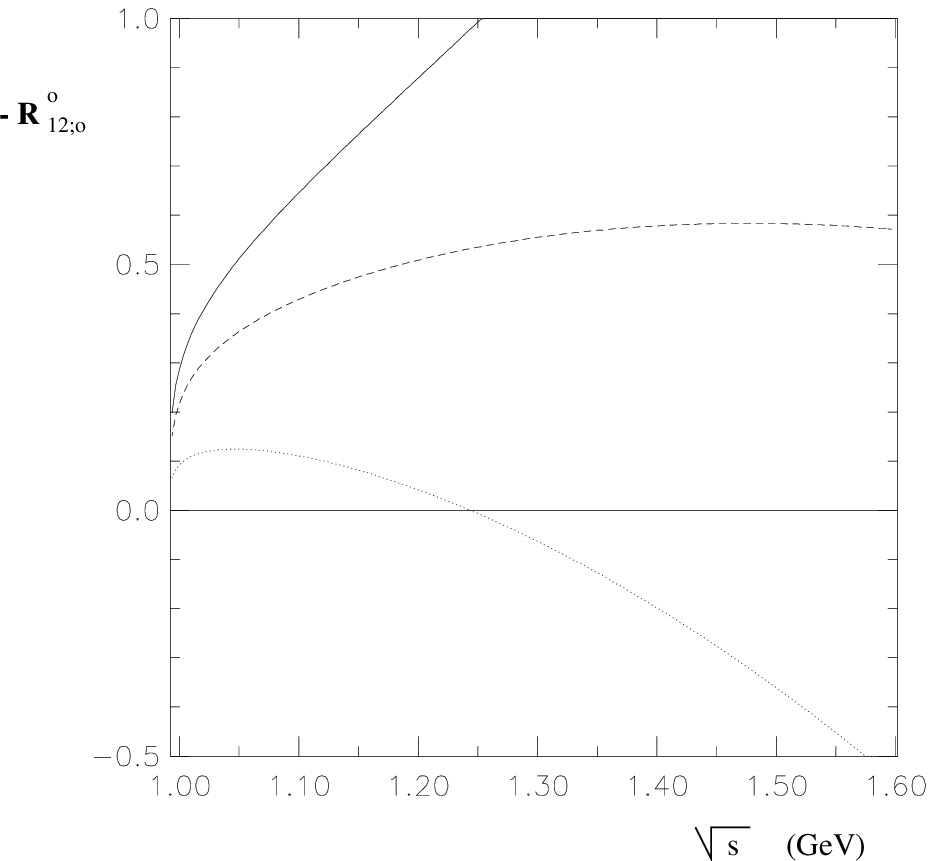}{Contributions to $\pi\pi \rightarrow K\overline{K}$
($R^0_{12;0}$). The solid line shows the current algebra result, the
dashed line represents the inclusion of $K^*(892)$, the dotted line
includes the $\sigma(550)$ too.}{figura10}{hptb}
shows the plots of  $R^0_{12;0}$ for the
current algebra part alone, the current algebra plus $K^*$ and the
current algebra plus $K^*$ plus $\sigma$ parts. Notice that
unitarity requires
\begin{equation}
| R^0_{12;0}|\leq\frac{\sqrt{1-{\eta^0_0}^2}}{2}\leq \frac{1}{2}\ .
\end{equation}
The current algebra result already clearly violates this bound at
$1.05~GeV$. As before, this is improved by the $K^*$ vector--meson
exchange contribution and further improved by the very important tail
of the $\sigma$ contribution. The sum of all three shows a structure
similar to the corresponding Fig.~\ref{figura2} in the $\pi\pi \rightarrow \pi\pi$ case.
The unitarity bound is not violated until about $1.55~GeV$. 

Next, let us consider the contribution of the $f_0(980)$ which, since
the resonance straddles the threshold, is expected to be important. We
need to know the effective coupling constant of the $f_0$ to $\pi\pi$
and to $K\overline{K}$. As we saw in Eq.~(\ref{f0(980)}), and the
subsequent discussion, the effective $\pi\pi$ coupling should be taken
as $\gamma_{f_0\pi\pi} e^{i\frac{\pi}{2}}$. Experimentally, only the
branching ratios for $f_0(980)\rightarrow \pi\pi$ and $f_0(980)\rightarrow
K\overline{K}$ are accurately known. We will adopt for definiteness
the value of $\gamma_{f_0\pi\pi}$ corresponding to the fit in the third
column of Table~\ref{tabella2} ($\Gamma_{tot}(f_0(980))=76~MeV$). It is more
difficult to estimate the $f_0(980)\rightarrow K\overline{K}$ effective
coupling constant since the central value of the resonance may
actually lie {\it below} the threshold. By taking account
\footnote{With $\Gamma_{tot}(f_0(980))=76~MeV$ we would have $\Gamma 
(f_0(980)\rightarrow K\overline{K}))=16.6~MeV$. Then $\gamma_{f_0 K
\overline{K}}$ is estimated from the formula:
\begin{displaymath}
16.6~MeV=|\gamma_{f_0 K \overline{K}}|^2\int_{2m_k}^{\infty}
\rho(M) |A(f_0(M)\rightarrow K\overline{K})|^2 \Phi (M)\,dM\ , 
\end{displaymath}
where $A(f_0(M)\rightarrow K\overline{K})$ is the reduced amplitude for an
$f_0$ of mass M to decay to $K\overline{K}$, $\Phi(M)$ is the phase
space factor and $\rho(M)$ is the weighting function given by 
\begin{displaymath}
\rho(M)=\sqrt{\frac{2}{\pi}}\frac{1}{\Gamma_{tot}}exp\left\{-2\left[\frac
{(M-M_0)^2}{\Gamma_{tot}^2}\right] \right\}\ .
\end{displaymath}
Here, $M_0$ is the central mass value of the $f_0(980)$.}
of the finite
width of the $f_0(980)$ we get the rough estimate 
$|\gamma_{f_0 K \overline{K}}|=10~GeV^{-1}\approx 4
|\gamma_{f_0\pi\pi}|$ for the choice in the third column,
$M_{f_0(980)}=990~MeV$. Of course, this estimate is very sensitive to
the exact value used for $M_{f_0(980)}$. It seems reasonable to take 
$\gamma_{f_0 K \overline{K}}$ to be purely real. The results of
including the $f_0(980)$ contribution, for both sign choices of $
 \gamma_{f_0 K \overline{K}}$, are shown in Fig.~\ref{figura11}.
\figinsert{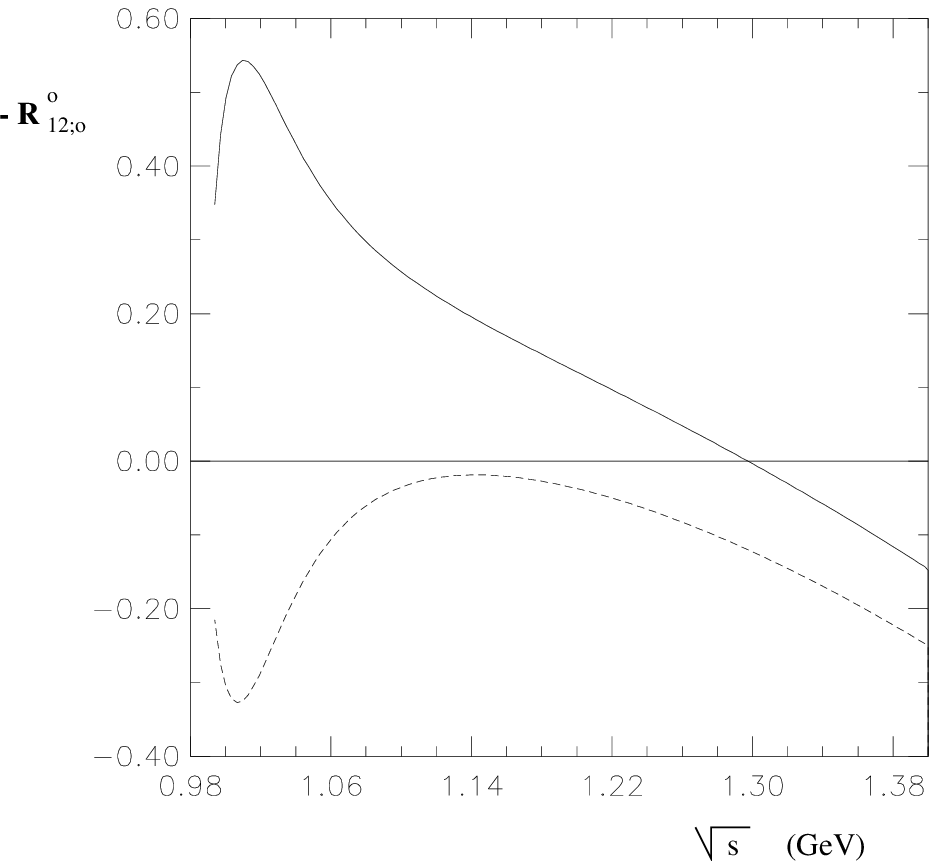}{Effect of $f_0(980)$ on $\pi\pi \rightarrow
K\overline{K}$. The solid curve corresponds to a negative $\gamma_{f_0
K \overline{K}}$ and the dashed one to a positive sign.}{figura11}{htpb} 
The unitarity
bounds are satisfied for the positive sign of $\gamma_{f_0 K
\overline{K}}$ but slightly violated for the negative sign choice.

Finally, let us consider the contributions to $\pi\pi \rightarrow
K\overline{K}$ from the members of the multiplets containing the {\it
next group} of particles. There will be a crossed channel contribution
from the strange excited vector meson $K^*(1410)$.  However it will be
very small since $K^*(1410)$ predominately couples to $K^*\pi$ and
has only a $7\%$ branching ratio to $K\pi$. In addition there will be
a crossed channel scalar $K^*_0(1430)$ diagram as well as a direct
channel scalar $f_0(1300)$ diagram contributing to $\pi\pi \rightarrow
K\overline{K}$. The $f_0(1300)$ piece is small because $f_0(1300)$ has
a very small branching ratio to $K\overline{K}$. Furthermore the
$K^*_0(1430)$ piece turns out also to be small; we have seen that the
crossed channel scalar gave a negligible contribution to $\pi\pi \rightarrow
\pi\pi$. The dominant {\it next group} diagrams involve the tensor
mesons. Near threshold, the crossed channel $K^*_2(1430)$ diagram is
the essential one since the direct channel $f_2(1270)$ contribution
for the $J=0$ partial wave is suppressed by a spin-2 projection
operator. Above $1270~MeV$ the $f_2(1270)$ contribution becomes
increasingly important although it has the opposite sign to the
crossed channel tensor piece.  
Figure~\ref{figura12} shows the net prediction for $R_{12;0}^0$ obtained with the inclusion of the main {\it next
group} contributions from the $K^*_2(1430)$ and $f_2(1270)$.
\figinsert{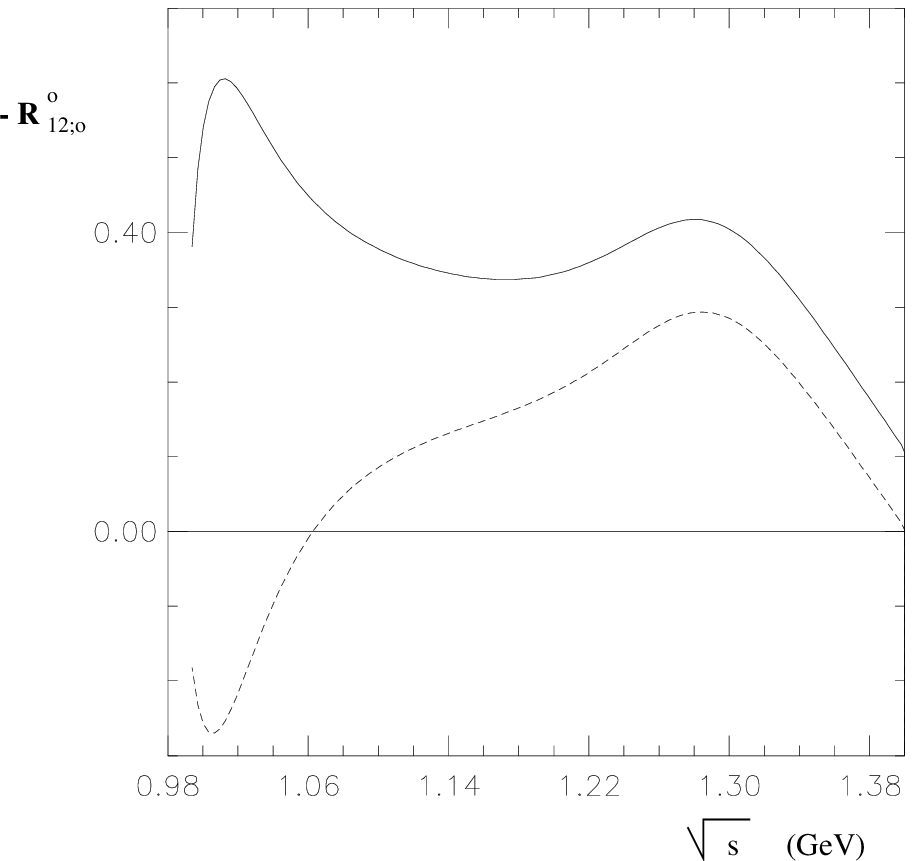}{Effects on $\pi\pi \rightarrow K \overline{K}$ due
to the {\it next group} of resonances for the two different sign
choices in Fig. \ref{figura11}.}{figura12}{htpb} 
Both assumed signs for $\gamma_{f_{0} K\overline{K}}$ are shown and other
parameters correspond to column 3 of Table~\ref{tabella2}. Clearly there is an
appreciable effect. Figure~\ref{figura13} 
\figinsert{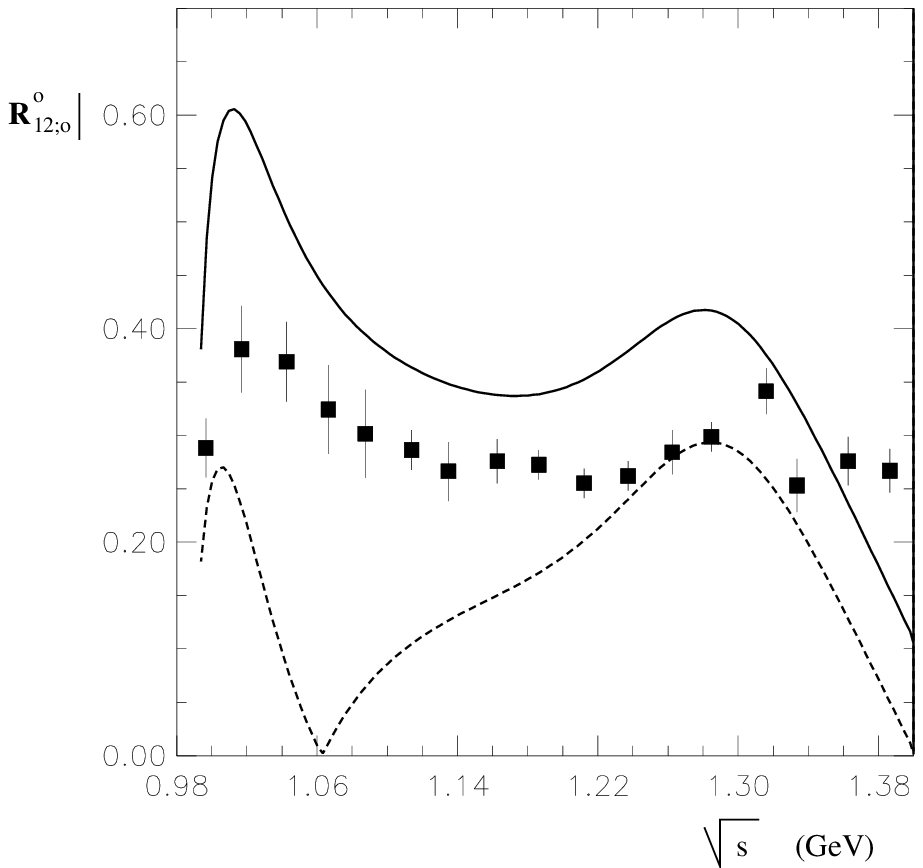}{ $|R_{12;0}^0|$ together with one experimental
determination \cite{Cohen:80} of
$\displaystyle{|T_{12;0}^0|}=\sqrt{(R_{12;0}^0)^2+(I_{12;0}^0)^2}$.
Signs for $\gamma_{f_0 K \overline{K}}$ as in Fig.~\ref{figura11}.}{figura13}{htpb}
shows the magnitude of $|R_{12;0}^0|$ together with one experimental
determination \cite{Cohen:80} of
$\displaystyle{|T_{12;0}^0|}=\sqrt{(R_{12;0}^0)^2+(I_{12;0}^0)^2}$. The
positive sign of  $\gamma_{f_0K\overline{K}}$ is favored but,
considering the uncertainty in  $|\gamma_{f_0K\overline{K}}|$ among
other things, we shall not insist on this. It seems to us that the
main conclusion is that the unitarity bound can be satisfied in the
energy range of interest. In this analysis we have shown that 
the {\it local cancellation} principle is satisfied and that the $\sigma$ 
plays an important role in the $\pi\pi\rightarrow K\overline{K}$ channel.

\section{The lonely $\sigma$}

In Reference~\cite{Tornqvist-Roos}, T\"ornqvist and Roos presented a model
of $\pi\pi$ scattering which supports the existence of the old
$\sigma$ meson at a pole position, $s_0^{{1}/{2}}=0.470-i0.250$~GeV. 
While this
model is constructed to satisfy unitarity, it does not explicitly take
crossing symmetry into account. In particular, one may question
\cite{Speth} the validity of neglecting the crossed-channel
$\rho$ meson exchange contributions, which are generally considered to
be important. It is actually very complicated, as noted by the authors
themselves, to examine this question in their model. 
We can investigate this 
issue in the framework of the 
 Chiral Resonance Model 
\cite{Sannino-Schechter,Sanninomrst,Harada-Sannino-Schechter,Harada-Sannino-Schechter-comment} We find that the consistent neglect of the $\rho$ exchange
does not destroy the existence of the $\sigma$ meson but merely 
modifies its
parameters so that they get close to the results of
Ref.~\cite{Tornqvist-Roos}.   

In the previous chapters \cite{Harada-Sannino-Schechter} a best fit 
to the real part of the $I=J=0$ 
partial amplitude $R^0_0$ was found for a mass 
$M_{\sigma}=559$ MeV, a width $G^{\prime}=370$ MeV 
and $G/G^{\prime}=0.29$ (pole position $s_0^{{1}/{2}}=0.585-i0.176$~GeV). 
It is an easy matter 
to neglect the $\rho$ meson contributions 
(including the associated contact term needed for chiral symmetry) 
and make a new fit. The resulting $R^0_0$ in comparison 
with the experimental data is shown in 
Fig.~\ref{figuracomment} and is about as good as the previous fit including 
the $\rho$ meson. (Of course, the $\rho$ meson is definitely present 
in nature.) The new fitted parameters are the mass 
$M_{\sigma}=378$ MeV, the width $G^{\prime}=836$ MeV and 
$G/G^{\prime}=0.08$ (pole position $s_0^{{1}/{2}}=0.495-i0.319$~GeV). 
The new mass and width are close 
to the values found in Ref.~\cite{Tornqvist-Roos}. 
We therefore would expect that including the $\rho$ exchange 
in their framework would raise their mass by roughly 
$100$ MeV and lower their width prediction. 
This behavior can be easily understood in a qualitative sense, 
since the addition of the $\rho$ raises the energy at 
which the unitarity bound is violated 
(see Fig.~\ref{figura1}). 
Of course, in the previous chapters, 
the question of whether the $\sigma$ and $f_0(980)$ 
are $q\bar{q}$, $q^2 \bar{q}^2$ states or 
some superposition is not directly addressed.  
\figinsert{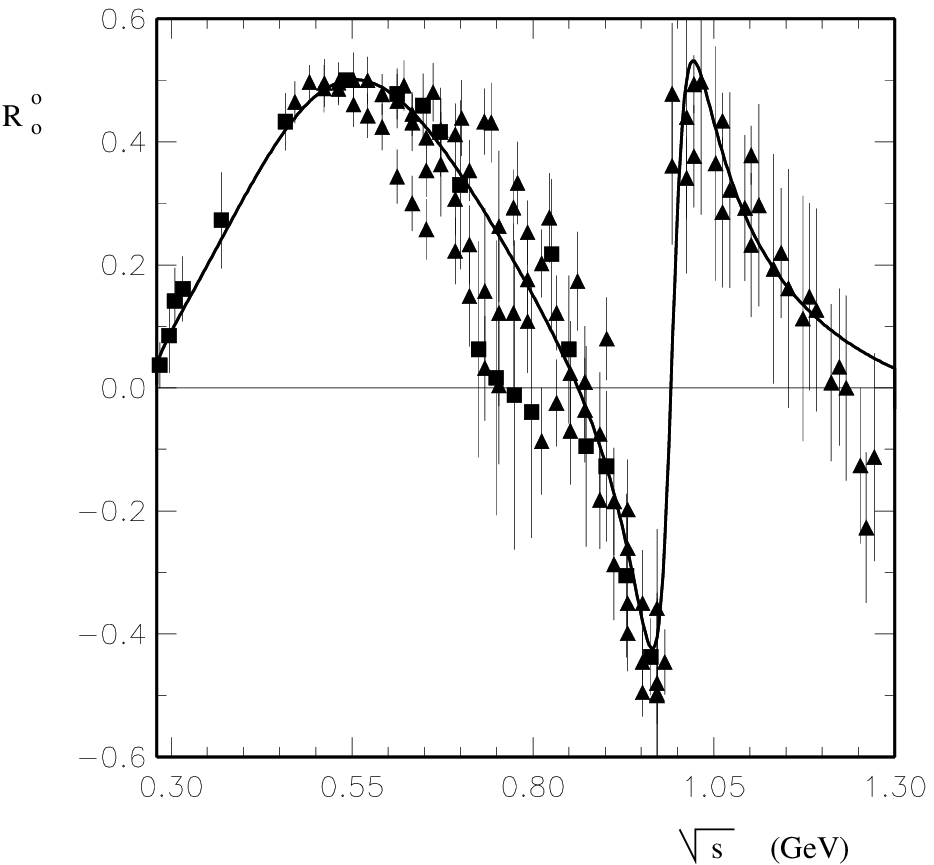}{The solid line is the {\it current algebra}
$+~\sigma +~f_0(980)$ result for $R^0_0$.}
{figuracomment}{htpb}

\section{Conclusions for the Chiral Resonance Model}

We have used the Chiral Resonance scheme to obtain 
a simple approximate analytic form for the real part
of the $\pi\pi-$scattering 
amplitude in the energy range from threshold to about
$1.2~GeV$. 
It satisfies both crossing symmetry and (more
non-trivially) unitarity in this range. 
Inspired by the leading $\displaystyle{1/N_c}$
approximation, we have written the amplitude as the sum of a contact
term and poles. Of course the leading $\displaystyle{1/N_c}$ amplitude can not be
directly compared with experiment since it is purely real (away from
the direct channel poles) and diverges at the pole positions.
Furthermore, an infinite number of poles and higher derivative
interactions are in principle needed. 
To overcome these problems we
have employed the following Chiral Resonance Model procedure:
\begin{itemize}
\item[a.]{We specialized to predicting the real part of the amplitude.}
\item[b.]{We postulated that including only resonances from threshold
to slightly more than the maximum energy of interest is sufficient. We
have seen that this {\it local cancellation} appears stable under the
addition of resonances in the $1300~MeV$ range. Beyond this range we
would expect still higher resonances to add in such a way as to
enforce unitarity at still higher energies. }
\item[c.]{In the effective interaction Lagrangian we included only
terms with the minimal number of derivatives consistent with the
assumed chiral symmetry.}
\item[d.]{The most subtle aspect concerns the method for regularizing
the divergences at the direct channel resonance poles. In the simplest
case of a single resonance dominating a particular channel (e.g. the
$\rho$ meson) it is sufficient to add the standard {\it width} term to
the denominator (e.g. the real part of Eq.~(\ref{Breit-Wigner})). For an
extremely broad resonance (like a needed low--energy scalar isosinglet)
the concept of {\it width} is not so clear and we employed the slight
modification of the Breit-Wigner amplitude given in
Eq.~(\ref{sigma-propagator}). Finally, for a relatively narrow resonance
in the presence of a non-negligible background, we employed the
regularization given in Eq.~(\ref{rescattering}) which includes the
background phase. Self-consistency is assured by requiring that the
background phase should be predicted by the model itself.}
\end{itemize}

All the regularizations introduced above are formally of higher than
leading order in the $\displaystyle{1/N_c}$ expansion (i.e. of order $1/N_c^2$ and
higher) and correspond physically to pion rescattering effects. 
This rescattering is schematically represented in Fig.~{\ref{coupling}}.
\figinsert{coupling.eps}{Effective resonance$-2\pi$ coupling due to 
the pions' rescattering effects. The latter has been shown schematically 
in a generic pertubative scheme. For simplicity we only considered the 
rescattering due to a four pion contact term ($\bullet$).}{coupling}{hpbt}
 In the case
of non-negligible background phase, there is an interesting difference
from the usual tree-level treatment of pole diagrams. The effective
squared coupling constant, $g^2_{R\pi\pi}$ of such a resonance to two
pions, is then not necessarily real and positive. Since this
regularization is interpreted as a rescattering effect it does not
mean that ghost fields are present in the theory. This formulation
maintains crossing symmetry which is typically lost when a
unitarization method is employed. 

In this analysis, the most non-trivial point is the satisfaction of
the unitarity bound for the predicted real part of the partial wave
amlitude, 
\begin{equation}
|R^I_l| \leq \frac{\eta^I_l}{2}\ ,
\end{equation}
where $\eta^I_l < 1 $ is the elasticity parameter. The well--known
difficulty concerns $R^0_0$. If $\eta^I_l (s)$ is known or calculated, 
the imaginary part $I^I_l (s)$ can be obtained, up to discrete
ambiguities, by Eq.~(\ref{imaginary}).

The picture of $\pi\pi$ scattering in the threshold to slightly more
than $1~GeV$ range which emerges from this model has four parts. Very
near threshold the current algebra contact term approximates
$R^0_0(s)$ very well. The imaginary part $I^0_0(s)$, which is formally
of order $1/N^2_c$ can be obtained from unitarity directly using 
Eq.~(\ref{imaginary}) or, equivalently, by chiral perturbation theory. 
At somewhat higher energies the most prominent feature is the $\rho$
meson pole in the $I=J=1$ channel. The crossed channel $\rho$ exchange
is also extremely important in taming the elastic unitarity violation
associated with the current algebra contact term (Fig.~\ref{figura1}). Even with
the $\rho$ present, Fig.~\ref{figura1} shows that unitarity is still violated, 
though much less drastically. This problem is overcome by introducing
a low mass $\approx 550~MeV$, extremely broad sigma meson. It also has
another desirable feature: $R^0_0(s)$ is boosted 
(see Fig.~\ref{figura3}) closer
to experiment in the $400-500~MeV$ range. The three parameters
characterizing this particle are essentially the only unknowns in the
model and were determined by making a best fit. In the $1~GeV$ region
it seems clear that the $f_0(980)$ resonance, interacting with the
predicted background in the {\it Ramsauer-Townsend}
effect manner, dominates the structure of the $I=J=0$ phase shift. The
inelasticity associated with the opening of the $K\overline{K}$
threshold has a relatively small effect. However we also presented a
preliminary calculation which shows that the present approach
satisfies the unitarity bounds in the inelastic $\pi\pi \rightarrow K
\overline{K}$ channel.

Other recent works 
\cite{Tornqvist:95,Morgan-Pennington:93,Harada,lnc,Ishida,GAM} which approach
the problem in different ways, also contain a low mass broad sigma. The
question of whether the lighter scalar mesons are of $q\overline{q}$ type or
{\it meson-meson} type has also been discussed
\cite{Tornqvist:95,Morgan-Pennington:93,Ishida}.
In our model it is difficult to decide this issue. Of course, it is
not a clean question from a field--theoretic standpoint. This question
is important for understanding whether the contributions
of such resonances are formally 
leading in the $\displaystyle{1/N_c}$ expansion. 
We are
postponing the answer as well as the answer to how to derive the
rescattering effects that were used to {\it regularize} the amplitude
near the direct channel poles as higher order in $\displaystyle{1/N_c}$ corrections. 
Presumably, the rescattering effects could some day be calculated as
loop corrections with a (very complicated) effective Wilsonian action.
This would be a generalization of the chiral perturbation scheme of pions.
Another aspect of the $\displaystyle{1/N_c}$ picture 
concerns the infinite number of
resonances which are expected to contribute already at leading order.
One may hope that the idea of {\it local cancellation} will help in
the development of a simple picture at high energies which might get
patched together with the present one. Is the simple high energy
theory a kind of string model ?

In this first part of the thesis we have shown 
\cite{Sannino-Schechter,Sanninomrst,Harada-Sannino-Schechter,Harada-Sannino-Schechter-comment} that it is possible 
to build a reasonable model for light meson interactions, which 
we have called the Chiral Resonance Model (ChRM). Using the previous scheme 
we have demonstrated that it is possible to understand
 $\pi\pi-$scattering up to the $1~GeV$ region by shoehorning together
poles and contact term contributions and employing a suitable
regularization procedure. It seems likely that any crossing symmetric
approximation will have a similar form. We will regard the ChRM as 
a leading order $\1N$ {\it mean field} approximation for the 
Quantum Chromodynamics.

%% file: chapterh1.tex
\chapter{Heavy Baryons in the Bound State Approach}
\section{Brief Introduction to the Heavy Physics}
In the second part of this thesis we will investigate 
the heavy 
baryon spectra. A heavy 
hadron  schematically consists of a heavy quark  
$Q$  of  spin $S_{\rm heavy}=1/2$
and a light cloud.  The latter describes the light degrees of  
freedom with 
total light spin $j_{l}$. There also exist heavy hadrons 
which contain a higher number of heavy quarks, but we will 
not consider them here.   
The total hadron spin is obtained 
by  adding together 
the heavy quark spin and the spin of the light cloud  
\begin{equation}
\mbox{\boldmath$J$}=\mbox{\boldmath$S$}_{\rm heavy} + 
\mbox{\boldmath$j$}_{l} \ .
\end{equation} 
A heavy baryon corresponds to an integer value of $j_{l}$, 
while a heavy meson 
to half--odd integer values of $j_l$. 

In the heavy quark mass limit ($m_Q/\Lambda\gg 1$) 
the heavy spin commutes with the QCD hamiltonian $H_{0}$ 
\begin{equation}
[H_{0},\mbox{\boldmath$S$}_{\rm heavy}]=0 \ .
\end{equation}
Hence the heavy spin is now a {\it good} quantum number leading 
to the {\it heavy spin} symmetry.  Since in the heavy limit 
the hamiltonian cannot depend on the mass of the heavy quark it is 
not possible to distinguish among heavy hadrons 
made with different heavy quarks. We will indicate this 
symmetry as {\it heavy flavor} symmetry. 
 Heavy spin symmetry predicts that for given $j_{l} \ne 0$ and 
fixed parity,   we have a degenerate doublet of total spin 
\begin{equation}
J=j_l \pm \frac{1}{2} \ .
\end{equation}
while for $j_l=0$ we have a heavy baryon with total spin $J=1/2$ 
which we can identify as $\Lambda_{Q}$, where $Q$ indicates the 
heavy quark contained in the heavy hadron. 
 Let us explicitly demonstrate that the splitting between the $J^P=1^-$ and 
$J^P=0^-$ states of a $Q\bar{q}$ meson must 
vanish in the limit of infinite heavy quark mass. Since the 
action of $S_{\rm heavy}^3$ on a $0^-$ state produces 
a $1^-$ state, i.e. $|M_{1^-}>= 2 S_{\rm heavy}^3 |M_{0^-}>$  we then have 
\begin{equation}
H_{0}|M_{1^-}>=m_{1^-}|M_{1^-}>
= 2 S_{\rm heavy}^3 H_{0}|M_{0^-}> = m_{0^-}|M_{1^-}>   \ ,
\end{equation}
implying that $m_{1^-} - m_{0^-}\rightarrow 0$ as 
 $m_{Q} \rightarrow \infty$. Experimentally \cite{pdg} we have 
\begin{equation}
\frac{m_{D^*}-m_D}{m_D}\approx 8\% \ , \qquad
\frac{m_{B^*}-m_B}{m_B}\approx 1\% \ ,
\end{equation}
which indicates the goodness of the $1/M$ expansion. We also notice 
that the heavy spin breaking is  $O(1/M)$ as can be understood 
by looking at a one gluon exchange diagram.

\section{Bound State Approach to the Heavy Baryon system}
There has been recent interest in studying heavy baryons (those with
the valence quark structure $qqQ$) in the bound state
picture~\cite{Callan-Klebanov,Blaizot-Rho-Scoccola} together with
heavy quark spin symmetry~\cite{Eichten-Feinberg}.
In this picture the heavy baryon is treated~\cite{%
Guralnik-Luke-Manohar,Rho,%
Gupta-Momen-Schechter-Subbaraman,Schechter-Subbaraman,%
Schechter-Subbaraman-Vaidya-Weigel,Oh-Park} 
as a heavy spin multiplet of
mesons ($Q\bar{q}$) bound in the background field of the nucleon 
($qqq$), which in turn arises as a soliton configuration of light 
meson fields.

A nice feature of this approach is that it permits, in principle, an
exact expansion of the heavy baryon properties in simultaneous powers
of $1/M$ and $1/N_{c}$.
In the simplest treatments, the light part of the chiral Lagrangian 
is made from only 
pion fields.
However it has been shown that the introduction of light vector
mesons~\cite{%
Gupta-Momen-Schechter-Subbaraman,Schechter-Subbaraman,%
Schechter-Subbaraman-Vaidya-Weigel}
substantially improves the accuracy of the model.
This is also true for the soliton treatment of the nucleon
itself~\cite{JJPSW,JPSSW,PW}.
Furthermore finite $M$ corrections as well as finite
$N_{c}$ (nucleon recoil) corrections are also important.
This has been recently demonstrated for the hyperfine splitting
problem~\cite{HQSSW:1,HQSSW:NPA} and it will be explained in some detail in
 the next 
chapter.

Since the bound state--soliton approach is somewhat involved
it may be worthwhile to point out a couple of its advantages.
In the first place, it is based on an effective chiral Lagrangian
containing physical parameters which are in principle subject to
direct experimental test.
Secondly, the bound state approach models a characteristic feature of
a confining theory.
When the bound system is suitably ``stretched'' it does not separate
into colored objects but into physical color singlet states.

\subsection{Effective Lagrangian for the Heavy-Light system}
Here we will review the chiral Lagrangian for 
the low lying heavy mesons in the heavy limit. In the 
next chapter we will see how to include next to leading  
corrections in $1/M$. The model is based on a chiral Lagrangian 
with two parts,
\begin{equation}
{\cal L} = {\cal L}_{\rm light} + {\cal L}_{\rm heavy} 
\end{equation}
The light part involves the chiral field
$U=\xi^2=\exp\left(2i\phi/F_\pi\right)$,
where $\phi$ is the $3\times3$ matrix of standard pseudoscalars. 
Relevant vector and pseudovector combinations are  
 $v_\mu $ and $ p_\mu $ defined in Eq.~(\ref{maurer}).  
In addition light vector mesons are included in a $3\times3$ matrix
field $\rho_\mu$.
The explicit form of ${\cal L}_{\rm light}$ is as in Ref.~\cite{HQSSW:NPA}. 

The heavy field $H$, which describes the heavy multiplet which contains the 
 heavy pseudoscalar  $P$ and a heavy  
vector meson $Q_{\mu}$ is 
\begin{eqnarray}
H&=&\frac{1+\gamma_{\mu}V^{\mu}}{2}\left[i \gamma_5
P^{\prime}+\gamma^{\alpha}{Q}^{~\prime}_ {\alpha}\right] \ , \nonumber \\
\overline{H}&=&\gamma_0 H^{\dagger} \gamma_0 \ ,
\label{Hdef}
\end{eqnarray}
where in the leading order in $1/M$ the pseudoscalar fluctuation field 
($P^{\prime}$) 
and the vector meson fluctuation ($Q^{\prime}_{\mu}$) are connected with the heavy fields via
\begin{equation}
P=e^{-iMV\cdot x}P^{\prime} \ , \quad 
Q_{\mu}=e^{-iMV\cdot x}Q^{\prime}_{\mu} \ .
\end{equation} 
$V_\mu$ is the super selected four
 velocity of the heavy meson.

 Under a heavy spin transformation $S$, $H$ transforms as  
\begin{equation} 
H\rightarrow S H \ ,
\end{equation}
while under a chiral transformation $H$ transforms as a matter field 
\begin{equation}
H \rightarrow H K^{\dagger} \ ,
\end{equation}
where $K$ is defined in Eq.~(\ref{nonlinear}). 
The heavy spin and chiral preserving lagrangian takes the form 
~\cite{Schechter-Subbaraman:2}
\begin{equation}
{\cal L}_{\rm heavy}/M
= i V_\mu \mbox{Tr}\, \left[ H D^\mu \bar{H} \right]
- d \, \mbox{Tr}\, \left[ H \gamma_\mu \gamma_5 p^\mu \bar{H}
\right] 
+ \frac{ic}{m_V} \mbox{Tr} \, \left[
H \gamma_\mu \gamma_\nu F^{\mu\nu}(\rho) \bar{H} \right]
\ ,
\label{Lag for H}
\end{equation}
 where 
 $D_\mu \equiv \partial_\mu + i \alpha \tilde{g} \rho_\mu - i (1-\alpha) v_\mu$, and $F_{\mu\nu}(\rho) = \partial_\mu
 \rho_\nu - \partial_\nu \rho_\mu + i \widetilde{g} 
 \left[ \rho_\mu \,,\, \rho_\nu \right]$.
 Furthermore, $m_V$ is the light vector meson mass while
 $d\simeq0.53$ and $c\simeq1.6$ are respectively the heavy
 meson--pion and magnetic type heavy meson--light vector meson coupling
 constants; $\alpha$ is a coupling constant whose value has not yet
 been firmly 
 established.

\section{Mechanics of Baryon States}
Following the Callan-Klebanov idea \cite{Callan-Klebanov}, we first find the classical 
solution of the light meson action and then obtain the classical 
approximation to the wave function in which this {\it baryon as 
soliton} is bound to a heavy meson (yielding a heavy baryon).
The {\it hedgehog} ansatz for the classical light baryon in the $SU(2)$ 
case simply corresponds to
\begin{eqnarray}
&&
\xi_{\rm c} (\mbox{\boldmath$x$}) = 
\exp \left[ \frac{i}{2} \hat{\mbox{\boldmath$x$}}\cdot
\mbox{\boldmath$\tau$} \, F(|\mbox{\boldmath$x$}|) \right]
\ ,
\nonumber\\
&&
\rho^a_{i{\rm c}} = 
\frac{1}{\sqrt{2}\tilde{g}\vert\mbox{\boldmath$x$}\vert}
\epsilon_{ika} \hat{x}_k G(\vert\mbox{\boldmath$x$}\vert)
\ ,
\nonumber\\
&&
\omega_{0{\rm c}} = \omega(\vert\mbox{\boldmath$x$}\vert)
\ ,
\nonumber\\
&&
\rho^a_{0{\rm c}} = \omega_{i{\rm c}} = 0 \ ,
\label{xi classical}
\end{eqnarray}
 where
 $\rho_{\mu{\rm c}} = \frac{1}{\sqrt{2}}
 \left( \omega_{\mu{\rm c}} + \tau^a \rho^a_{\mu{\rm c}} \right)$
 and $\tilde{g}$ is a coupling constant.
The appropriate boundary conditions are
\begin{eqnarray}
&&
F(0) = -\pi \ , \quad G(0) = 2 \ , \quad \omega^\prime(0) = 0 \ ,
\nonumber\\
&&
F(\infty) = G(\infty) = \omega(\infty) = 0 \ , 
\end{eqnarray}
which correspond to unit baryon number.
The wave function for the heavy meson bound to the background Skyrmion
field (\ref{xi classical}) is conveniently presented in the rest
frame, $\mbox{\boldmath$V$}=0$.
In this frame
\begin{equation}
\bar{H}_{\rm c} \rightarrow
\left( \begin{array}{cc}
0 & 0 \\ \bar{h}_{lh}^a & 0 
\end{array}\right) 
\ ,
\label{H: matrix}
\end{equation}
with $a$, $l$, $h$ representing respectively the isospin, light spin
and heavy spin bivalent indices.
The calculation simplifies if we deal with a radial wave function
obtained after removing the factor
$\hat{\mbox{\boldmath$x$}}\cdot\mbox{\boldmath$\tau$}$:
\begin{equation}
\bar{h}_{lh}^a = \frac{u(|\mbox{\boldmath$x$}|)}{\sqrt{M}}
\left(
  \hat{\mbox{\boldmath$x$}}\cdot\mbox{\boldmath$\tau$}
\right)_{ad}
\psi_{dl,h} 
\ ,
\label{h: classical}
\label{Wf:h bar}
\end{equation}
where $u(|\mbox{\boldmath$x$}|)$ is a radial wave function, assumed 
to be very sharply peaked near $|\mbox{\boldmath$x$}|=0$ for large $M$ 
(i.e. $r^2|u(r)|^2 \approx \delta(r)$).
The heavy spinor $\chi_h$ can be trivially factored out 
\begin{equation}
\psi_{dl,h}= \psi_{dl}\chi_h
\end{equation}
in this expression
as a manifestation of the heavy quark symmetry.
We perform a partial wave analysis of the generalized ``angular'' wave
function $\psi_{dl}$:
\begin{equation}
\psi_{dl}\left(g,g_3;r,k\right) = 
\sum_{r_3,k_3} C_{r_3,k_3;g_3}^{r,k;g}
Y^{r_3}_r \xi_{dl}(k,k_3) \ .
\label{def:wave function}
\end{equation}
Here $Y^{r_3}_r$ stands for the standard spherical harmonic
representing orbital angular momentum $r$ while $C$ denotes the
ordinary Clebsch--Gordan coefficients.
$\xi_{dl}(k,k_3)$ represents a wave function in which the ``light
spin'' and isospin (referring to the ``light cloud'' component 
of the heavy meson) are added vectorially to give
\begin{equation}
\mbox{\boldmath$K$} = \mbox{\boldmath$I$}_{\rm light}
+ \mbox{\boldmath$S$}_{\rm light} \ ,
\label{def: k}
\end{equation}
with eigenvalues $\mbox{\boldmath$K$}^2=k(k+1)$.
The total light ``grand spin''
\begin{equation}
\mbox{\boldmath$g$} = \mbox{\boldmath$r$} +
\mbox{\boldmath$K$} \ , 
\label{def: g1}
\end{equation}
is a {\it good} quantum number in the heavy limit.
It is also convenient to define the total {\it grand spin} operator 
\begin{equation}
\mbox{\boldmath$G$} = \mbox{\boldmath$g$} +
\mbox{\boldmath$S$}_{\rm heavy} \ , 
\end{equation}
where $\mbox{\boldmath$S$}_{\rm heavy}$ is the spin of the heavy quark 
within the heavy meson. 
Substituting the wave--function (\ref{Wf:h bar}) into 
$\int\,d^3x\,{\cal L}_{\rm heavy}$ given in Eq.~(\ref{Lag for H})
yields the potential operator
\begin{eqnarray}
V &=& \int d \Omega \ \psi^{\ast} 
\left\{
  \mbox{\boldmath$\sigma$}\cdot\mbox{\boldmath$\tau$} 
  \Delta_1 + 1\, \Delta_2
\right\} \psi
\nonumber\\
&=&
\int \, d\Omega\ \psi^{\ast}
\left\{
  4 \Delta_1 \mbox{\boldmath$S$}_{\rm light} \cdot
  \mbox{\boldmath$I$}_{\rm light} 
  + 1\, \Delta_2
\right\}
\psi
\nonumber\\
&=&
2 \Delta_1 \, \left[ k(k+1) - \frac{3}{2} \right] + \Delta_2
\ ,
\label{potential: H}
\end{eqnarray}
where $\int\,d\Omega$ is the solid angle integration and 
Eq.~(\ref{def: k}) was used in the last step.
In addition
\begin{eqnarray}
\Delta_1 &=& \frac{1}{2} d\, F'(0) - \frac{c}{m_V\tilde{g}}
G^{\prime\prime}(0) \ ,
\nonumber\\
\Delta_2 &=& - \frac{\alpha\tilde{g}}{\sqrt{2}} \omega(0)
\ .
\label{binding}
\end{eqnarray}
The $\Delta_2$ term is relatively small~\cite{%
Schechter-Subbaraman,Schechter-Subbaraman-Vaidya-Weigel,HQSSW:NPA}.
Both terms in $\Delta_1$ are positive with the second one (due to
light vectors) slightly larger.
There are just the two possibilities $k=0$ and $k=1$.
It is seen that the $k=0$ states, for any orbital angular momentum
$r$, will be bound with binding energy 
$3\Delta_1$.
The $k=1$ states are unbound in this limit.
The parity of the bound state wave function is 
\begin{equation}
P_B = \left( -1 \right)^r \ ,
\label{baryon parity}
\end{equation}
which emerges as a product of $\left(-1\right)^r$ for $Y^{r_3}_r$ in 
Eq.~(\ref{def:wave function}), $-1$ for the 
$\hat{\mbox{\boldmath$x$}}\cdot\mbox{\boldmath$\tau$}$ factor in
Eq.~(\ref{Wf:h bar}) and $-1$ due to the fact that 
the mesons ($P,Q$) bound to the soliton have negative parity.

\section{Collective Quantization}
\label{collective-quantization}
In the soliton approach, the particle states with definite rotational 
and flavor quantum numbers appear after the so--called 
{\it rotational collective modes} are introduced and the theory is 
quantized. This is conveniently done by first finding the time 
independent parameters which leave the theory invariant, then 
those {\it collective} parameters are allowed to depend on time.
The collective angle--type coordinate $A(t)$ is introduced~\cite{Ad83} as 
\begin{eqnarray}
\xi(\mbox{\boldmath$x$},t) 
&=&
A(t) \xi_{\rm c} (\mbox{\boldmath$x$}) A^{\dag}(t)
\ ,
\nonumber\\
\mbox{\boldmath$\tau$}\cdot\mbox{\boldmath$\rho$} 
\left( \mbox{\boldmath$x$}\,,\,t\right)
&=&
A(t) \mbox{\boldmath$\tau$}\cdot\mbox{\boldmath$\rho$}_{\rm c}
\left(\mbox{\boldmath$x$}\right) A^{-1}(t) \ ,
\nonumber\\
\bar{H}(\mbox{\boldmath$x$},t) 
&=&
A(t) \bar{H}_{\rm c} (\mbox{\boldmath$x$}) 
\ ,
\label{def: collective}
\end{eqnarray}
where $\xi_{\rm c}$ and $\mbox{\boldmath$\rho$}_{\rm c}$
are defined in Eq.~(\ref{xi classical}) and
$\bar{H}_{\rm c}$ in Eqs.~(\ref{H: matrix}) and
(\ref{h: classical}). For our purposes the important variable is the
``angular--velocity'' $\mbox{\boldmath$\Omega$}$ defined by
\begin{equation}
A^{\dag} \dot{A} = \frac{i}{2} 
\mbox{\boldmath$\tau$} \cdot \mbox{\boldmath$\Omega$}
\ ,
\label{def: omega}
\end{equation}
which measures the time dependence of the collective coordinates
$A(t)$. 
It should furthermore be mentioned that, due to the collective 
rotation, the vector meson field components which vanish classically 
($\rho^a_0$ and $\omega_i$) get induced:
\begin{eqnarray}
\omega_i=-\frac{\sqrt{2}}{r}\varphi(r)\epsilon_{ijk}\Omega_j{\hat r}_k
\quad {\rm and}\quad
\rho_0^k=-\frac{1}{\sqrt{2}}\left(
\xi_1(r)\Omega_k+\xi_2(r){\hat{\mbox{\boldmath $r$}}}\cdot
\mbox{\boldmath $\Omega$}{\hat r}_k \right)\ .
\label{induced}
\end{eqnarray}
Substituting Eq.~(\ref{def: collective}) and Eq.~(\ref{induced}) into 
$\int\,d^3x\, \left({\cal L}_{\rm light} + {\cal L}_{\rm heavy}\right)$ gives 
an additional contribution to the lagrangian of the general form 
\begin{equation}
L_{coll}=\frac{1}{2}\alpha^2 \mbox{\boldmath$\Omega$}^2 - 
\chi \mbox{\boldmath$\Omega$}\cdot \mbox{\boldmath$G$} \ , 
\label{simple l coll}
\end{equation}
in which $\alpha$ and $\chi$ represent spatial integrals over 
the profiles in Eq.~(\ref{xi classical}).
The 
induced radial functions $\varphi(r)$, $\xi_1(r)$ and $\xi_2(r)$ are 
obtained from a variational approach to $\alpha^2$ 
\cite{Me89}.
For each bound state solution $\bar{H}_{\rm c}$, there will be a tower
of states characterized by a soliton angular momentum 
$\mbox{\boldmath$J$}^{\rm sol}$ and the total isospin 
$\mbox{\boldmath$I$}$ satisfying $I=J^{\rm sol}$.
The soliton angular 
momentum is computed from this collective Lagrangian as
\begin{equation}
\mbox{\boldmath $J$}^{\rm sol} = 
\frac{\partial L_{\rm coll}}{\partial \mbox{\boldmath $\Omega$}}
\ ,
\label{Jsol}
\end{equation}
while the total baryon angular momentum is the sum
\begin{equation}
\mbox{\boldmath$J$} =
\mbox{\boldmath$g$} + \mbox{\boldmath$J$}^{\rm sol} + 
\mbox{\boldmath$S$}_{\rm heavy}
\ .
\label{baryon spin}
\end{equation}

The rotational collective Hamiltonian is obtained by performing 
the standard Legendre transform
\begin{equation}
H_{coll} = \frac{1}{2\alpha^2} \left(\mbox{\boldmath$J$}^{\rm sol} + 
\chi \mbox{\boldmath$G$} \right)^2 \ .
\end{equation}
The moment of inertia $\alpha^2$ is identified from the light soliton 
sector as $\alpha^{-2}=\frac{2}{3}(m_{\Delta} - m_{N})$ in terms 
of the nucleon and $\Delta$ masses. Equation (\ref{simple l coll}) can be 
simplified by noting that the total angular momentum is given by 
$\mbox{\boldmath$J$}=
\mbox{\boldmath$J$}^{\rm sol} + \mbox{\boldmath$G$}$. Then we 
deduce the heavy baryon mass formula \cite{Callan-Klebanov} 
\begin{equation}
H_{coll} = \frac{1}{3} (m_{\Delta} - m_{N}) 
\left[(1-\chi)I(I+1) + \chi J(J+1) + \cdots \right] \ ,
\label{hcollective1}
\end{equation}
where the ellipsis stands for the 
 $\chi(\chi-1) G(G+1)$ term which does not split the heavy 
baryon masses. In the heavy limit ${\cal L}_{\rm heavy}$ 
leads to $\chi=0$. Thus we have the final results 
\begin{eqnarray}
m(\Sigma^*_Q) - m(\Sigma_{Q})&=& 0 \ ,  \\ 
m(\Sigma_Q) - m(\Lambda_{Q})&=&  \frac{2}{3} (m_{\Delta} - m_{N}) 
\label{theretical 1}\ ,
\end{eqnarray} 
wherein the subscript $Q$ denotes the heavy baryon which contains the 
heavy quark $Q$. It may be interesting to compare the 
 experimental determination \cite{pdg} for 
$m(\Sigma_c) - m(\Lambda_{c}) \simeq  170~{\rm MeV}$ whith the 
theoretical prediction in Eq.~(\ref{theretical 1}) which provides 195 MeV.
 The result $m(\Sigma^*_Q) - m(\Sigma_{Q})= 0$ is, 
of course, expected in the heavy quark spin symmetry limit. However 
if we consider heavy spin violating term (i.e. $\chi \ne 0$) we get 
\begin{equation}
m(\Sigma^*_Q) - m(\Sigma_{Q})=\chi \, (m_{\Delta} - m_{N}) \ .
\end{equation} 
The next chapter will be devoted to the calculation of the hyperfine 
splitting parameter $\chi$.

%% file: chapterh2.tex
\chapter{Heavy Baryon Hyperfine Splitting}

\section{Introduction }
We have already noticed that a  
compelling feature of the heavy soliton approach is 
that it permits, in principle,
an exact expansion of the heavy baryon properties in simultaneous 
powers of $1/M$, $1/N_{c}$ and, since it is based on a chiral 
Lagrangian, number of derivatives acting on the light components 
of the heavy system. In practice there are obstacles related to the 
large number of unknown parameters which must be introduced.
Rather than treating the light soliton in a model with many 
derivatives of the light pseudoscalar fields it turns out to be 
much more efficient to use the light vector mesons. Based on a 
model~\cite{Schechter-Subbaraman} of the light vector interactions with the 
heavy multiplet, the leading order (in the $1/N_{c}$ and $1/M$ 
expansions) heavy baryon mass splittings have been discussed 
\cite{Schechter-Subbaraman-Vaidya-Weigel}, obtaining satisfactory agreement with experiment. 
Actually the need for light vector mesons is not surprising since, in 
the soliton approach, they are necessary to explain, for example, 
the neutron--proton mass difference~\cite{JJPSW} and the nucleon
axial singlet matrix element~\cite{JPSSW}.

In the present chapter we focus our attention on the hyperfine splitting,
which is of subleading order both in $1/M$ and $1/N_{c}$.
This is a more complicated calculation and also involves using a 
cranking procedure~\cite{Ad83} to obtain physical states which
carry good spin and isospin quantum numbers. The first calculation 
of the heavy baryon hyperfine splitting in the perturbative
bound state framework was carried out by Jenkins and 
Manohar~\cite{Guralnik-Luke-Manohar}
who got the formula
\begin{equation}
m(\Sigma_Q^{\ast}) - m(\Sigma_Q) =
\frac{\left( m(\Delta) - m(N) \right) \left(M^{\ast} - M\right)}{
4d\,F'(0)}
\ ,
\label{eq:1.1}
\end{equation}
where $M^{\ast}-M$ is the heavy vector--heavy pseudoscalar mass 
difference, $d$ is the light pseudoscalar--heavy meson coupling 
constant and $F'(0)$ is the slope of the Skyrme ``profile function''
at the origin.  This formula is obtained
(see also section 5) by using the leading order 
in number of derivatives (zero) and leading order in $1/M$ heavy 
spin violation term.  Therefore it is expected to provide the 
dominant contribution. Unfortunately, on evaluation, it is found 
to provide only a small portion of the experimental 
$\Sigma_c^{\ast}$--$\Sigma_c$ masss difference. This naturally 
suggests the need for including additional higher order in derivative 
heavy spin violation terms. However, there are many possible terms 
with unknown coefficients so that the systematic perturbative approach 
is not very predictive.

To overcome this problem we employ a relativistic Lagrangian 
model~\cite{Schechter-Subbaraman} which uses ordinary heavy pseudoscalar and 
vector fields rather than the heavy ``fluctuation'' field 
multiplet~\cite{Eichten-Feinberg}. This model reduces to the heavy multiplet 
approach in leading order and does not contain any new parameters.
We will show that \cite{HQSSW:1} such a model 
(considered, for simplicity, to contain only light pseudoscalars;
{\it i.e.}, the light part is the original Skyrme model~\cite{Sk61})
yields a ``hidden'' heavy spin violation which is not manifest 
from the form of the Lagrangian itself. This hidden part involves 
two derivatives and is actually more important numerically than the 
zero derivative ``manifest'' piece which leads to Eq~(\ref{eq:1.1}).
However this new result is still not sufficient to bring the 
predicted $\Sigma_c^{\ast}$--$\Sigma_c$ mass difference into 
agreement with experiment. The prediction for this 
difference is actually 
correlated to those for 
 $\Sigma_c$--$\Lambda_c$ and $\Delta$--$N$, the 
  $\Delta$ - nucleon  mass difference by \cite{Callan-Klebanov}:
\begin{equation}
m\left(\Sigma_c^{\ast}\right)-m\left(\Sigma_c\right)
=m\left(\Delta\right)-m\left(N\right)-\frac{3}{2}\left[
m\left(\Sigma_c\right)-m\left(\Lambda\right)\right] \ .
\label{ck63}
\end{equation}
This formula depends only on the collective quantization procedure 
being used rather than the detailed structure of the model. 
If $m\left(\Sigma_c\right)-m\left(\Lambda\right)$ and 
$m\left(\Delta\right)-m\left(N\right)$ are taken to agree with 
experiment, Eq~(\ref{ck63}) predicts $41$~MeV rather than the 
experimental value of $66$~MeV. This means that it is impossible 
to exactly predict, in models of the present type, all three 
mass differences which appear in Eq~(\ref{ck63}). The goodness 
of the overall fit must be judged by comparing all three 
quantities with experiment. Our focus, of course, is the left hand side 
of Eq~(\ref{ck63}) which is of order $1/M$ while the right hand 
side involves the difference of two order $M^0$ quantities. 
A similar calculation in the model with 
only light pseudoscalars was carried out by Oh and Park~\cite{Oh-Park}.
However, they did not make a $1/M$ expansion in order to reveal the 
hidden violation terms. They also introduced a one--derivative 
``manifest'' heavy spin violation term with a new relatively large 
unknown constant in order to improve the agreement with experiment.

In the present chapter we show that it is not necessary to introduce
any new violation terms to agree with experiment if a chiral 
Lagrangian including light vectors is employed. Typical results are 
summarized, compared with experiment and compared with the Skyrme 
model for the light sector in Table~\ref{tab:1.1}.
\begin{table}[htbp]
\begin{center}
\begin{tabular}{|c||c|c|c|c|}
\hline
mass difference & expt. & present model
& present model + CM & Skyrme \\
\hline\hline
$\Lambda_c-N$ & 1345 & 1257 & 1356 & 1553 \\
$\Lambda_b-\Lambda_c$ & 3356$\pm$50 & 3164 & 3285 & 3215 \\
$\Lambda'_c-\Lambda_c$ & 308 & 249 & 342 & 208 \\
$\Sigma_c-\Lambda_c$ & 168 & 172 & 158 & 185 \\
$\Sigma_c^{\ast}- \Sigma_c$ & 66 & 42 & 63 & 16\\
\hline
\end{tabular}
\end{center}
\caption{
Typical results for the present model (including light vectors)
compared with model with light pseudoscalars only (``Skyrme'' column)
and compared with experiment. No ``manifest'' heavy spin violation 
effects other than $M^{\ast}\neq M$ have been included. The column 
``present model + CM'' simply takes into account recoil corrections 
by replacing the heavy meson mass by the reduced mass. $\Lambda'_c$ 
denotes a negative parity, spin $1/2$ state. The quantity $\alpha$ 
in Eqs~(\ref{covderpq}) was taken to be zero.
All masses in MeV.}
\label{tab:1.1}
\end{table}
A much more detailed discussion is given later in the text.
We  notice from the last row, that the model with light 
vectors gives a very satisfactory account of the 
$\Sigma_c^{\ast}$--$\Sigma_c$ hyperfine splitting in 
contrast to the model without light vectors. There are also 
noticeable effects when the use of the heavy meson reduced 
mass is taken as a simple approximation for kinematical corrections.
Similarly, the first four rows of Table~\ref{tab:1.1} show that 
the other predictions of the model with light vectors agree well 
with experiment.  

\subsection{Relativistic Lagrangian for the Heavy Mesons.}
For the sector of the model describing the light pseudoscalar
and vector mesons we adopt the chirally invariant Lagrangian
discussed in detail in the literature \cite{Ka84,Ja88,HQSSW:NPA}.
We  now present the relativistic Lagrangian, which 
describes the coupling between the light and heavy mesons 
\cite{Schechter-Subbaraman}
\begin{eqnarray}
{\cal L}_H&=&D_\mu P D^\mu {\overline P}
-\frac{1}{2}Q_{\mu\nu} {\overline Q}^{\mu\nu}
-M^2P{\overline P} +M^{*2}Q_\mu {\overline Q}^{\mu}
\nonumber \\ &&
+2iMd\left(Pp_\mu {\overline Q}^{\mu}-Q_\mu p^\mu {\overline P}\right)
+\frac{d}{2}\epsilon^{\alpha\beta\mu\nu}
\left[Q_{\nu\alpha}p_\mu {\overline Q}_\beta +
Q_\beta p_\mu {\overline Q}_{\nu\alpha}\right]
\label{lagheavy} \\ &&
+\frac{2 icM}{m_V}\left\{
2Q_\mu F^{\mu\nu}\left(\rho\right){\overline Q}_\nu
-\frac{i}{M}\epsilon^{\alpha\beta\mu\nu}\left[
D_\beta PF_{\mu\nu}\left(\rho\right){\overline Q}_\alpha
+Q_\alpha F_{\mu\nu}\left(\rho\right)D_\beta {\overline P}
\right]\right\}.
\nonumber
\end{eqnarray}
Here we have allowed the mass $M$ of the heavy pseudoscalar meson
$P$ to differ from the mass $M^*$ of the heavy vector meson $Q_{\mu}$.
Note that the heavy meson fields are conventionally defined as {\it row}
vectors in isospin space. The covariant derivative introduces the 
additional parameter $\alpha$:
\begin{equation}
D_\mu \left({\overline P},{\overline Q}_{\alpha}\right) 
=\left(\partial_\mu +i\alpha \widetilde{g} \rho_\mu
-i\left(1-\alpha\right)v_\mu\right)
\left({\overline P},{\overline Q}_{\alpha}\right)\ .
\label{covderpq} 
\end{equation}
The covariant field tensor of the heavy vector meson is then 
defined as
\begin{eqnarray}
{\overline Q}_{\mu\nu}=
D_\mu {\overline Q}_\nu-D_\nu {\overline Q}_\mu \ .
\label{heavyft}
\end{eqnarray}
The coupling constants $d,c$ and $\alpha$, which appear in the 
Lagrangian (\ref{lagheavy}), have still not been very accurately 
determined. In particular there is no direct experimental evidence 
for the value of $\alpha$, which would be unity if a possible 
definition of light vector meson dominance for the electromagnetic 
form factors of the heavy mesons were to be adopted \cite{Ja95}. We 
will later adjust $\alpha$ to the spectrum of the heavy baryons. The 
other parameters in (\ref{lagheavy}) will be taken to be:
\begin{eqnarray}
d&=&0.53\ , \quad c=1.60\ ;
\nonumber \\
M&=&1865{\rm MeV}\ , \quad M^*=2007\,{\rm MeV}\ , \qquad {\rm D-meson}\ ;
\nonumber \\
M&=&5279{\rm MeV}\ , \quad M^*=5325\,{\rm MeV}\ , \qquad {\rm B-meson}.
\label{heavypara}
\end{eqnarray}

It should be stressed that the assumption of infinitely large
masses for the heavy mesons has not been made in (\ref{lagheavy}).
However, a model Lagrangian which was only required to exhibit the 
Lorentz and chiral invariances would be more general than the 
relativistic Lagrangian (\ref{lagheavy}). 
 Actually the 
coefficients of the various Lorentz and chirally invariant pieces 
in the relativistic Lagrangian (\ref{lagheavy}) have precisely been 
arranged to yield the spin--flavor symmetric model (\ref{Lag for H}) 
in the heavy quark limit \cite{Schechter-Subbaraman:2}.

\section{An apparent puzzle}

In this paragraph we will resolve an apparent puzzle which arises 
when calculating the corrections to the hyperfine splitting
using the relativistic lagrangian presented 
in Eq.~(\ref{lagheavy}). Here we will neglect the light 
vector contributions which is equivalent to set $\alpha=0$ and 
$c=0$ in Eq.~(\ref{lagheavy}).  
Then the heavy Lagrangian in Eq.~(\ref{lagheavy}) becomes 
\begin{eqnarray}
{\cal L}(P,Q_\mu) &=&
+D_\mu P D^\mu {\overline P} - M^2 P {\overline P}
-\frac{1}{2} Q_{\mu\nu} \overline{Q}^{\mu\nu}
+ {M^{\ast}}^2 Q_\mu {\overline Q}^\mu
\nonumber \\
&&{}
+ 2 i M d 
  \left( P p_\mu {\overline Q}^\mu - Q^\mu p_\mu {\overline P} \right) 
\nonumber \\
&&{}
+ d' \epsilon^{\alpha\beta\mu\nu}
  \left(
    D_\alpha Q_\beta p_\mu {\overline Q}_\nu
    - Q_\alpha p_\beta D_\mu {\overline Q}_\nu
  \right)
\ .
\label{P Q lag}
\end{eqnarray}
Here we have modified the coefficient of the fifth term in 
Eq.~(\ref{lagheavy}) to include a new source of heavy spin breaking. 
First let us consider the calculation of the hyperfine splitting in
the heavy field approach.
This, of course, arises at first sub-leading order in $1/M$ and
violates the heavy spin symmetry.
Thus we must add to Eq.~(\ref{Lag for H})
suitable heavy spin violating terms~\cite{Guralnik-Luke-Manohar}:
\begin{equation}
{\cal L}'_{\rm heavy}/M =
\frac{M-M^{\ast}}{8} {\rm Tr} 
\left[H \sigma_{\mu\nu} \overline{H} \sigma^{\mu\nu} \right]
+ \frac{(d-d')}{2} {\rm Tr} 
\left[ H p^\mu \overline{H} \gamma_\mu \gamma_5 \right]
+ \cdots 
\ .
\label{breaking terms}
\end{equation}
The first term has no derivatives while the second term has one
derivative. The hyperfine splitting is related to a collective 
Lagrangian parameter (see section 5.3 for details) $\chi$
with a proportionality factor of the $\Delta$-$N$ mass difference:
\begin{equation}
m(\Sigma_Q^{\ast}) - m(\Sigma_Q) =
\left[ m(\Delta) - m(N) \right] \chi 
\ .
\label{mass diff}
\end{equation}
(At present only $\Sigma_c$ is well established experimentally.)
For Eq.~(\ref{breaking terms}) we have
\begin{equation}
\chi = \frac{M^{\ast}-M}{4dF'(0)} 
+ \frac{d-d'}{4d}
\ .
\label{chi: 1}
\end{equation}
The first term was obtained in Ref.~\cite{Guralnik-Luke-Manohar}
while the second seems to be new.
Notice that $(M^{\ast}-M)$ and $(d-d')$ behave as $1/M$.
These quantities are the same as the ones appearing in the ordinary 
field Lagrangian (\ref{P Q lag}).
It would thus seem that ${\cal L}'_{\rm heavy}$ in 
Eq.~(\ref{breaking terms}) neatly summarizes the heavy spin violation
in Eq.~(\ref{P Q lag}).

Now let us consider the calculation of $\chi$ from Eq.~(\ref{P Q lag})  
directly based on exact numerical solution of the associated coupled
differential equations.
We content ourselves with the graphical presentation of some 
results\footnote{
For the Skyrme model parameters we use the experimental value of
$F_\pi$ and $e_{\rm Sk}=6.0$.
This results in a profile with $F'(0)=1.20$\,GeV.}
and relegate the details to Ref.~\cite{HQSSW:NPA}\footnote{%
Similar calculations were done in Ref.~\cite{Oh-Park}
but they did not consider the $M=M^{\ast}$, $d=d'$ case. }.
\begin{figure}[htbp]
\begin{center}
\ \epsfbox{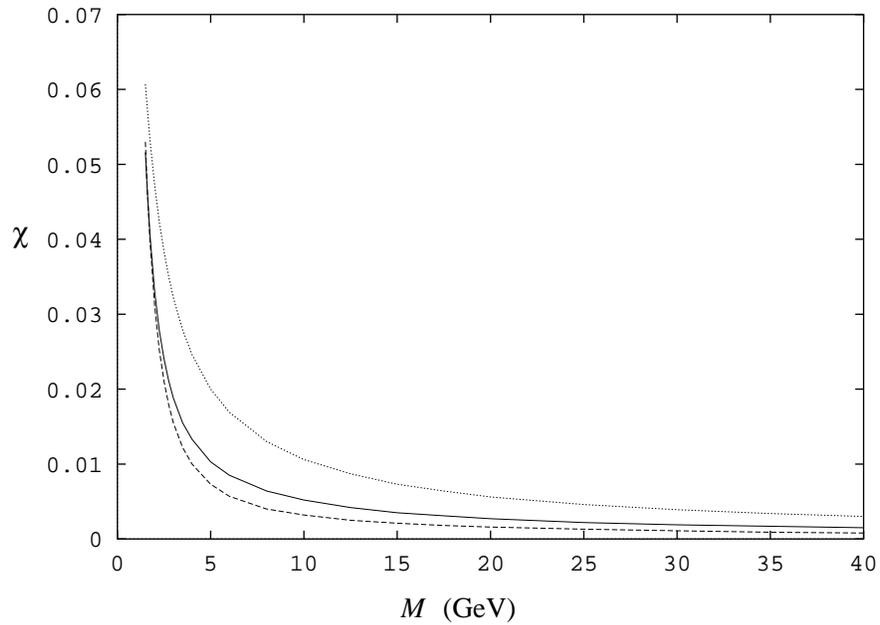} \\
\end{center}
\caption{
$\chi$ vs.~$M$ computed by numerical integration.
Solid line $M^{\ast}=M$, $d'=d$;
dotted line $M^{\ast}\neq M$, $d'=d$,
dashed line $M^{\ast}=M$, $d'\neq d$.
}
\label{fig: 1}
\end{figure}
Figure~\ref{fig: 1} shows $\chi$ plotted against $M$ for three cases:
i)~$M^{\ast}=M$, $d'=d=0.53$, 
ii)~$M^{\ast}-M\simeq(0.258\mbox{GeV})^2/M$ (a fit to experiment),
$d'=d=0.53$,
iii)~$M^{\ast}=M$, $d'-d=(0.0991\mbox{GeV})/M$ (an arbitrary choice
which sets the coupling constant splitting to be 10\% at the $D$ meson
mass).
We immediately notice that $\chi$ does not vanish when there is no
manifest heavy spin violation, i.e.,
$M=M^{\ast}$, $d=d'$.
In fact the dominant part of the contribution to $\chi$ for realistic
heavy meson masses is already present in this case.
By subtracting out this piece we note that the signs of the
contributions due to $M^{\ast}\neq M$ and $d'\neq d$ agree with those
predicted in Eq.~(\ref{chi: 1}).
It is interesting to note that all three curves in Fig.~\ref{fig: 1}
fall off as $1/M$ for $M\geq10$\,GeV.
But our main task is to understand the source of the puzzling non-zero
contribution in case i.
It is clear that the ordinary field Lagrangian (\ref{P Q lag})
must contain heavy spin violating pieces which are not manifest.
We will now explore this in detail by rewriting Eq.~(\ref{P Q lag}) 
in terms of the ``fluctuation field'' $H$ and expanding it in powers
of $1/M$.

\subsection{Expansion of Lagrangian}

Since the effects of $M\neq M^{\ast}$ and $d\neq d'$ were taken into
account in Eq.~(\ref{chi: 1}) it is sufficient to expand 
Eq.~(\ref{P Q lag}) with $M^{\ast}=M$ and $d'=d$.
To describe the heavy particle moving with four--velocity $V_\mu$,
we introduce the factorization 
\begin{equation}
P = e^{-iMV\cdot x} P' \ , \qquad 
Q_\mu = e^{-iMV\cdot x} \widetilde{Q}_\mu
\ .
\label{def: P Q}
\end{equation}
$P'$ is the pseudoscalar ``fluctuation field''.
$\widetilde{Q}_\mu$ is not exactly the vector fluctuation field
since $V\cdot\widetilde{Q}$ is not constrained to be zero.
We therefore introduce the correct fluctuation field $Q'_\mu$ by
\begin{equation}
\widetilde{Q}_\mu = Q'_\mu + V_\mu V\cdot\widetilde{Q} 
\label{def: Q}
\ ,
\end{equation}
which shows that $V\cdot Q'=0$.
Substituting Eqs~(\ref{def: P Q}) and (\ref{def: Q}) into the
Lagrangian (\ref{lagheavy}) gives, in addition to the leading terms of
order $M$, the presently interesting terms of order $M^0$:
\begin{eqnarray}
{\cal L} (P,Q) &=&
(\mbox{\rm order}\ M) -
P' D^2 \overline{P'} + Q'_\mu D^2 \overline{Q'}^\mu
- Q'^\mu D_\nu D_\mu \overline{Q'}^\nu
\nonumber\\
&&{}
-d \epsilon^{\alpha\beta\mu\nu}
\left(
 D_\alpha Q'_\beta p_\mu \overline{Q'}_\nu -
 Q'_\alpha p_\beta D_\mu \overline{Q'}_\nu
\right)
\nonumber\\
&&{}
+ M^2 V\cdot\widetilde{Q} V\cdot\overline{\widetilde{Q}}
+  iM \left(
 D_\mu Q'^\mu V\cdot\overline{\widetilde{Q}} -
 V\cdot\widetilde{Q} D_\mu \overline{Q'}^\mu
\right)
\nonumber\\
&&{}
+  2iMd \left(
 P' V\cdot p V\cdot\overline{\widetilde{Q}} -
 V\cdot\widetilde{Q} V\cdot p \overline{P'}
\right)
+ \cdots
\ ,
\label{P Q lag 2}
\end{eqnarray}
where the three dots stand for terms of order $1/M$.
In contrast to the massless fields $P'$ and $Q'$,
$V\cdot\widetilde{Q}$ is seen to have the large mass $M$.
We thus integrate it out using the equation of motion
\begin{equation}
V\cdot\widetilde{Q} =
-\frac{i}{M} D_\mu Q'^\mu 
-\frac{2id}{M} P' V\cdot p 
\ .
\label{EOM for Q}
\end{equation}
Substituting Eq.~(\ref{EOM for Q}) back into Eq.~(\ref{P Q lag 2})
gives
\begin{eqnarray}
{\cal L} (P,Q) &=&
(\mbox{\rm order}\ M) -
P' D^2 \overline{P'} + Q'_\mu D^2 \overline{Q'}^\mu
- i Q'_\mu F^{\mu\nu}(v) \overline{Q'}_\nu
\nonumber\\
&&{}
- 2d \left(
 P' V\cdot p D_\mu \overline{Q'}^\mu +
 D_\mu Q'^\mu V\cdot p \overline{P'}
\right)
\nonumber\\
&&{}
-id \epsilon^{\alpha\beta\mu\nu}
\left(
 D_\alpha Q'_\beta p_\mu \overline{Q'}_\nu -
 Q'_\alpha p_\beta D_\mu \overline{Q'}_\nu
\right)
\nonumber\\
&&{}
- 4 d^2 P' \left(V\cdot p\right)^2 \overline{P'}
+ \cdots
\ ,
\label{P Q lag 3}
\end{eqnarray}
where $F_{\mu\nu}(v) = \partial_\mu v_\nu - \partial_\nu v_\mu 
-  i[v_\mu,v_\nu]$.
In order to extract the heavy spin violating pieces it is convenient
to rewrite the order $M^0$
Lagrangian in terms of the heavy multiplet
field  
$H$ Eq.~(\ref{Hdef}).
After some algebraic calculation we find
\begin{eqnarray}
{\cal L} (H) &=&
{\cal L}_{\rm heavy}
+ \frac{1}{2} {\rm Tr} \left[ H D^2 \overline{H} \right]
+ i \frac{1}{8} {\rm Tr} \left[
\left[ H , \gamma_\mu \gamma_\nu \right] F^{\mu\nu}(v) \overline{H}
\right]
\nonumber\\
&&{}
+ d
\Biggl[
\frac{i}{2} {\rm Tr} 
\left[ D_\mu H \gamma^\mu \gamma_5 (V \cdot p) \overline{H} \right]
- \frac{i}{4} {\rm Tr} \left[ 
\gamma \cdot D H \gamma_\mu \gamma_5 p^\mu \overline{H} 
\right]
\nonumber\\
&&\qquad{}
- \frac{i}{4} {\rm Tr} \left[
\gamma \cdot D H p_\mu \overline{H} \gamma^\mu \gamma_5
\right]
- \frac{1}{8} {\rm Tr} \left[
\sigma_{\mu\nu} D_\alpha H \gamma^\alpha V\cdot p \gamma_5 
\sigma^{\mu\nu} \overline{H}
\right]
+ {\rm h.c.}
\Biggr]
\nonumber\\
&&{}
+ d^2 \left[
\frac{1}{2} {\rm Tr} \left[
H \left( V\cdot p \right)^2 \overline{H}
\right]
+ \frac{1}{4} {\rm Tr} \left[
\sigma_{\mu\nu} H \sigma^{\mu\nu} \left( V\cdot p \right)^2
\overline{H}
\right]
\right]
+\cdots
\ ,
\label{H lag}
\end{eqnarray}
where ${\cal L}_{\rm heavy}$ is given in 
Eq.~(\ref{Lag for H}) with $c=\alpha=0$. At this stage we see 
that Eq.~(\ref{H lag}) actually contains pieces
which are not manifestly invariant under the heavy spin
transformations $H\rightarrow SH$, 
$\overline{H} \rightarrow \overline{H}S^{\dag}$. 
These pieces involve two derivatives.

\section{Hyperfine splitting from the Hidden Terms}

We now sketch the computation of the portion of $\chi$ in 
Eq.~(\ref{mass diff}) which results from the ``hidden'' heavy spin
violation in Eq.~(\ref{lagheavy}) that has been made explicit in 
Eq.~(\ref{H lag}).
For this purpose one needs to collectively quantize the Lagrangian as 
described in paragraph 4.4.
To leading order in $M$, the ``angular
part'' of the ground state wave function in Eq.~(\ref{Wf:h bar}) is~\cite{%
Gupta-Momen-Schechter-Subbaraman,Schechter-Subbaraman}
\begin{equation}
\psi^{(1)}_{dl,h} = \frac{1}{\sqrt{8\pi}} \epsilon_{dl}
\delta_{2h}\ .
\label{psi1a}
\end{equation}
The specific value of the index $h$ results from the choice
$G_3=G=1/2$ where $G$ is the ``grand spin''.
To next leading order in $M$ the ground state wave function receives a
heavy spin violating admixture of
\begin{equation}
\psi^{(2)}_{dl,h} = \frac{1}{\sqrt{4\pi}}\left[
\sqrt{\frac{2}{3}} \delta_{d1} \delta_{l1} \delta_{h1} + 
\frac{1}{\sqrt{6}} \left(
 \delta_{d2} \delta_{l1} + \delta_{d1} \delta_{l2}
\right)\right] \delta_{h2}
\ .
\label{psi2a}
\end{equation}
Finally, the hyperfine splitting parameter $\chi$ is recognized by
expanding the collective Lagrangian~\cite{Callan-Klebanov},
in powers of $\mbox{\boldmath $\Omega$}$ and picking up the
linear piece $L_{\rm coll} = (\chi/2) \Omega_3 + \cdots$.
Noting that the $\Delta$--nucleon mass difference is given by 
the moment of inertia, which relates the angular velocity to 
the spin operator~\cite{Ad83}, this piece of the Lagrangian 
yields Eq.~(\ref{mass diff}) after canonical quantization of 
the collective coordinates~\cite{Callan-Klebanov}.
There are two types of contribution to $\chi$.
The first type, from the heavy spin violating terms proportional to
$d$ in Eq.~(\ref{H lag}), corresponds to the evaluation of heavy spin
violating operators in the ground state (\ref{psi1a}).
The second type corresponds to the evaluation of heavy spin
conserving operators in the ground state which includes an admixture 
of Eq.~(\ref{psi2a}) due to the 
${\rm Tr}\left[ \gamma_\mu \gamma_\nu H F^{\mu\nu}(v) \overline{H} \right]$
term in Eq.~(\ref{H lag}).
The net result for the ``hidden'' part of $\chi$ is
\begin{equation}
\chi = \frac{F'(0)}{4M}
\left( d - \frac{1}{2d} \right)
\ .
\label{chi: 2}
\end{equation}
This equation is expected to hold for large $M$.
To this should be added the ``manifest'' part given in 
Eq.~(\ref{chi: 1}).

It is important to compare Eq.~(\ref{chi: 2}) with the result for
$\chi$ obtained by the exact numerical solution for the model based 
on Eq.~(\ref{P Q lag}). This is gotten as an integral over the 
properly normalized radial functions $\Phi(r),\ldots,\Psi_2(r)$ which 
appear in the P--wave solution of the bound state equation~%
\cite{Schechter-Subbaraman-Vaidya-Weigel}:
\begin{eqnarray}
{\overline P}&=&A(t)\frac{\Phi(r)}{\sqrt{4\pi}}\
\hat{\mbox{\boldmath $r$}}\cdot
\mbox{\boldmath $\tau$}\rho e^{i\epsilon t}, \qquad
{\overline Q}_0=\frac{i}{\sqrt{4\pi}}A(t)
\Psi_4(r)\rho e^{i\epsilon t},
\nonumber \\
{\overline Q}_i&=&\frac{1}{\sqrt{4\pi}}A(t)\left[
i\Psi_1(r)\hat r_i+\frac{1}{2}\Psi_2(r)\epsilon_{ijk}
\hat r_j\tau_k\right]\rho e^{i\epsilon t}\ .
\label{pwavean}
\end{eqnarray}
The spinor $\rho$ labels the grand spin of the bound heavy
meson. The choice $G_3=+1/2$ corresponds to $\rho=(1,0)^{\dag}$.
The heavy limit bound state wave function in Eq.~(\ref{psi1a})
corresponds to the special choice
\begin{equation}
\Phi(r)\propto u(r) \ ,\quad 
\Psi_1(r)=-\Phi(r)\ ,\quad \Psi_2(r)=-2\Phi(r)
\quad {\rm and} \quad \Psi_4(r)=0
\ .
\label{pwhl}
\end{equation}
The numerical
solution to the bound state equations exactly exhibits 
these relations for $M,M^{\ast}\to\infty$~%
\cite{Schechter-Subbaraman-Vaidya-Weigel}.

\begin{figure}[htbp]
\begin{center}
\ \epsfbox{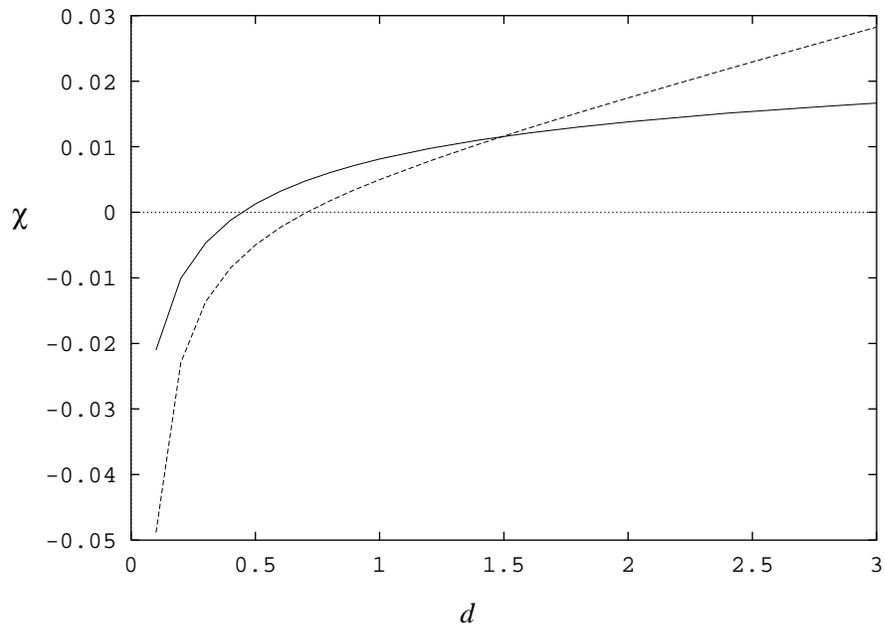}\\
\end{center}
\caption{
The $d$ dependence of $\chi$ for $M=M^{\ast}=30$\,GeV and $d=d'$.
Solid line is the exact numerical calculation.
Dashed line is the large $M$ perturbation formula given in 
Eq.~(\ref{chi: 2}).
}
\label{fig: 2}
\end{figure}

Equation~(\ref{chi: 2}) has an interesting $d$-dependence and vanishes
at $d=1/\sqrt{2}$, which actually is not too far from the experimental
value of this quantity.
In Fig.~\ref{fig: 2} we compare the $d$-dependence of the exact 
numerical calculation with the perturbative result of 
Eq.~(\ref{chi: 2}).
It is seen that the large $M$ perturbation approach works 
reasonably well and the gross structure of the hyperfine splitting 
is reproduced. For a detailed comparison of the two treatments 
it is important to note that for fixed $M=M^{\ast}$ the 
binding of the heavy meson increases with $d$. In particular 
this implies that the wave function is only reasonably 
localized for large enough $d$. As a strong localization is 
a basic feature of the perturbative approach it is easy to 
understand why this calculation does not yield the exact 
(numerical) result for small $d$. In fact, as $d$ increases 
the agreement expectedly improves. However, upon further increase 
of $d$ 
(at finite $M,M^{\ast}$),
the numerical solution to the bound state equations shows 
noticeable deviations from the heavy limit relations (\ref{pwhl}),
which causes the moderate differences at larger $d$.

We have solved \cite{HQSSW:1} the apparent puzzle associated with the use of a model
Lagrangian containing ordinary fields for computing the hyperfine
splitting parameter $\chi$ by carefully expanding the Lagrangian in
powers of $1/M$.
The key point was the need to preserve the constraint 
$V\cdot Q\,'=0$ for the heavy vector fluctuation field.

Of course, such a model Lagrangian (which has been used in many
calculations) is not exactly QCD.
Nevertheless it seems reasonable since it automatically has the
correct relativistic kinematics and satisfies the heavy spin symmetry
at leading order.
We have seen (Eq.~(\ref{H lag})) that at next order in $1/M$, it
predicts the coefficients of many terms which otherwise would be
unspecified by heavy spin symmetry.

It is amusing to note that these $1/M$ suppressed terms involve two
derivatives and are actually more important for the computation of
$\chi$ than the zero derivative term in Eq.~(\ref{chi: 1}).
This is readily understandable since the dynamical scale in this
calculation is the binding energy, 
$m(B)+m(N)-m(\Lambda_b)\simeq620$\,MeV which is rather large for
neglecting light vector mesons, higher derivatives etc.
[See, for example, Ref.~\cite{Harada-Sannino-Schechter}.]

We are regarding the Lagrangian (\ref{lagheavy}) with 
$\alpha=c=0$ as an illustrative
model rather than as a realistic one for comparison with experiment.
As indicated earlier it seems necessary to include, in addition
to finite $M$ corrections, the effects of light vector mesons as well
as nucleon recoil.
The discussion of $\chi$ in this more complicated model
and further details of the present calculation will
be given in the next paragraph \cite{HQSSW:NPA}.

\section{Perturbative Approach and the Vector contribution.}

The perturbative approach can illuminate several aspects of the
hyperfine splitting problem. This is due to the heavy quark symmetry 
which is naturally exploited by making an expansion in powers of 
$1/M$ using the heavy field formalism. Our starting Lagrangian 
(\ref{lagheavy}) has been set up in such a way as to yield a 
heavy quark symmetric result as $M\rightarrow\infty$ when 
$M=M^{\ast}$ is assumed, {\it cf.} Eq.~(\ref{Lag for H}). The 
perturbative $1/M$ expansion is more general (presumably exact) but 
less predictive.  Thus the $1/M$ expansion provides a useful 
calibration in the large $M$ limit. Since it deals with perturbation 
matrix elements it provides us with a convenient classification of 
the various sources of hyperfine splitting. The method is also 
advantageous in that it can be extended, without too much algebraic 
work, to different channels of interest. On the other hand, once the 
particular channels of interest are settled on, it is clearly more 
convenient to employ the exact numerical solution, which efficiently 
sums up a class of $1/M$ corrections.

The leading order Lagrangian (\ref{Lag for H}) can be supplemented by 
terms which manifestly break the heavy quark symmetry to leading order
($M^0$ with the present normalization) as follows:
\begin{eqnarray}
\frac{1}{M}{\cal L}'_H &=&
\frac{M-M^{\ast}}{8} \mbox{\rm Tr}
\left[ H \sigma_{\mu\nu} \overline{H} \sigma^{\mu\nu} \right]
+\frac{(d-d')}{2} \mbox{\rm Tr}
\left[ H p_\mu \overline{H} \gamma^\mu \gamma_5 \right]
\nonumber\\
&& \quad
{}+i \frac{(c-c')}{m_V} \mbox{\rm Tr}
\left[ \gamma_\mu \gamma_\nu H F^{\mu\nu}(\rho) \overline{H} \right]
\nonumber \\
&&{}
- \tilde{\alpha} V^\beta \mbox{\rm Tr}
\left[
  H \sigma_{\mu\nu} \left( \tilde{g} \rho_\beta + v_\beta \right)
  \overline{H} \sigma^{\mu\nu}
\right]
\ .
\label{heavy next}
\end{eqnarray}
Here the $(M-M^{\ast})$ term measures the heavy spin violation due to
the heavy pseudoscalar -- heavy vector mass difference.
The $(d-d')$ term measures the heavy spin violation induced by choosing
different coefficients for the fifth and sixth terms in
Eq.~(\ref{lagheavy}), while the
$(c-c')$ term  corresponds to choosing different coefficients for the
last and next--to last terms in Eq.~(\ref{lagheavy}). Finally the 
$\tilde{\alpha}$ term corresponds to the leading term
obtained by using different values of $\alpha$ in
Eqs~(\ref{covderpq}) for $P$ and $Q$.
Note that $(M-M^{\ast})$, $(d-d')$, $(c-c')$ and 
$\tilde{\alpha}$ all behave as $1/M$.

In addition to the terms in Eq.~(\ref{heavy next}), which manifestly
break the heavy quark symmetry, there are, in fact, ``hidden'' 
violation terms contained in Eq.~(\ref{lagheavy}). The explicit 
expression for the hidden terms in the model without light vectors is 
given in Eq.~(\ref{H lag}) \cite{HQSSW:1}. These were shown 
to exist (for the model without light vectors) in Ref.~\cite{HQSSW:1} 
and arise from performing a detailed $1/M$ expansion of the 
relativistic Lagrangian. 
In Reference~\cite{HQSSW:NPA} 
the numerical study has confirmed that this is also true 
when light vector mesons are included. 
In the last paragraph we have shown 
({\it cf.}  Fig.~\ref{fig: 2})~\cite{HQSSW:1}) 
that the dependence on $d$ of the hyperfine 
splitting computed from these hidden terms using the perturbative 
approach generally matched the exact numerical calculation. Hence we 
shall not explicitly isolate the extra hidden terms due to the 
addition of the light vectors but shall content ourselves with the 
numerical treatment given in Ref.~\cite{HQSSW:NPA}.

Here we discuss the  computation of $\chi$ in some detail.
We have already noticed that 
the dynamics of the model dictates that the bound-states occur for
$k=0$, in which case $\xi_{dl}(0,0) = \epsilon_{dl}/\sqrt{2}$. 
{}For reader's convenience we display again the relevant wave functions.
The bound-state wave--function simply is
\begin{equation}
\psi_{dl,h}(0,0,0,0) = \frac{1}{\sqrt{8\pi}} \epsilon_{dl} 
\chi_{h}
\label{psi:bound}
\ .
\end{equation}
The $k=1$ unbound wave--function with no orbital excitation ($r=0$) is
\begin{equation}
\psi_{dl,h}(1,g_3,0,1) = \frac{1}{\sqrt{4\pi}} 
\xi_{dl}(1,g_3) \chi_{h}
\label{psi:unbound}
\ .
\end{equation}
When violations of the heavy quark symmetry are included,
$g$ is no longer a good quantum number.
The grand spin, is a good quantum number
\begin{equation}
\mbox{\boldmath$G'$} = \mbox{\boldmath$g$} + 
\mbox{\boldmath$S$}_{\rm heavy}
\ .
\end{equation}
In the notation of Eqs~(\ref{psi:bound}) and (\ref{psi:unbound})
we have the grand spin eigenstates
\begin{eqnarray}
\psi^{(1)}_{dl,h} (G'=G'_3=1/2)
&=& \frac{1}{\sqrt{8\pi}} \epsilon_{dl} \delta_{2h}
\ ,
\label{psi 1}\\
\psi^{(2)}_{dl,h} (G'=G'_3=1/2)
&=&
\frac{1}{\sqrt{4\pi}}
\left[
  \sqrt{\frac{2}{3}} \delta_{d1} \delta_{l1} \delta_{h1} + 
  \frac{1}{\sqrt{6}} \left(
    \delta_{d2} \delta_{l1} + \delta_{d1} \delta_{l2}
  \right) \delta_{h2}
\right]
\ .
\label{psi 2}
\end{eqnarray}
Note that in Eq.~(\ref{psi 1}) the $G'_3=+1/2$ wave function is
$\delta_{2h}$ since the index $2$ corresponds to $+1/2$ for the
anti--quark wave--function. The two states (\ref{psi 1}) and 
(\ref{psi 2}) differ with respect to their 
$\mbox{\boldmath $g$}$ and $\mbox{\boldmath $K$}$ labels.

Now let us consider the potential for the bound-state wave--function 
in the presence of the first heavy quark symmetry violating term in 
Eq.~(\ref{heavy next}). Substituting the $G^\prime$--eigenstates 
$\psi^{(1)}$ and $\psi^{(2)}$ from Eqs~(\ref{psi 1}) and 
(\ref{psi 2}) into Eq.~(\ref{Lag for H}) and the first term of 
Eq.~(\ref{heavy next}) yields, after a spatial integration, the 
potential matrix in the $\psi^{(1)}$--$\psi^{(2)}$ space:
\begin{equation}
V = - \frac{d\,F'(0)}{2}
\left( \begin{array}{cc}
3 & 0 \\ 0 & -1 
\end{array} \right)
+ \frac{M-M^{\ast}}{4}
\left( \begin{array}{cc}
0 & \sqrt{3} \\ \sqrt{3} & 2
\end{array} \right)
\ ,
\label{potential}
\end{equation}
where $F(r)$ is defined in Eq.~(\ref{xi classical}) and $F' = dF/dr$.
The first matrix shows that $\psi^{(1)}$ is bound while $\psi^{(2)}$
is unbound in the heavy spin limit. Since the second matrix gives 
mixing between $\psi^{(1)}$ and $\psi^{(2)}$ the latter must be 
included in the presence of effects which break the heavy quark 
symmetry. The diagonalized bound wave function is seen to be
\begin{equation}
\psi^{(1)} - \frac{\sqrt{3}}{8} \frac{M-M^{\ast}}{d\,F'(0)} \psi^{(2)}
\ .
\label{wave mixing}
\end{equation}
This is the proper wave--function to be ``cranked'' in order to
generate the heavy spin violation. Using it in Eq.~(\ref{def: collective}), 
which is then substituted into the $\alpha=0$ limit of the first term 
of Eq.~(\ref{Lag for H}), contributes a term in the collective Lagrangian
\begin{equation}
\frac{\chi}{2} \Omega_3 \qquad {\rm where} \qquad
\chi = \frac{M^{\ast}-M}{4d\,F'(0)} \ .
\label{chi:1st}
\end{equation}
By using the Wigner--Eckart theorem we may express this for states of
either $G'_3$ as the matrix element of the operator 
$\chi \mbox{\boldmath$\Omega$} \cdot \mbox{\boldmath$G'$}$.
For convenience we have chosen to consider our wave--function as 
representing the conjugate particle in Eq.~(\ref{h: classical}). Hence 
the matrix element of $\mbox{\boldmath$G'$}$ in this section differs 
by a minus sign from that of $\mbox{\boldmath$G$}$ defined in 
$\mbox{\boldmath$J$} = \mbox{\boldmath$G$} +
\mbox{\boldmath$J$}^{\rm sol}$, with $J_i^{\rm sol} \equiv 
\left(\partial L/ \partial \Omega_i \right)$,
which is the appropriate one when we form the total heavy baryon
spin. 
Then the collective Lagrangian, $L_{\rm coll}$ may be written
(see section 4.4)
\begin{equation}
L_{\rm coll} = \frac{1}{2} \alpha^2 \mbox{\boldmath$\Omega$}^2
- \chi \mbox{\boldmath$\Omega$} \cdot \mbox{\boldmath$G$}
\label{collective Lag}
\end{equation}
which again leads to the Hamiltonian (\ref{hcollective1}) and hence to 
the well known formula {\it cf.} Eq.~(\ref{eq:1.1})
\begin{equation}
m(\Sigma_Q^{\ast}) - m(\Sigma_Q) =
\left[ m(\Delta) - m(N) \right] \, \chi
\ .
\label{formula for splitting}
\end{equation}
The purpose in deriving this again was to explain 
the perturbative method and our notation.

Next we shall give some new perturbative ``manifest'' contributions 
to $\chi$ from Eq.~(\ref{heavy next}).
When all these terms are included the potential $V$ in
Eq.~(\ref{potential}) is modified so that the properly diagonalized
wave--function which replaces Eq.~(\ref{wave mixing}) becomes
\begin{equation}
\psi^{(1)} + \epsilon \psi^{(2)} \ ,
\label{wave mix:2}
\end{equation}
with
\begin{equation}
\epsilon = 
\frac{ 
  - \frac{\sqrt{3}}{4} \left( M - M^{\ast} \right)
  + \frac{\sqrt{3}}{4} \left( d - d' \right) \, F'(0)
  + \sqrt{3} \tilde{\alpha} \omega(0)
  + \frac{\sqrt{3}\left(c-c'\right)}{m_V} \frac{G''(0)}{\widetilde{g}}
}{
  2 d\, F'(0) - \frac{ 2 c G''(0)}{\widetilde{g} m_V }
}
\ .
\end{equation}
There are two types of contribution to $\chi$. The first type is 
analogous to Eq.~(\ref{chi:1st}) and arises when 
Eq.~(\ref{wave mix:2}) is cranked and substituted into 
Eq.~(\ref{Lag for H}). The second type is obtained by 
substituting the leading order wave function $\psi^{(1)}$ into the 
$(c-c')$ and $\tilde{\alpha}$ terms in Eq.~(\ref{heavy next}).
The complete expression for $\chi$ resulting from the ``manifest''
heavy spin violation is
\begin{eqnarray}
\lefteqn{
\chi = \epsilon 
\left[
  \frac{2}{\sqrt{3}} \left( 1 - \frac{4}{3}\alpha \right) +
  \frac{2\sqrt{2}}{3\sqrt{3}} \alpha \widetilde{g} 
\left( \xi_1(0) - \xi_2(0) \right)
  - 8 \sqrt{\frac{2}{3}} \frac{c}{m_V} \varphi''(0)
\right]
}
\nonumber\\
&&
{}+
\frac{\tilde{\alpha}}{3}
\left[ 8 - 2 \sqrt{2} \widetilde{g} \left( \xi_1(0) - \xi_2(0) \right) \right]
- 4\sqrt{2} \frac{c-c'}{m_V} \varphi''(0)
\ .
\label{manifest chi}
\end{eqnarray}
The quantities $\xi_1(0)$, $\xi_2(0)$ and
$\varphi''(0)$ are defined in Eq.~(\ref{induced}).
This formula may be useful for quickly estimating the effects of
heavy spin violation in the coupling constants, which were not
explicitly given in the previous discussion. Unfortunately there 
is no determination of the magnitude of these effects from the 
mesonic sector at present. In the previous paragraphs \cite{HQSSW:1} 
the discussion of the 
``hidden'' heavy contributions to $\chi$ was given for the 
Lagrangian with only light pseudoscalars. 

The hyperfine splitting just discussed is for the ground state or
P--wave heavy baryons. It is of some interest to briefly consider 
the negative parity heavy baryons with one unit of orbital 
excitation (S--wave).  In the heavy spin limit these bound states 
correspond to the $r=1$ and $k=0$ choice in 
Eq.~(\ref{h: classical}):
\begin{equation}
\psi_{dl,h}(1,g_3,1,0) = 
\frac{\epsilon_{dl}}{\sqrt{2}} Y_1^{g_3} \chi_h \ .
\end{equation}
The spin, $\mbox{\boldmath$J$}_{\rm light}$ of the ``light cloud''
part of the heavy baryon is gotten by adding this $g=1$ piece to the 
soliton spin $\mbox{\boldmath$J$}^{\rm sol}$. For the $I=0$ (which 
implies $J^{\rm sol}=0$) heavy baryons one finds $J_{\rm light}=1$ 
and the degenerate multiplet
\begin{equation}
\left\{ \Lambda_Q'(1/2) \,,\, \Lambda_Q'(3/2) \right\}
\ .
\label{Lambda prime}
\end{equation}
For the $I=J^{\rm sol}=1$ heavy baryons, $J_{\rm light}$ can be 
either $0$, $1$ or $2$ and we find the degenerate heavy multiplets
\begin{eqnarray}
&& \qquad \Sigma_Q'(1/2) \ ,
\nonumber\\
&& \left\{ \Sigma_Q'(1/2) \,,\, \Sigma_Q'(3/2) \right\} \ ,
\nonumber\\
&& \left\{ \Sigma_Q'(3/2) \,,\, \Sigma_Q'(5/2) \right\}
\ .
\label{excited Sigmas}
\end{eqnarray}
In general, the situation is even more complicated and the 
subject will be fully investigated in the next chapter.  
At present there are 
experimental candidates\cite{pdg} for a negative parity spin 
$1/2$ baryon $\Lambda_c'$ at $2593.6\pm1.0$\,MeV and a negative 
parity spin $3/2$ baryon $\Lambda_c'$ at $2626.4\pm0.9$\,MeV.

Since experimental information is available, it is especially
interesting to consider the splitting between the two $\Lambda_Q'$ 
states in Eq.~(\ref{Lambda prime}). This splitting stems from the 
violation of the heavy quark symmetry. For the $\Lambda_Q$ type 
states the total spin coincides with the grand spin 
$\mbox{\boldmath$G$}$ so that Eq.~(\ref{Lambda prime}) may be
alternatively considered a $G=1/2$, $G=3/2$ multiplet. Since the 
good quantum number is $G$, we may in general expect the hyperfine
parameter $\chi$ to depend on $G$. The collective Hamiltonian takes 
the form
\begin{equation}
H_{\rm coll} = \frac{ \left(\mbox{\boldmath$J$}^{\rm sol} + 
\chi_G \mbox{\boldmath$G$} \right)^2}{ 2\alpha^2}
\ .
\label{collective Hamiltonian}
\end{equation}
On general grounds we see that for the case of the $\Lambda_Q'$'s the
collective Hamiltonian contribution to the hyperfine splitting will be
suppressed. Setting $\mbox{\boldmath$J$}^{\rm sol}=0$ in 
Eq.~(\ref{collective Hamiltonian}) shows that the hyperfine splitting
is of order $(\chi^2)$ or equivalently of order $(1/M^2)$.
Unlike the ground state which involves only the $G=1/2$ P--wave
channel, there is another possibility for hyperfine splitting here.
It is allowed for the $G=1/2$ and $G=3/2$ bound state energies to
differ from each other. In the Lagrangian with only light 
pseudoscalars this does not happen and the 
$\Lambda_Q'(1/2) - \Lambda_Q'(3/2)$ splitting is of order
$1/M^2$. However when light vectors are added, there are ``hidden''
$1/M$ terms, which violate the heavy quark symmetry as {\it e.g.}
\begin{equation}
i\,\,
\mbox{\rm Tr} \, 
\left[
  \sigma_{\alpha\mu} H \gamma_\nu F^{\mu\nu}(\rho)
  D^\alpha \overline{H}
\right] \, + \, \mbox{\rm h.c.}
\ .
\end{equation}
This term is likely to generate splitting for the multiplet 
(\ref{Lambda prime}) to order $1/M$ by giving different binding
energies to the $G=1/2$ and $G=3/2$ channels.
It would be very interesting to investigate this in more detail.

Finally, we add a remark concerning an amusing conceptual feature in
the computation of hyperfine splitting among the five $\Sigma_Q'$'s in
Eq.~(\ref{excited Sigmas}). The total angular momentum of each state
is given by
\begin{equation}
\mbox{\boldmath$J$} = 
\underbrace{
  \mbox{\boldmath$J$}^{\rm sol} + \mbox{\boldmath$g$} 
}_{\mbox{\boldmath$J$}_{\rm light}}
\!\!\!\!\!
\overbrace{
  \ \ \ + \mbox{\boldmath$S$}_{\rm heavy}
}^{\mbox{\boldmath$G$}}
\ ,
\label{recoupling formula}
\end{equation}
where we are now considering each operator to be acting on the
wave--function rather than its complex conjugate. We have illustrated 
two different intermediate angular momenta which can alternatively 
be used to label the final state. 
If $\mbox{\boldmath$J$}_{\rm light}$ is used, we get the
heavy-spin multiplets in Eq.~(\ref{excited Sigmas}). 
On the other hand, when the hyperfine splitting is turned on, the 
choice $\mbox{\boldmath$G$}$ is convenient, because it remains a 
good quantum number. According to the laws of quantum mechanics, we 
cannot simultaneously use both to specify the states, since the 
commutator
\begin{equation}
\left[ 
  \, \mbox{\boldmath$J$}_{\rm light}^2 \ , \ 
  \mbox{\boldmath$G$}^2 \,
\right]
= 4 i \mbox{\boldmath$J$}_{\rm light} 
\mbox{\boldmath$\cdot$}
\left(\mbox{\boldmath$S$}_{\rm heavy}
\mbox{\boldmath$\times$} \mbox{\boldmath$g$}\right)
\label{commutator}
\end{equation}
is generally non--vanishing. This means that we cannot uniquely 
trace the splitting of, say, 
the $\left\{\Sigma^\prime_Q(1/2),\Sigma^\prime_Q(3/2)\right\}$
heavy multiplet in Eq~(\ref{excited Sigmas}), as hyperfine
splitting interactions are turned on. Physically, this causes 
a mixing between the $\Sigma^\prime_Q$'s of the same spin. Rather,
we must look at the whole pattern of the five masses. On the other hand,
the problem simplifies for the computation of the ground state
hyperfine splitting in Eq.~(\ref{formula for splitting}).
In that case the bound state wave function is characterized by 
$\mbox{\boldmath$g$}=0$. Thus the commutator in 
Eq.~(\ref{commutator}) vanishes, and it is ``trivially'' possible to 
track the hyperfine splitting as a mass difference.

\section{Numerical Results}

In this section we will briefly comment on the numerical results 
obtained in \cite{HQSSW:NPA} for the masses of the heavy baryons 
within the relativistic Lagrangian model discussed 
above. The numerical procedure requires the solution 
of coupled inhomogeneous differential equations; the details are 
provied in Ref.~\cite{HQSSW:NPA}. In particular we will concentrate on the spin and isospin 
splitting in the realistic case of finite heavy meson masses 
(\ref{heavypara}). It should be noted that sizable quantum 
corrections occur for the classical soliton mass $M_{\rm cl}$ 
\cite{Me96}. It seems that these corrections are (approximately) equal 
for all baryons. Hence we will only consider mass differences between 
various baryons. In that case the absolute value of the classical mass 
$M_{\rm cl}$ is redundant. The parameters in the light sector can  
completely be determined from properties of the corresponding mesons and by 
the $\frac{1}{2}^+$ and $\frac{3}{2}^+$  light baryons \cite{HQSSW:NPA}.
The corresponding moment of inertia is 
$\alpha^2=5.00{\rm GeV}^{-1}$.  In the preceding paragraphs 
Ref.~\cite{HQSSW:1} we have shown 
(in the case without light vectors) that a major fraction of 
the P--wave hyperfine constant is due to terms in the relativistic 
Lagrangian (\ref{lagheavy}) which do not manifestly break the heavy 
spin symmetry rather than to terms, which explicitly break this 
symmetry; as for example $M\ne M^*$. 
The numerical results in Ref.~\cite{HQSSW:NPA} provide a quantitative 
extimate of the hidden contribution, when the light vectors are also included,  by performing the calculation using identical 
masses from the charm sector {\it i.e.} $M=M^*=1.865{\rm GeV}$ and 
furthermore $\alpha=0.3$. All other parameters are as in 
Eq.~(\ref{heavypara}). This results in $\chi_P=0.080$ 
($\chi$ for the P--wave). From table 
\ref{tab_1} we recognize that this is about 80\% of the value 
obtained using the physical masses $M=1.865{\rm GeV}$ and 
$M^*=2.007{\rm GeV}$. In the case of the S--wave the hidden piece is 
even more dominant. For the symmetric choice of the mass parameters 
one finds $\chi_S=0.175$ ($\chi$ for the S--waves) 
which is more than 90\% of the value 
displayed in Table \ref{tab_1} for $\alpha=0.3$. It is also possible 
to show \cite{HQSSW:NPA} that the light vector model predicts a 
substantially larger $\chi_P$.

Since the contribution of the manifest 
$(M^{\ast}-M)$ breaking term is relatively small
it is reasonable to expect that the others will be small too.

Let us next discuss the spectrum of the baryons containing a single 
heavy quark. For this case we assume the realistic masses as in 
Eq.~(\ref{heavypara}). In Table \ref{tab_1} the numerical results for 
the lowest S-- and P--wave bound states in the charm sector are 
displayed. As already noted in Ref.~\cite{Schechter-Subbaraman} the 
binding energy ($\omega$) 
 decreases with growing coupling constant $\alpha$. This is the case 
for both the P-- and S--wave channels. For $M\rightarrow\infty$ the 
heavy limit (see Eq.~(\ref{binding}))
\begin{eqnarray}
\omega\longrightarrow\frac{3}{2}dF^\prime(0)
+\frac{3c}{\tilde{g}m_V}G^{\prime\prime}(0)
+\frac{\alpha \tilde{g}}{\sqrt{2}}\omega(0)=3\Delta_1 -\Delta_2
\label{ebpw}
\end{eqnarray}
will be attained. We note that the hyperfine parameters in these two channels 
behave oppositely as functions of $\alpha$.
\begin{table}
\centerline{
\begin{tabular}{l | c c c c c | c | c}
$\alpha$ & -0.1 & 0.0 & 0.1 & 0.2 & 0.3 & Expt. & Skyrme \\
\hline
$\omega_P$    & 564 & 544 & 522 & 500 & 478 & & 243\\
$\chi_P$      & 0.147 & 0.140 & 0.131 & 0.123 & 0.114 & & 0.053 \\
$\omega_S$    & 316 & 298 & 281 & 264 & 247 & & 57\\
$\chi_S$      & 0.172 & 0.181 & 0.189 & 0.197 & 0.205 & &0.346 \\  
\hline
$\Sigma_c$     & 171 & 172 & 174 & 175 & 177 & 168 & 185\\
$\Sigma_c^*$   & 215 & 214 & 213 & 212 & 211 & 233 & 201\\
$\Lambda_c^\prime $ & 250 & 249 & 245 & 242 & 238 & 308 & 208 \\
$\Sigma_c^\prime $ & 415 & 413 & 408 & 402 & 397 & ? & 335\\
$\Sigma_c^{\prime *} $ & 468 & 467 & 464 & 461 & 458 & ? & 437 \\
\hline
$N$ & -1237 & -1257 & -1278 & -1299 & -1321 & -1345 & -1553\\
\hline
$\Lambda_b$ & 3160 & 3164 & 3167 & 3170 & 3173 & $3356\pm50$
&3215 
\end{tabular}
}
\caption{\label{tab_1}
Parameters for heavy baryons and mass differences 
with respect to $\Lambda_c$. Primes indicate negative parity
baryons, {\it i.e.} S--wave bound states. All energies are in MeV.}
\end{table}
Here we have chosen to measure the mass differences with respect to 
the lightest charmed baryon, $\Lambda_c$. Hence the mass differences
with respect to $\Sigma_c$ and $\Sigma_c^*$ directly reflect the 
$\alpha$--dependence of hyperfine parameter $\chi_P$ while the 
corresponding dependence of the binding energy $\omega_P$ can be 
extracted from the splitting relative to the nucleon. In addition 
the splitting with respect to the negative parity charmed baryons 
reflects the $\alpha$--dependence of the S--wave channel 
binding energy $\omega_S$. Finally the mass difference to 
$\Lambda_b$ contains the energy eigenvalues and hyperfine 
parameters computed with the $B$ and $B^*$ meson masses in 
Eq.~(\ref{heavypara}).

While the mass difference to the nucleon is improved with
a positive value for $\alpha$, the agreement for the 
mass differences between the heavy baryons slightly 
deteriorates when increasing this parameter. Nevertheless,
fair agreement with the experimental data is 
achieved for quite a range of $\alpha$.

Table \ref{tab_1} also contains the model predictions when 
the background soliton is taken from the basic Skyrme model 
\cite{Sk61,Ad83} which does not include the light vector mesons. 
Here we have adjusted the only free parameter 
($e_{\rm Skyrme}=4.25$) to reproduce the $\Delta$--nucleon mass 
difference. From the $\Lambda_c$--nucleon mass difference we observe 
that in comparison with the nucleon the masses of the heavy baryons 
are predicted about $200~{MeV}$ too large. This confirms the 
above statement that the spectra of both the light and the heavy 
baryons can only be reasonably reproduced when light vector mesons 
are included. This conclusion can already be drawn from the too 
small binding energies \cite{Schechter-Subbaraman-Vaidya-Weigel}. 
The hyperfine corrections 
make only minor changes in the $\Lambda_b$--$\Lambda_c$ splitting.

In Table \ref{tab_2} we display the analogous predictions for 
the bottom sector. According to the heavy spin symmetry the 
binding energies of the P-- and S--wave channels approach each 
other. Hence the mass differences between the even and odd parity 
baryons containing a bottom quark correspondingly decrease. 
From Table \ref{tab_1} we can infer that $\chi_P$ decreases less quickly 
with the heavy meson mass than $\chi_S$.
\begin{table}
\centerline{
\begin{tabular}{l | c c c c c }
$\alpha$ & -0.1 & 0.0 & 0.1 & 0.2 & 0.3 \\
\hline
$\omega_P$    & 811 & 786 & 762 & 737 & 713 \\
$\chi_P$      & 0.055 & 0.053 & 0.050 & 0.048 & 0.045 \\
$\omega_S$    & 639 & 617 & 595 & 573 & 552 \\
$\chi_S$      & 0.043 & 0.046 & 0.049 & 0.052 & 0.055 \\  
\hline
$\Sigma_b$     & 189 & 189 & 190 & 190 & 191  \\
$\Sigma_b^*$   & 206 & 205 & 205 & 205 & 205  \\
$\Lambda_b^\prime $ & 171 & 168 & 167 & 164 & 161  \\
$\Sigma_b^\prime $ & 363 & 359 & 358 & 354 & 351  \\
$\Sigma_b^{\prime *} $ & 375 & 373 & 371 & 369 & 367  \\
\hline
$N$ & -4397 & -4422 & -4446 & -4471 & -4494
\end{tabular}
}
\caption{\label{tab_2}
Parameters for heavy baryons and mass differences 
with respect to $\Lambda_b$. Primes indicate negative parity
baryons, {\it i.e.} S--wave bound states. All energies are in MeV.
The empirical value for the relative position of the nucleon
is $4701\pm50{\rm MeV}$ \cite{pdg}.}
\end{table}
Except for $\Lambda_b$ no empirical data for the masses of these 
baryons are known at present. These results for the mass differences 
among the bottom baryons are predictions of the model which can, in 
the future, be compared with experiment. As could have been inferred 
from the next to last row in Table \ref{tab_1} the absolute position 
of the bottom multiplet is about $200\pm50{\rm MeV}$ too low. 
On the absolute scale this apparently is only a 5\% deviation from 
the data. Certainly a larger value $\alpha\approx1$,
which corresponds to a model for light vector resonance dominance 
of the heavy meson form factor~\cite{Ja95}, would yield an 
excellent agreement for the mass difference between $\Lambda_b$ 
and the nucleon. On the other hand such a choice would slightly
spoil the nice picture for the charm multiplet.

The preceding calculations are based on the 
$N_{c}\rightarrow\infty$ limit in which the nucleon is 
infinitely heavy. From a common sense point of view this is 
peculiar since the nucleon is actually lighter than the heavy 
mesons being bound to it in the model. 
Hence, for comparison with experiment it is desirable to estimate
kinematic effects associated with the nucleon's motion.
These are expected~\cite{Schechter-Subbaraman} to lower the binding energy of 
the heavy baryons which have up to now come out too high
(see $\Lambda_c$--$N$ mass difference in Table~\ref{tab_1}, for example.).
In order to estimate these kinematical effects in the bound state 
approach we have substituted the reduced masses 
\begin{eqnarray}
\frac{1}{M}\longrightarrow\frac{1}{M_{\rm cl}}
+\frac{1}{M}
\quad {\rm and} \quad
\frac{1}{M^*}\longrightarrow\frac{1}{M_{\rm cl}}
+\frac{1}{M^*}
\label{reduced}
\end{eqnarray}
into the bound state equations. In a non--relativistic treatment
this corresponds to the elimination of the center of mass motion 
\cite{Schechter-Subbaraman}. 
The results for the spectrum of the heavy baryons 
obtained from the replacement (\ref{reduced}) are in displayed in 
Table \ref{tab_4}. 
\begin{table}
\centerline{
\begin{tabular}{l | c c c c c c c c | c}
$\alpha$ & 0.0 & -0.1 & -0.2 & -0.3 & -0.4 & -0.5 & -0.6 & -0.7 & Expt. \\
\hline
$\omega_P$    & 450 & 469 & 488 & 508 & 527 & 546 & 566 & 585 & \\
$\chi_P$      & 0.212 & 0.232 & 0.246 & 0.260 & 0.273 & 0.286 & 
0.299 & 0.312 & \\
$\omega_S$    & 123 & 134 & 146 & 158 & 171 & 184 & 197 & 210 & \\
$\chi_S$      & 0.410 & 0.399 & 0.387 & 0.374 & 0.361 & 0.346 & 
0.331 & 0.315 & \\  
\hline
$\Sigma_c$     & 158 & 154 & 151 & 148 & 145 & 143 & 140 & 138 & 168 \\
$\Sigma_c^*$   & 221 & 223 & 225 & 226 & 227 & 229 & 230 & 231 & 233 \\
$\Lambda_c^\prime $ & 342 & 346 & 353 & 359 & 363 & 367 & 
371 & 375 & 308 \\
$\Sigma_c^\prime $  & 460 & 468 & 475 & 484 & 490 & 497 & 
505 & 512 & ? \\
$\Sigma_c^{\prime *} $ & 583 & 587 & 591 & 596 & 599 & 601 & 
605 & 607 & ? \\
\hline
$N$ & -1356 & -1338 & -1320 & -1302 & -1283 & -1265 & -1246 & 
-1228 & -1345 \\
\hline
$\Lambda_b$ & 3285 & 3282 & 3280 & 3278 & 3275 & 3272 & 3271 & 
3269 & $3356 \pm 50$
\end{tabular}
}
\caption{\label{tab_4}
Parameters for heavy baryons and mass differences with 
respect to $\Lambda_c$. Primes indicate negative parity states, 
{\it i.e.} S--wave bound states. All energies are in MeV. In this 
calculation the reduced masses (\ref{reduced}) enter the bound state 
equations from which the binding energies are extracted. The 
physical meson masses 1865MeV and 5279MeV are used when computing 
the mass differences to the nucleon and the $\Lambda_b$ from these 
binding energies. Radially excited states are omitted because 
they are only very loosely bound, if at all.
The empirical data are taken from the PDG \cite{pdg}, see also 
\cite{Br97}.}
\end{table}
Again we consider $\alpha$ as a free parameter. We notice that 
there is a remarkable improvement in the prediction for the 
$\Lambda_b$ mass, which was previously the worst one. The changes 
in some of the mass parameters can approximately be compensated 
by a suitable re--adjustment of $\alpha$. 
For $\alpha\approx0.0$ to $-0.4$ the agreement with the 
existing data is quite reasonable. When using the reduced meson 
masses the $\Sigma_c$ baryon is always predicted a bit too 
light while it is too heavy when the physical masses are 
substituted in the bound state equation. For $\Lambda_c^\prime$
the situation is opposite. While the use of the physical meson masses
gives too small a mass, the substitution of the reduced masses 
gives too large a prediction for the mass of this baryon. These 
results indicate that kinematical corrections are indeed 
important. 

It is interesting to see how far the heavy quark approach 
can be pushed to lighter quarks. To answer this question we 
have considered the strange quark. In the corresponding kaon 
sector the P--wave is only very loosely bound when the physical 
masses are substituted. On the other hand sizable binding 
energies are obtained when the reduced masses are used 
\cite{Schechter-Subbaraman-Vaidya-Weigel}. 
This behavior is somewhat different from the charm and bottom 
sector and can be understood by noting that the difference 
$M^*-M$ is considerably reduced when using (\ref{reduced}). In the 
heavy sectors (charm and bottom) this difference is small in any 
event.
\begin{table}
\centerline{
\begin{tabular}{l | c c c c c c c c | c}
$\alpha$ & 0.0 & -0.1 & -0.2 & -0.3 & -0.4 & -0.5 & -0.6 & -0.7 & Expt. \\
\hline
$\omega_P$    &  80 &  94 & 109 & 124 & 140 & 155 & 171 & 188 & \\
$\chi_P$      & 0.346 & 0.371 & 0.394 & 0.417 & 0.439 & 0.460 & 
0.479 & 0.498 & \\
\hline
$\Sigma$     & 131 & 126 & 121 & 117 & 112 & 108 & 104 & 100 & 77 \\
$\Sigma^*$   & 235 & 237 & 239 & 242 & 244 & 246 & 248 & 250 & 269 \\
\hline
$N$ & -366 & -354 & -341 & -327 & -313 & -300 & -285 & 
-269 & -177
\end{tabular}
}
\caption{\label{tab_5}
Same as Table \ref{tab_4} for even parity baryons
in the kaon sector.}
\end{table}
The resulting spectrum for the strange baryons is shown in Table 
\ref{tab_5}. The comparison with the experimental data shows that 
even the use of the reduced masses does not provide sufficient 
binding. In the S--wave channel the situation is worse, even when 
the reduced masses are substituted bound states are not detected unless
$\alpha\le-1.0$. The failure of the present approach in the
strange sector strongly suggests that for these baryons a 
chirally invariant set--up \cite{We96} is more appropriate.

%% file: chapterh3.tex
\chapter{Generalization of the Bound State Model}

\section{Introduction}

Here we shall investigate the spectrum of excited states in the bound
state--soliton framework.
Some aspects of this problem have already been 
treated~\cite{CW,Schechter-Subbaraman,Oh-Park,HQSSW:1,HQSSW:NPA}.
We will deal with an aspect which does not seem to have been
previously discussed in the literature.
This emerges when one compares the excited heavy baryon spectrum with
that expected in the constituent quark model (CQM)~\cite{CI}.
We do not have in mind specific dynamical treatments of the CQM 
but rather just its general geometric structure.
Namely we shall just refer to the counting of states which follows
from considering the baryon as a three body system obeying
Fermi--Dirac statistics.  We shall restrict our attention to the 
physical states for $N_{c}=3$.
In this framework the CQM counting of the heavy excited baryon
multiplets has been recently discussed~\cite{Koerner}.
At the level of two light flavors there are expected to be seven
negative parity first excited $\Lambda$--type heavy baryons and seven
negative parity first excited $\Sigma$--type heavy baryons.
On the other hand a similar counting~\cite{%
Schechter-Subbaraman,HQSSW:NPA} in the bound state treatments 
mentioned above yields only two of the $\Lambda$--type and five of
the $\Sigma$--type.  Thus there are seven missing first excited states.
One thought is that these missing states should be unbound and thus
represent new dynamical information with respect to the simple
geometrical picture. There is certainly not enough data for the 
charmed baryons to decide this issue.  However for the strange 
baryons there are ten established particles for these fourteen 
states. Hence it is reasonable to believe that these states exist 
for the heavy baryons too. In the CQM one may have two different 
sources of orbital angular momentum excitation; for example the 
relative angular momentum of the two light quarks, $L_I$ and the 
angular momentum, $L_E$ of the diquark system with respect to the
heavy quark.  The parity of the heavy baryon is given by 
$P=\left(-1\right)^{L_I+L_E}$. However, in the bound state models 
considered up to now there is only room for one relative angular 
momentum, $r$ associated with the wave function of the heavy meson 
with respect to the soliton. The parity is given by 
$P=\left(-1\right)^r$.  Both models agree on the counting of the 
``ground'' states ($L_I=L_E=r=0$).  Also the counting of the states 
with ($L_I=0$, $L_E=1$) agrees with those of $r=1$ in the bound state 
model.  However, the bound state model has no analog of the ($L_I=1$, 
$L_E=0$) states and, in general, no analog of the higher $L_I\neq0$ 
states either.

It is clear that we must find a way of incorporating a new angular
momentum quantum number in the bound state picture. One might imagine 
a number of different ways to accomplish this goal.  Here we will 
investigate a method which approximates a three body problem by an 
effective two body problem. Specifically we will consider binding 
excited heavy mesons with orbital angular momentum $\ell$ to the
soliton.
The excited heavy mesons may be interpreted as bound states of 
the original heavy meson and a surrounding light meson cloud.
Then the baryon parity comes out to be $\left(-1\right)^{r+\ell}$.
This suggests a correspondence (but not an identity) $r
\leftrightarrow L_E$, $\ell \leftrightarrow L_I$ and additional new
states.
An interesting conceptual point of the model is that it displays a
correspondence between the excited heavy mesons and the excited heavy
baryons.

Almost immediately one sees that the model is considerably more
complicated than the previous one in which the single heavy field
multiplet $H$ is bound to the soliton.
Now, for each value of $\ell\neq0$, there will be two different higher 
spin heavy multiplets which can contribute.
In fact there is also a mixing between multiplets with different
$\ell$, which is therefore not actually a good quantum number for the
model (unless the mixing is neglected).

Thus we will make a number of approximations which seem reasonable for
an initial analysis.
For one thing we shall neglect the light vector mesons even though we
know they may be important.
We shall also neglect the possible effects of higher spin light
mesons, which one might otherwise consider natural when higher spin
heavy mesons are being included.
Since there is a proliferation of interaction terms among the light
and heavy mesons we shall limit ourselves to those with the minimum
number of derivatives.
Finally, $1/M$ and nucleon recoil corrections will be neglected.
The resulting model is the analog of the initial one used previously.
Even though the true picture is likely to be more involved than our
simplified model, we feel that the general scheme presented here will
provide a useful guide for further work.

We would like to stress that this bound state model goes beyond the
kinematical enumeration of states and contains dynamical information.
Specifically, the question of which states are bound depends on the
magnitudes and signs of the coupling constants.
There is a choice of coupling constants yielding a natural pattern of
bound states which includes the missing ones.
It turns out that it is easier to obtain the precise missing state
pattern for the $\Lambda$--type heavy particles.
Generally, there seem to be more than just the missing $\Sigma$--type
heavy baryons present.
However we show that the collective quantization, which is anyway
required in the bound state approach, leads to a splitting which may
favor the missing heavy spin multiplets.

This chapter is organized in the following way.
Section~\ref{sec:preliminaries} starts with a review of the CQM
geometrical counting of excited heavy baryon multiplets.
It continues with a quick summary of the treatment of heavy baryons in
the existing bound state models.
The comparison of the mass spectrum in the two different approaches
reveals that there is a large family of ``missing'' excited states.
This is discussed in general terms in section~\ref{sec: missing} where
a proposal for solving the problem by considering the binding of heavy
excited mesons to the Skyrmion is made.
A correspondence between the angular momentum variables of the CQM and
of the new model is set up.
A detailed treatment of the proposed model for the case of the first
excited heavy baryons is given in section~\ref{sec:4}.
This includes discussion of the heavy meson bound state wave function,
the classical potential energy as well as the energy corrections due
to quantization of the collective variables of the model.
It is pointed out that there is a possible way of choosing the
coupling constants so as to bind all the missing states.
The generalization to the excited heavy baryon states of arbitrary
spin is given in section~\ref{sec:5}.
This section also contains some new material on the interactions of
the heavy meson multiplets with light chiral fields.
Section~\ref{sec:6} contains a discussion of the present status of the
model introduced here.
Finally, some details of the calculations are given in
Appendices~\ref{app:a} and \ref{app:b}.

\section{The Enigma of the Excited Missing States}

\label{sec:preliminaries}

In this section, for the reader's convenience, we will briefly discuss
which heavy baryon states are predicted by the CQM as well as some 
relevant material needed for the bound state approach to the heavy
baryon states. 

It is generally agreed that the geometrical structure of the CQM 
provides a reasonable guide for, at least, counting and labeling 
the physical strong interaction ground states. When radial excitations
or dynamical aspects are considered the model predictions are 
presumably less reliable. In the CQM the heavy baryons consist of two 
light quarks ($q$) and a heavy quark ($Q$) in a color singlet state.
Since the color singlet states are antisymmetric on interchange of the
color labels of any two quarks, the overall wave function must,
according to Fermi-Dirac statistics, be fully symmetric on interchange
of flavor, spin and spatial indices.
Here we will consider the case of two light flavors.
For counting the states we may choose coordinates~\cite{Koerner}
so that the total angular momentum of the heavy baryon,
$\mbox{\boldmath$J$}$ is decomposed as
\begin{equation}
\mbox{\boldmath$J$} = \mbox{\boldmath$L$}_I + \mbox{\boldmath$L$}_E
+ \mbox{\boldmath$S$} + \mbox{\boldmath$S$}_H \ ,
\label{baryon spin:CQM}
\end{equation}
where $\mbox{\boldmath$L$}_I$ represents the relative orbital angular
momentum of the two light quarks, $\mbox{\boldmath$L$}_E$ the orbital
angular momentum of the light diquark center of mass with respect to
the heavy quark, $\mbox{\boldmath$S$}$ the total spin of the diquarks
and $\mbox{\boldmath$S$}_H$ the spin of the heavy quark. In the 
``heavy'' 
limit where the heavy quark becomes infinitely
massive $\mbox{\boldmath$S$}_H$ 
completely decouples. The parity of the heavy baryon is given by
\begin{equation}
P_B = \left( -1 \right)^{L_I + L_E} \ .
\label{baryon parity:CQM}
\end{equation}
Since we are treating only the light degrees of freedom as identical
particles it is only necessary to symmetrize the diquark product wave
function with respect to the $\mbox{\boldmath$L$}_I$,
$\mbox{\boldmath$S$}$ and isospin $\mbox{\boldmath$I$}$ labels.
Note that the diquark isospin $\mbox{\boldmath$I$}$ equals 
the baryon isospin. 
There are four possible ways to build an overall wave function
symmetric with respect to these three labels:
\begin{eqnarray}
&&\mbox{a)} \quad I=0 \ , \quad S=0 \ , \quad L_I = \mbox{even} \ ,
\nonumber\\
&&\mbox{b)} \quad I=1 \ , \quad S=1 \ , \quad L_I = \mbox{even} \ ,
\nonumber\\
&&\mbox{c)} \quad I=0 \ , \quad S=1 \ , \quad L_I = \mbox{odd} \ ,
\nonumber\\
&&\mbox{d)} \quad I=1 \ , \quad S=0 \ , \quad L_I = \mbox{odd} \ .
\label{condition:CQM}
\end{eqnarray}
There is no kinematic restriction on $L_E$.\footnote{%
We are adopting a convention where bold--faced angular momentum
quantities are vectors and the regular quantities stand for their
eigenvalues.}

Let us count the possible baryon states. The 
$L_I=L_E=0$ heavy baryon ground state consists of $\Lambda_Q$
($J^P=\frac{1}{2}^+$) from a) and the heavy spin multiplet
$\left\{\Sigma_Q\left(\frac{1}{2}^+\right) \ , \ 
\Sigma_Q\left(\frac{3}{2}^+\right) \right\}$ from b).
It is especially interesting to consider the first orbitally excited
states.  These all have negative parity with either ($L_E=1$, $L_I=0$) or
($L_E=0$, $L_I=1$).
For $L_E=1$, a) provides the heavy spin multiplet 
$\left\{\Lambda_Q\left(\frac{1}{2}^-\right)\ , 
\ \Lambda_Q\left(\frac{3}{2}^-\right)\right\}$ and b) provides
$\Sigma_Q\left(\frac{1}{2}^-\right)$,
$\left\{\Sigma_Q\left(\frac{1}{2}^-\right)\ ,
\ \Sigma_Q\left(\frac{3}{2}^-\right)\right\}$,
$\left\{\Sigma_Q\left(\frac{3}{2}^-\right)\ ,
\ \Sigma_Q\left(\frac{5}{2}^-\right)\right\}$.
For $L_I=1$ c) provides $\Lambda_Q\left(\frac{1}{2}^-\right)$,
$\left\{\Lambda_Q\left(\frac{1}{2}^-\right)\ ,\
\Lambda_Q\left(\frac{3}{2}^-\right)\right\}$,
$\left\{\Lambda_Q\left(\frac{3}{2}^-\right)\ ,\
\Lambda_Q\left(\frac{5}{2}^-\right)\right\}$,
while d) provides
$\left\{\Sigma_Q\left(\frac{1}{2}^-\right)\ ,
\ \Sigma_Q\left(\frac{3}{2}^-\right)\right\}$.
Altogether there are fourteen different isotopic spin multiplets at
the first excited level.
The higher excited levels can be easily enumerated in the same way.
For convenient reference these are listed in Table~\ref{table:1}.
\begin{table}[htbp]
\begin{center}
\begin{tabular}{c|cc}
\multicolumn{1}{c}{\ } & $L_E=0$ & $L_E=1$ \\
\hline
 $L_I=0$ 
& \multicolumn{1}{c}{$\begin{array}{c}
\Lambda_Q\left(\frac{1}{2}^+\right) \\
\left\{\Sigma_Q\left(\frac{1}{2}^+\right) \,,\,
\Sigma_Q\left(\frac{3}{2}^+\right)\right\}
\end{array}$ }
& \multicolumn{1}{c}{$\begin{array}{c}
\left\{\Lambda_Q\left(\frac{1}{2}^-\right) \,,\,
\Lambda_Q\left(\frac{3}{2}^-\right)\right\} \\
\Sigma_Q\left(\frac{1}{2}^-\right) \\
\left\{\Sigma_Q\left(\frac{1}{2}^-\right) \,,\,
\Sigma_Q\left(\frac{3}{2}^-\right)\right\} \\
\left\{\Sigma_Q\left(\frac{3}{2}^-\right) \,,\,
\Sigma_Q\left(\frac{5}{2}^-\right)\right\}
\end{array}$ }
\\
\hline
 $L_I=1$
& \multicolumn{1}{c}{$\begin{array}{c}
\Lambda_Q\left(\frac{1}{2}^-\right) \\
\left\{\Lambda_Q\left(\frac{1}{2}^-\right) \,,\,
\Lambda_Q\left(\frac{3}{2}^-\right)\right\} \\
\left\{\Lambda_Q\left(\frac{3}{2}^-\right) \,,\,
\Lambda_Q\left(\frac{5}{2}^-\right)\right\} \\
\left\{\Sigma_Q\left(\frac{1}{2}^-\right) \,,\,
\Sigma_Q\left(\frac{3}{2}^-\right)\right\}
\end{array}$ }
& $\cdots$ 
\\
\hline
 $\vdots$ & & 
\\
\hline
 $L_I=2n-1$
& \multicolumn{1}{c}{$\begin{array}{c}
\left\{\Lambda_Q\left(\left(2n-\frac{5}{2}\right)^-\right) \,,\,
\Lambda_Q\left(\left(2n-\frac{3}{2}\right)^-\right)\right\} \\
\left\{\Lambda_Q\left(\left(2n-\frac{3}{2}\right)^-\right) \,,\,
\Lambda_Q\left(\left(2n-\frac{1}{2}\right)^-\right)\right\} \\
\left\{\Lambda_Q\left(\left(2n-\frac{1}{2}\right)^-\right) \,,\,
\Lambda_Q\left(\left(2n+\frac{1}{2}\right)^-\right)\right\} \\
\left\{\Sigma_Q\left(\left(2n-\frac{3}{2}\right)^-\right) \,,\,
\Sigma_Q\left(\left(2n-\frac{1}{2}\right)^-\right)\right\}
\end{array}$ }
& $\cdots$ 
\\
\hline
 $L_I=2n$
& \multicolumn{1}{c}{$\begin{array}{c}
\left\{\Lambda_Q\left(\left(2n-\frac{1}{2}\right)^+\right) \,,\,
\Lambda_Q\left(\left(2n+\frac{1}{2}\right)^+\right)\right\} \\
\left\{\Sigma_Q\left(\left(2n-\frac{3}{2}\right)^+\right) \,,\,
\Sigma_Q\left(\left(2n-\frac{1}{2}\right)^+\right)\right\} \\
\left\{\Sigma_Q\left(\left(2n-\frac{1}{2}\right)^+\right) \,,\,
\Sigma_Q\left(\left(2n+\frac{1}{2}\right)^+\right)\right\} \\
\left\{\Sigma_Q\left(\left(2n+\frac{1}{2}\right)^+\right) \,,\,
\Sigma_Q\left(\left(2n+\frac{3}{2}\right)^+\right)\right\} 
\end{array}$ }
& $\cdots$ 
\\
\hline
 $\vdots$ & & 
\\
\end{tabular}
\end{center}
\caption{
Examples of the heavy baryon multiplets predicted by the CQM.
}
\label{table:1}
\end{table}

It is natural to wonder whether all of these states should actually
exist experimentally. This is clearly a premature question for the $c$
and $b$ baryons.  However an indication for the first excited states 
can be gotten from the
ordinary hyperons (or $s$ baryons).  In this case there are six well
established candidates~\cite{PDG} for the $\Lambda$'s 
[$\Lambda(1405)$, $\Lambda(1520)$, $\Lambda(1670)$, $\Lambda(1690)$,
$\Lambda(1800)$ and $\Lambda(1830)$]; only one $\frac{3}{2}^-$ state
has not yet been observed.  For the $\Sigma$'s there are four well
established candidates [$\Sigma(1670)$, $\Sigma(1750)$, $\Sigma(1775)$
and $\Sigma(1940)$]; two $\frac{1}{2}^-$ states and one
$\frac{3}{2}^-$ state have not yet been observed.  
Thus it seems plausible to expect that all fourteen of the first 
excited negative parity heavy baryons do indeed exist.
We might also expect higher excited states to exist.

What is the situation in the bound state approach?
To study this we shall briefly summarize the usual approach~\cite{%
Schechter-Subbaraman,Oh-Park,HQSSW:NPA}
to the excited heavy baryons in the bound state picture.
In this model the heavy baryon is considered to be a heavy meson
bound, via its interactions with the light mesons, to a nucleon
treated as a Skyrme soliton.
 The total baryon angular momentum is the sum (see Eq.~(\ref{baryon spin}))
\begin{equation}
\mbox{\boldmath$J$} =
\mbox{\boldmath$g$} + \mbox{\boldmath$J$}^{\rm sol} + 
\mbox{\boldmath$S$}_{\rm heavy}
\ ,
\end{equation}
where $\mbox{\boldmath$S$}_{\rm heavy}$ is the spin of the heavy quark
within the heavy meson.

Now we can list the bound states of this model.
First consider the $r=0$ states.
According to Eq.~(\ref{baryon parity}), they have positive parity.
Since Eq.~(\ref{potential: H}) shows that $k=0$ for binding,
Eq.~(\ref{def: g1}) tells us that the light ``grand spin'' $g=0$.
Equation~(\ref{baryon spin}) indicates (noting $I=J^{\rm sol}$) that
there will be a  $\Lambda_Q\left(\frac{1}{2}^+\right)$ state as well
as a $\left\{\Sigma_Q\left(\frac{1}{2}^+\right),\ 
\Sigma_Q\left(\frac{3}{2}^+\right)\right\}$
heavy spin multiplet.
Actually the model also predicts a whole tower of states with 
increasing isospin. Next there will be an $I=2$ 
heavy spin multiplet with spins and parity $\frac{3}{2}^+$ and 
$\frac{5}{2}^+$, and so forth. Clearly the isospin zero and one 
states correspond exactly to the $L_I=L_E=0$ ground states of the 
constituent quark model. The isotopic spin two states would also be 
present if we were to consider the ground state heavy baryons in 
a constituent quark model with number of colors, $N_{c}=5$.
This is consistent with the picture~\cite{Ad83}
of the Skyrme model as a
description of the large $N_{c}$ limit.

Next, consider the $r=1$ states.
These all have negative parity and (since the bound states have $k=0$)
light grand spin, $g=1$.
The $J^{\rm sol}=I=0$ choice yields a heavy multiplet
$\left\{\Lambda_Q\left(\frac{1}{2}^-\right)\ ,
\ \Lambda_Q\left(\frac{3}{2}^-\right)\right\}$
while the $J^{\rm sol}=I=1$ choice yields the three heavy multiplets
$\left\{\Sigma_Q\left(\frac{1}{2}^-\right)\right\}$,
$\left\{\Sigma_Q\left(\frac{1}{2}^-\right)\ , 
\Sigma_Q\left(\frac{3}{2}^-\right)\right\}$
and
$\left\{\Sigma_Q\left(\frac{3}{2}^-\right)\ , 
\Sigma_Q\left(\frac{5}{2}^-\right)\right\}$. These three multiplets
are associated with the intermediate sums
$|\mbox{\boldmath $g$}+\mbox{\boldmath $J$}^{\rm sol}|=0,1,2$,
respectively. It is evident that the seven states obtained have the 
same quantum numbers as the seven constituent quark states with 
$L_I=0$ and $L_E=1$.  Proceeding in the same way, it is easy to see 
that the bound states with general $r$ agree with those states in 
the constituent quark model which have $L_I=0$ and $L_E=r$.
This may be understood by rewriting Eqs.~(\ref{baryon spin}) and
(\ref{def: g1}) as
\begin{equation}
\mbox{\boldmath$J$} =
\mbox{\boldmath$r$} + \mbox{\boldmath$J$}^{\rm sol} +
\mbox{\boldmath$S$}_{\rm heavy}
\ ,
\end{equation}
where $k=0$ for the bound states was used.
Comparing this with the $L_I=0$ limit of the constituent quark model
relation (\ref{baryon spin:CQM}) shows that there seems to be a
correspondence
\begin{eqnarray}
\mbox{\boldmath$S$}_{\rm heavy}
&\leftrightarrow&
\mbox{\boldmath$S$}_H
\ ,
\nonumber\\
\mbox{\boldmath$r$}
&\leftrightarrow&
\mbox{\boldmath$L$}_E
\ ,
\nonumber\\
\mbox{\boldmath$J$}^{\rm sol}
&\leftrightarrow&
\mbox{\boldmath$S$}
\ .
\label{correspondence:1}
\end{eqnarray}
This correspondence is reinforced when we notice that 
$I=J^{\rm sol}$ in the bound state model and,
for the relevant cases a) and b) in Eq.~(\ref{condition:CQM}) of the
constituent quark model, $I=S$ also.
We stress that Eq.~(\ref{correspondence:1}) is a correspondence rather
than an exact identification of the same dynamical variables in
different models. It should be remarked that in the exact heavy and
large 
$N_c$ limits the heavy baryons for all values of $r=g$ will have the 
same mass. When finite $1/M$ corrections are taken into account, there 
will always be, in addition to other things, a ``centrifugal term'' in 
the effective potential of the form 
${g(g+1)}/({2M\vert\mbox{\boldmath$x$}\vert^2})$,
which makes the states with larger values of $g$, heavier.
It should also be remarked that the above described ordering of 
heavy baryon states in the bound state approach applies only
to the heavy limit, where $\mbox{\boldmath $S$}_{\rm heavy}$ 
decouples. For finite heavy quark masses, multiplets are 
characterized by the total grand spin 
$\mbox{\boldmath $g$}+
\mbox{\boldmath $S$}_{\rm heavy}$. Then states like
$\Lambda_Q\left(\frac{1}{2}^-\right)$ and 
$\Lambda_Q\left(\frac{3}{2}^-\right)$ no longer constitute 
a degenerate multiplet.

\section{The Planetary Conjecture}
\label{sec: missing}

It is clear that the bound state model discussed above 
contains only half
of the fourteen negative parity, first excited states predicted by the
CQM. The states with $L_I\neq0$ are all missing.
Since the enumeration of states in the CQM was
purely kinematical one might at first think that the bound state
model (noting that the dynamical condition $k=0$ was used) is
providing a welcome constraint on the large number of expected states.
However, experiment indicates that this is not likely to be the case.
As pointed out in the last section, there are at present good
experimental candidates for ten out of the fourteen negative parity,
first excited ordinary hyperons.
Thus the missing excited states appear to be a serious problem for the
bound state model.

\begin{figure}[htbp]
\begin{center}
\ \epsfbox{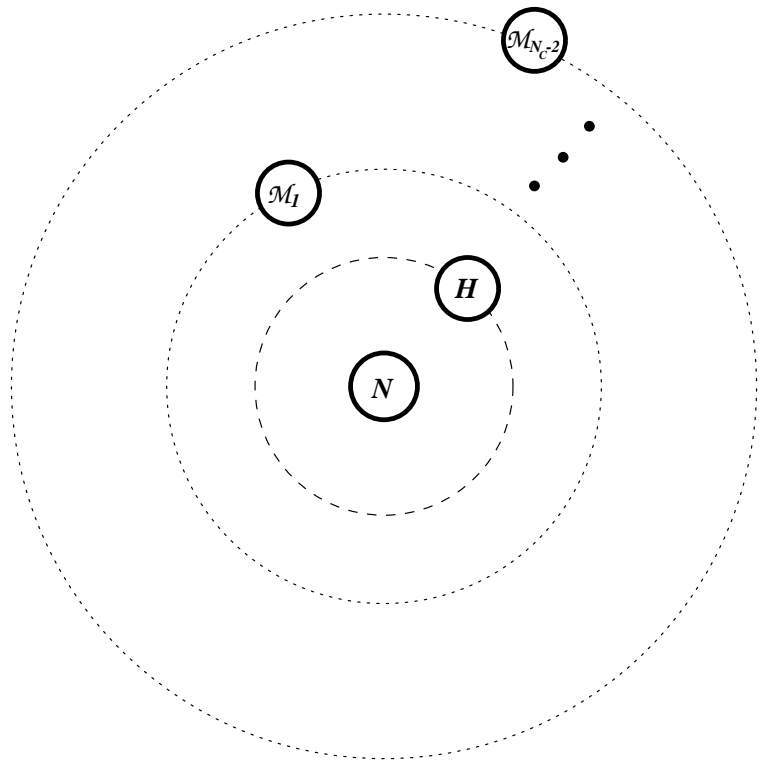}
\end{center}
\caption{Schematic planetary picture for large $N_{c}$ excited
heavy baryons in the bound state approach.}
\label{fig:1}
\end{figure}

\begin{figure}[htbp]
\begin{center}
\ \epsfbox{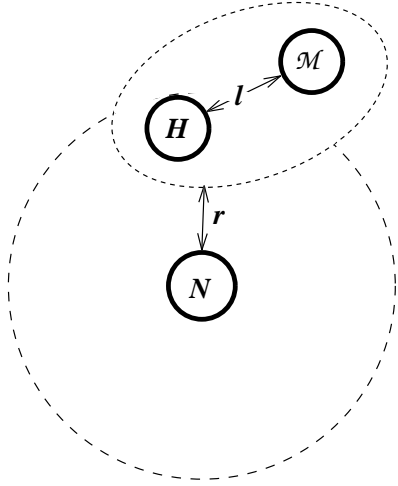}
\end{center}
\caption{Schematic picture of the ``two body'' approximation for the
$N_{c}=3$ excited heavy baryons.}
\label{fig:2}
\end{figure}

The goal of the present work is to find a suitable extension of the
bound state model which gives the same spectrum as the CQM.
Reference to Eq.~(\ref{baryon spin:CQM}) suggests that we introduce a
new degree of freedom which is related in some way to the light
diquark relative angular momentum $\mbox{\boldmath$L$}_I$.
To gain some perspective, and because we are working in a Skyrme model
overall framework, it is worthwhile to consider the heavy baryons in a
hypothetical world with $N_{c}$ quark colors. In such a case 
there would be $N_{c}-1$ relative angular momentum variables and 
we would require $N_{c}-2$ additional degrees of freedom.
Very schematically we might imagine, as in Fig.~\ref{fig:1},
one heavy meson $H$ and $N_{c}-2$ light mesons ${\cal M}_i$ orbiting
around the nucleon.
One might imagine a number of different schemes for treating the
inevitably complicated bound state dynamics of such a system.
Even in the $N_{c}=3$ case it is much simpler if we can manage to
reduce the three body problem to an effective two body problem.
This can be achieved, as schematically indicated in Fig.~\ref{fig:2},
if we link the two ``orbiting'' mesons together in a state which
carries internal angular momentum.
The ``linked mesons'' will be described mathematically by a single 
excited heavy meson multiplet field. One may alternatively 
consider these ``linked mesons'' as bare heavy mesons surrounded
by a light meson cloud. Such fields are usually classified by the 
value\footnote{ Actually if we want to picture the linked mesons as 
literally composed of a meson--meson pair, we should assign relative 
orbital angular momentum $\ell-1$ to these bosonic constituents and
allow for both light pseudoscalars and vectors.}, $\ell$ of the
relative orbital angular momentum of a $\bar{q}Q$ pair which describes
it in the CQM.  
We will not attempt to explain the binding of these two  
mesons but shall simply incorporate the ``experimental'' higher spin 
meson fields into our chiral Lagrangian. Different $\ell$ excitations
will correspond to the use of different meson field multiplets.
{}From now on we will restrict our attention to $N_{c}=3$.

Taking the new degree of freedom $\mbox{\boldmath$\ell$}$ into account
requires us to modify the previous formulas describing the heavy
baryon.
Now the parity formula (\ref{baryon parity}) is modified to 
\begin{equation}
P_B = \left( -1 \right)^{\ell+r} \ ,
\label{new baryon parity}
\end{equation}
which is seen to be compatible with the CQM relation 
(\ref{baryon parity:CQM}).
Now Eq.~(\ref{baryon spin}) holds but with the light grand spin 
$\mbox{\boldmath$g$}$ modified to,
\begin{equation}
\mbox{\boldmath$g$} = 
\mbox{\boldmath$r$} + \mbox{\boldmath$K'$} \ .
\label{def: g}
\end{equation}
Note that $\mbox{\boldmath$K$}$ in Eq.~(\ref{def: k}) has been
incorporated in
\begin{equation}
\mbox{\boldmath$K'$} = \mbox{\boldmath$I$}_{\rm light} +
\mbox{\boldmath$S$}_{\rm light} + \mbox{\boldmath$\ell$}
\ .
\label{def: k prime}
\end{equation}
The new correspondence between the bound--state picture variables and
those of the CQM is:
\begin{eqnarray}
\mbox{\boldmath$S$}_{\rm heavy}
&\leftrightarrow&
\mbox{\boldmath$S$}_H
\ ,
\nonumber\\
\mbox{\boldmath$r$}
&\leftrightarrow&
\mbox{\boldmath$L$}_E
\ ,
\nonumber\\
\mbox{\boldmath$\ell$}
&\leftrightarrow&
\mbox{\boldmath$L$}_I
\ ,
\nonumber\\
\mbox{\boldmath$I$}_{\rm light} +
\mbox{\boldmath$S$}_{\rm light} +
\mbox{\boldmath$J$}^{\rm sol}
&\leftrightarrow&
\mbox{\boldmath$S$}
\ .
\label{correspondence:2}
\end{eqnarray}
Previously $\mbox{\boldmath$I$}_{\rm light} + 
\mbox{\boldmath$S$}_{\rm light} = \mbox{\boldmath$K$}$ had zero
quantum numbers on the bound states; now the picture is a little more
complicated.
We will see that the dynamics may lead to new bound states which are in
correspondence with the CQM.
Equation~(\ref{correspondence:2}) should be interpreted in the sense
of this correspondence.

It is easiest to see that the lowest new states generated agree with
the CQM for $\ell=\mbox{even}$, which corresponds to negative parity
heavy mesons. In this case $k=0$ or equivalently $k'=\ell$ may be 
favored dynamically. Then the last line in Eq.~(\ref{correspondence:2}) 
indicates that $J^{\rm sol}$, which can take on the 
values $0$ and $1$, corresponds to the light diquark spin $S$ in the CQM.
This leads to the CQM states of type a) and b) in Eq.~(\ref{condition:CQM}).
This is just a generalization of the discussion for the ground state
given in section~\ref{sec:preliminaries}. Now let us discuss how 
the states corresponding to c) and d) can be constructed in the bound 
state scenario. Apparently we require $\ell=\mbox{odd}$, {\it i.e.} 
positive parity heavy mesons. For $I=0$ we also have $J^{\rm sol}=0$. 
Hence the last line in Eq.~(\ref{correspondence:2}) requires $k=1$ 
for $S=1$. To generate states of type d) also $k=1$ would be needed in 
order to accommodate $I=J^{\rm sol}=1$ and $S=0$. Actually for the 
case $k=1$ and $J^{\rm sol}=1$ states with $S=0,1,2$ would 
be possible. The states with $S=1,2$ should be ruled out by the 
dynamics of the model.

One may perhaps wonder whether we are pushing the bound state picture
too far; since things seem to be getting more complicated why not just
use the constituent quark model?
Apart from the intrinsic interest of the soliton approach there are
two more or less practical reasons for pursuing the approach.
The first is that the parameters of the underlying chiral Lagrangian
are, unlike parameters such as the constituent quark masses and
inter--quark potentials of the CQM, physical ones and in principle
subject to direct experimental test.
The second reason is that the bound state approach actually models the
expected behavior of a confining theory; namely, when sufficient
energy is applied to ``stretch'' the heavy baryon it does not come
apart into a heavy quark and two light quarks but rather into a
nucleon and a heavy meson.
The light quark--antiquark pair which one usually imagines popping out
of the vacuum when the color singlet state has been
suitably stretched, was there all the time, waiting to play a role, in
the bound state picture.
The model may therefore be useful in treating reactions of this sort.

\section{ Model for the Missing First 
Excited States}
\label{sec:4}

Before going on to the general orbital excited states it may be
helpful to 
see how the dynamics could work out for explaining the missing seven 
$\Lambda_{Q}$ 
and $\Sigma_{Q}$ type, negative parity, excited states. In the {\it new} 
bound state picture these correspond to the choices\footnote{%
Actually, $\ell$ was 
introduced for convenience in making a comparison with the
constituent quark  
model. It is really hidden in the heavy mesons which, strictly
speaking, are  
specified by the light cloud angular momentum 
$\mbox{\boldmath$J$}_{\rm light}$ 
and parity. We can perform the calculation without mentioning $\ell$.
} 
$\ell=1$, $r=0$. As discussed, we are considering that the orbital
angular momentum $\ell$ is ``locked--up''
in suitable excited heavy mesons. As in Eq.~(\ref{def:wave function}),
$r$ appears as a parameter in the new heavy meson wave--function. The 
treatment of the excited heavy mesons in the effective theory context,
has been given already by Falk and Luke~\cite{Falk-Luke}. For a 
review see \cite{Ca97}. The case (for orbital angular 
momentum=1) where the light cloud spin of the heavy meson is $1/2$ is 
described by the heavy multiplet 
\begin{equation}
{\cal{H}} = \frac{1 +  \gamma_\mu V^\mu}{2}
\left(S +  \gamma_{5}\gamma_\nu A^\nu  \right) \ ,
\label{def:curH}
\end{equation}
where $S$ is the fluctuation field for a scalar ($J^P=0^+$) 
particle and $A_{\mu}$, satisfying $V_{\mu}A^{\mu}=0$, similarly corresponds 
to an axial ($J^P=1^+$) particle. The case where the light cloud 
spin is $3/2$ is described by 
\begin{equation}
{\cal{H_{\mu}}}=
\frac{1 + \gamma_\alpha V^\alpha}{2}\left(T_{\mu\nu}\gamma^{\nu} -
i \sqrt{\frac{3}{2}}B^{\nu}\gamma_5\left[g_{\mu\nu} -
\frac{1}{3}\gamma_{\nu}  
\left(\gamma_{\mu}-V_{\mu}\right)\right]\right)
\label{def:curHm}
\end{equation}
satisfying the {\it Rarita-Schwinger} constraints 
${\cal H}_{\mu}\gamma^{\mu}={\cal H}_{\mu} V^{\mu}=0$. The field 
$T_{\mu\nu}=T_{\nu\mu}$ (with $V^{\mu}T_{\mu\nu}=T_{\mu}^{\mu}=0$) is a
spin $2$  
tensor ($J^P=2^+$) and $B_{\mu}$ (with $V_{\mu}B^{\mu}=0$) is 
another axial ($J^P=1^+$). Currently, experimental
candidates exist  
for the tensor and an axial.

In order to prevent the calculation from becoming too complicated we
will adopt the approximation of
leaving out the light vector mesons.
This is a common approximation used by workers in the field but it
should be kept in mind that the effect of the light vectors is
expected to be substantial.

The kinetic terms of the effective chiral Lagrangian (analogous
to the  first term of Eq.~(\ref{Lag for H})) are:
\begin{equation}
{\cal L}_{\rm kin}
= -i {\cal M}V^{\mu}\mbox{Tr}\, 
\left[ {\cal H} D_\mu \bar{{\cal H}} \right]
- i {\cal M} \, V^{\mu}\mbox{Tr}\, 
\left[ {\cal H}^{\alpha} D_\mu \bar{\cal H}_{\alpha}
\right] \ ,
\label{Lag for curlH}
\end{equation} 
where ${\cal M}$ is a characteristic heavy mass scale for the excited
mesons.  
For simplicity\footnote{%
A more general approach is to replace ${\cal M}$ on the 
 right hand side of Eq.~(\ref{Lag for curlH}) by the same ${M}$ used
in Eq.~(\ref{Lag for H}) and to add the splitting terms 
$-2M(M_S-M)\mbox{Tr}\,\left[ {\cal H}\bar{\cal H}\right] - 
2M(M_T-M)\mbox{Tr}\,\left[{\cal H}_{\mu}\bar{\cal H}^{\mu}\right]$.} 
we are neglecting mass differences between the $\ell=1$ heavy mesons.
The interaction terms involving only the ${\cal H}$ and 
${\cal H}_{\mu}$ fields,  to lowest order in derivatives, are 
\begin{eqnarray}
{\cal L}_{\rm int}/{\cal M}&=& -  d_{\rm S} \mbox{Tr}\,
\left[{\cal H}\gamma^{\mu}\gamma_5 p_{\mu}\bar{\cal H}\right] 
-  d_{\rm T} \mbox{Tr}\, \left[{\cal H}^{\mu}\gamma^{\alpha}
\gamma_5 p_{\alpha}\bar{\cal H}_{\mu}\right] \nonumber \\
 &+&\left[
 f_{\rm ST}\mbox{Tr}\,
\left[{\cal H}\gamma_{5} p^{\mu}\bar{\cal H}_{\mu}\right] + 
\mbox{h.c.}
\right] \ . 
\label{Int for curlH}
\end{eqnarray}
These generalize the second term in Eq.~(\ref{Lag for H}) and 
$d_{\rm S}$, $d_{\rm T}$  
 and $f_{\rm ST}$ (which may be complex) are the heavy meson--pion
coupling  
constants. Similar terms which involve $\ell\neq 1$ multiplets are not
needed  
for our present purpose but will be discussed in the next section.

As in  section \ref{sec:preliminaries},
the wave--functions for the excited heavy 
mesons bound to the background Skyrmion are conveniently presented in
the rest frame $\mbox{\boldmath$V$}=0$. 
The analogs of Eq.~(\ref{H: matrix}) become  
\begin{equation}
\bar{{\cal H}}_{\rm c} \rightarrow
\left( \begin{array}{cc}
\bar{f}_{lh}^a & 0 \\ 
0 & 0 
\end{array}\right) 
\ , \quad
\left(\bar{{\cal H}}_{i}\right)_{\rm c} \rightarrow
\left( \begin{array}{cc}
0 & 0 \\ \bar{f}_{i,lh}^a & 0 
\end{array}\right)
\ ,
\label{curlH: matrix}
\end{equation}
and $\left(\bar{{\cal H}}_0\right)_{\rm c}\rightarrow 0$. Now the 
wave--functions in Eq.~(\ref{curlH: matrix}) are expanded as:
\begin{eqnarray}
\bar{f}_{lh}^a &=& 
\frac{u\left( \vert\mbox{\boldmath$x$}\vert\right)}{\sqrt{{\cal M}}} 
\left(\hat{\mbox{\boldmath$x$}}\cdot
\mbox{\boldmath$\tau$}_{ad}\right) 
\Phi_{ld}\left(k^{\prime},k^{\prime}_3;r\right) 
\chi_{h} \ , \nonumber \\
\bar{f}_{i,lh}^a &=& 
\frac{u\left( \vert\mbox{\boldmath$x$}\vert\right)}{\sqrt{{\cal M}}} 
\left(\hat{\mbox{\boldmath$x$}}\cdot
\mbox{\boldmath$\tau$}_{ad}\right)\Phi_{i,ld}
\left(k^{\prime},k^{\prime}_3;r\right)\chi_{h} \ ,
\label{f: classical}
\end{eqnarray}
where $u$ stands for a sharply peaked radial wave--function which 
may differ for the two cases. Other notations are as in 
Eq.~(\ref{h: classical}). Note that the constraint 
$\gamma^{\mu}\bar{\cal H}_{\mu}=0$ implies that 
\begin{equation}
\left(\sigma_i\right)_{ll'}\Phi_{i,l'd}=0  \ .
\label{rarita}
\end{equation}
It is interesting to see explicitly how the extra angular 
momentum $\ell=1$ is ``locked--up'' in the heavy meson
wave--functions.  
For the ${\cal H}$ wave--function, the fact that 
$\mbox{\boldmath $J$}_{\rm light}=\mbox{\boldmath$\ell$}+
\mbox{\boldmath$S$}_{\rm light}$ 
takes the value $1/2$ leads, using Eq.~(\ref{def: k prime}), 
to the possible 
values $k^{\prime}=0$ or $1$. The corresponding wave--functions are 
\begin{equation}
\Phi_{ld}\left(k^{\prime}=k^{\prime}_3=0\right)=
\frac{\epsilon_{ld}}{\sqrt{8\pi}}\ , 
\quad
\Phi_{ld}\left(k^{\prime}=k_3=1\right)=
\frac{\delta_{l1}\delta_{d1}}{\sqrt{4\pi}}\ , 
\label{curlH wave--function}
\end{equation}
where, for the present case, we are taking $r=0$. 
For the $\bar{\cal H}_i$ 
wave--function it is important to satisfy 
$j_l=\vert\mbox{\boldmath$J$}_{\rm light}\vert$ = $3/2$ condition 
(\ref{rarita}). This may be accomplished by combining with suitable 
Clebsch-Gordan coefficients an $\ell=1$ wave--function with the 
$S_{\rm light}=1/2$ 
spinor to give 
\begin{eqnarray}
\Phi_{i,ld}\left(k^{\prime}=k^{\prime}_3=2\right)&=&w^{(+1)}_i \delta_{l1} 
\delta_{d1} \ , \nonumber \\
\Phi_{i,ld}\left(k^{\prime}=k^{\prime}_3=1\right)&=& 
\frac{\sqrt{3}}{2}w^{(+1)}_i \delta_{l1} \delta_{d2} - \frac{1}{2\sqrt{3}} 
w^{(+1)}_{i} \delta_{l2}\delta_{d1} - \frac{1}{\sqrt{6}} w^{(0)}_{i}
\delta_{l1}\delta_{d1} 
\ , 
\label{curlHmu wave--function}
\end{eqnarray}
where 
$w^{(\pm 1)}_{j}=\frac{\mp 1}{\sqrt{8 \pi}} 
\left(\delta_{j1}\pm i\delta_{j2}\right)$ and 
$w^{(0)}_i=\frac{\delta_{i3}}{\sqrt{4\pi}}$ is a spherical
decomposition.

The main question is: 
Which of the channels contain bound states?
Note that, for the reduced space in which
$\hat{\mbox{\boldmath$x$}}\cdot \mbox{\boldmath$\tau$}$ 
has been removed 
as in Eq.~(\ref{f: classical}), 
$k^{\prime}$ is a good quantum number. 
Furthermore, because the wave--function 
$u\left(\vert\mbox{\boldmath$x$}\vert\right)$ is sharply peaked,
the relevant matrix elements are actually 
independent of the orbital angular momentum $r$.
The classical potential for each $k^{\prime}$ channel may be
calculated  by setting $r=0$ and substituting the appropriate reduced
wave--functions from Eqs.~(\ref{curlH wave--function}) and 
(\ref{curlHmu wave--function}) into the  
interaction Lagrangian (\ref{Int for curlH}). 
(see Appendix~\ref{app:a} for more details.)
The $k^{\prime}=0$ channel 
gets a contribution only from the $d_{\rm S}$ term in 
Eq.~(\ref{Int for curlH}) 
while the $k^{\prime}=2$ channel receives a contribution only from 
the $d_{\rm T}$ term. On the other hand, all three terms contribute to
the  
$k^{\prime}=1$ channel. The resulting potentials are:
\begin{eqnarray}
V\left(k^{\prime}=0\right)&=& - \frac{3}{2}d_{\rm S} F^{\prime}(0) 
\label{kprime=0}\ , \\
V\left(k^{\prime}=2\right)&=& - \frac{1}{2}d_{\rm T} F^{\prime}(0) 
\label{kprime=2}\ , 
\end{eqnarray}
\begin{equation}
V\left(k^{\prime}=1\right)=
\left( \begin{array}{cc}
\langle{\cal H }|V|{\cal H}\rangle & 
 \langle{\cal H }|V|{\cal H}_{\mu}\rangle \\ 
\langle{\cal H}_{\mu} |V| {\cal H}\rangle  &  
 \langle{\cal H}_{\mu} |V| {\cal H}_{\mu}\rangle
\end{array}\right)
= 
\left( \begin{array}{cc}
\frac{1}{2}d_{\rm S} &  \sqrt{\frac{2}{3}}f_{\rm ST} \\ 
 \sqrt{\frac{2}{3}}f_{\rm ST}^*  & \frac{5}{6} d_{\rm T} 
\end{array}\right) F'(0)
\ .
\label{kprime=1}
\end{equation}
The classical criterion for a channel to contain a bound state is 
that its potential be negative. Since $F^{\prime}(0)>0$ we require for 
bound states in the $k^{\prime}=0$ and $k^{\prime}=2$ channels 
\begin{equation}
d_{\rm S} > 0 \ , \quad d_{\rm T}>0 \ ,
\label{dst constraints}
\end{equation}
respectively\footnote{
In a more general picture where $\ell=3$ excited heavy 
mesons are included, the $k^{\prime}=2$ channel will also be described
by a potential matrix. Then the criterion for $d_{\rm T}$ is modified.
(See next section.)
}. 
For bound 
states in the $k^{\prime}=1$ channel we must examine the signs of the 
eigenvalues of Eq.~(\ref{kprime=1}).  Assuming that 
Eq.~(\ref{dst constraints}) holds (as will be seen to be desirable) it
is easy to see that there is, at most, {\it one} 
$k^{\prime}=1$ bound state. The condition  
for this bound state to exist is 
\begin{equation}
\left\vert{f_{\rm ST}}\right\vert^2 > \frac{5}{8} \, d_{S}\, d_{T} \ .
\label{fst constraints}
\end{equation}
The (primed) states which diagonalize Eq.~(\ref{kprime=1}) are simply 
related to the original ones by 
\begin{equation}
\left( \begin{array}{c}
\Phi\\ 
 \Phi_{i}
\end{array}\right)=
\left( \begin{array}{cc}
\cos\theta&\sin\theta\\ 
- p^* \sin\theta&  p \cos\theta
\end{array}\right)
\left( \begin{array}{c}
\Phi^{\prime}\\ 
 \Phi_{i}^{\prime}
\end{array}\right) \ ,
\label{rotation}
\end{equation}
\begin{equation}
\tan2\theta=
\frac{4 \sqrt{6} \left\vert f_{\rm ST}\right\vert}{
5 d_{\rm T} - 3 d_{\rm S}} \ ,
\label{def: theta}
\end{equation}
where $p$ is the phase of $f_{\rm ST}$. $\Phi$ and 
$\Phi_{i}$ are shorthand 
notations\footnote{Strictly speaking, to put $\Phi_{ld}$ on a 
parallel footing to $\Phi_{i,ld}$ we should replace 
 $\Phi_{ld} \rightarrow 
 \sqrt{\frac{3}{8}}\left(P^{3/2}\right)_{ik;ll^{\prime}} 
 \left(\tau_{k}\right)_{dd^{\prime}}\Phi_{l^{\prime}d^{\prime}}$ with
the spin $3/2$ projection operator, 
 $\left(P^{3/2}\right)_{ik;ll^{\prime}}
 = \frac{2}{3}\left(\delta_{ik} \delta_{ll^{\prime}} - 
 \frac{i}{2}\epsilon_{jik}
 \left(\sigma_{j}\right)_{ll^{\prime}}\right)$
 (see Appendix~\ref{app:a}).
}
for the appropriate wave--functions.
Clearly, the results for which states are bound depend on the numerical 
values and signs of the coupling constants. At the moment there is 
no purely experimental information on these quantities. However, it is
 very interesting to observe that if Eqs.~(\ref{dst constraints}) and
(\ref{fst constraints}) hold, 
then the missing first excited $\Lambda_{Q}$ 
states are bound. To see this note that the heavy baryon spin is given 
by Eq.~(\ref{baryon spin}) with $\mbox{\boldmath$g$}$ defined 
in Eqs.~(\ref{def: g}) and 
(\ref{def: k prime}). For the $\Lambda_{Q}$--type states, noting 
that $I=J^{\rm sol}=0$ in the Skyrme approach gives the baryon spin as 
\begin{equation}
\mbox{\boldmath$J$}=\mbox{\boldmath$g$}+
\mbox{\boldmath$S$}_{\rm heavy}
\quad
\left(\Lambda_{Q}~{\rm states}\right) \ .
\label{J baryon}
\end{equation}
The $r=0$ choice enables us to set $g=k^{\prime}$. With just the 
three attractive channels $k^{\prime}=0$, 
$k^{\prime}=1$ and $k^{\prime}=2$ we 
thus end up with the missing first three excited $\Lambda_{Q}$ heavy 
multiplets $\Lambda_{Q}\left(\frac{1}{2}^{-}\right)$ ,
 $\left\{\Lambda_{Q}\left(\frac{1}{2}^{-}\right), 
\Lambda_{Q}\left(\frac{3}{2}^{-}\right)\right\}$ and 
 $\left\{\Lambda_{Q}\left(\frac{3}{2}^{-}\right), 
\Lambda_{Q}\left(\frac{5}{2}^{-}\right)\right\}$. 
It should be stressed that this counting involves dynamics rather than
pure  
kinematics. For example, it may be seen from
Eqs.~(\ref{kprime=0})--(\ref{kprime=1})  
that it is dynamically impossible to have four bound heavy multiplets 
($k^{\prime}=0,~ k^{\prime}=2$ and two $k^{\prime}=1$ channels).  
The missing first excited $\Sigma_{Q}$--type states comprise the
single  heavy multiplet 
 $\left\{\Sigma_{Q}\left(\frac{1}{2}^{-}\right), 
\Sigma_{Q}\left(\frac{3}{2}^{-}\right)\right\}$. At 
the classical level there are apparently more bound multiplets present. 
However, we will now see that the introduction of collective
coordinates,  
as is anyway required in the Skyrme model~\cite{Sk61}
to generate states with good 
isospin quantum number, will split the heavy multiplets from each other. 
Thus, deciding which states are bound actually requires 
a more detailed analysis.

We need to extend Eq.~(\ref{def: collective}) 
in order to allow the $\ell=1$ heavy meson 
fields to depend on the collective rotation variable $A(t)$:
\begin{equation}
\bar{{\cal H}}(\mbox{\boldmath$x$},t)=
A(t) \bar{{\cal H}}_{\rm c} (\mbox{\boldmath$x$}) 
\ ,
\quad
\bar{{\cal H}}_{i}(\mbox{\boldmath$x$},t)=A(t)
\bar{{\cal H}}_{i{\rm c}} (\mbox{\boldmath$x$}) 
\ ,
\label{def: curl collective}
\end{equation}
where $\bar{{\cal H}}_{\rm c}$ and $\bar{{\cal H}}_{i{\rm c}}$ 
are given in Eq.~(\ref{curlH: matrix}). 
Note, again, that the matrix $A(t)$ acts on 
the isospin indices. We also have $\bar{{\cal H}}_{0{\rm c}}=0$ 
due to the 
rest frame constraint $V^{\mu}\bar{{\cal H}}_{\mu {\rm c}}=0$. 
Now substituting Eq.~(\ref{def: curl collective}) as well as the first
of Eq.~(\ref{def: collective}) into the heavy field 
Lagrangian\footnote{%
Note that Eq.~(\ref{Lag for curlH}) contributes but 
Eq.~(\ref{Int for curlH}) does not contribute.} 
yields~\cite{Callan-Klebanov} the collective Lagrangian\footnote{%
In Eq.~(\ref{def: Lcoll}) $k^{\prime}$ is defined to 
operate on the heavy particle wave--functions rather than on their
conjugates.  
This is required when the heavy meson is coupled to the Skyrme
background field since $\Lambda_{Q}$ is made as 
$\left(qqq\right)\left(\bar{q}Q\right)$ 
rather than $\left(qqq\right)\left(\bar{Q}q\right)$. For convenience 
in Eqs.~(\ref{h: classical}) and (\ref{f: classical}) we have 
considered the conjugate wave--functions 
(since they are usual in the light sector). This has been compensated 
by the minus sign in the second term of Eq.~(\ref{def: Lcoll}).}
\begin{equation} 
L_{\rm coll}=\frac{1}{2}\,\alpha^2 \mbox{\boldmath$\Omega$}^2 - 
\chi\left(k^{\prime}\right) 
\mbox{\boldmath$K^{\prime}$}\cdot\mbox{\boldmath$\Omega$} \ ,
\label{def: Lcoll}
\end{equation}
where $\mbox{\boldmath$\Omega$}$ is defined in Eq.~(\ref{def: omega})
and  $\alpha^2$ is the Skyrme model moment of inertia. In the vector 
meson model the induced fields ($\rho^a_0$ and $\omega_i$) are 
determined from a variational approach to $\alpha^2$. The 
quantities $\chi\left(k^{\prime}\right)$ are given by 
(see Appendix~\ref{app:b}).
\begin{equation}
\chi\left(k^{\prime}\right)=
\left\{\begin{array}{ll}
0 & k^{\prime}=0 \\
\frac{1}{4}\left(3\,\cos^2\theta - 1\right) 
& k^{\prime}=1 \\
\frac{1}{4} & k^{\prime}=2 
\end{array}\right. \ ,
\label{def: chi}
\end{equation}
where the angle $\theta$ is defined in Eq.~(\ref{def: theta}). 
(Note that if light vector mesons are included the expressions
for $\chi$ would be more involved as the induced fields will also
contribute.)
In writing Eq.~(\ref{def: chi}) it was assumed that the first state in
Eq.~(\ref{rotation}) 
(i.e. $\Phi^{\prime}$ rather than $\Phi^{\prime}_i$) is the bound one;
the collective Lagrangian is constructed as an expansion around the
bound 
state solutions. We next determine from Eq.~(\ref{Jsol}), the
canonical (angular)  
momentum $\mbox{\boldmath$J$}^{\rm sol}$ as 
$\alpha^2\,\mbox{\boldmath$\Omega$} - \chi\left(k^{\prime}\right)
 \mbox{\boldmath$K$}^{\prime}$. The usual Legendre transform then
leads to the collective Hamiltonian
\begin{equation}
H_{\rm coll}=\frac{1}{2\alpha^2} \left(\mbox{\boldmath$J$}^{\rm sol}+
\chi\left(k^{\prime}\right)\,\mbox{\boldmath$K^{\prime}$}\right)^2 \ .
\label{Hcoll}
\end{equation}
Again we remark that 
$J^{\rm sol}=I$. 
It is useful to define the light part of the total heavy baryon spin
as  
\begin{equation}
\mbox{\boldmath$j$}= \mbox{\boldmath$r$} + 
\mbox{\boldmath$K$}^{\prime}+\mbox{\boldmath$J$}^{\rm sol} \ ,
\label{baryon jlight}
\end{equation}
and rewrite Eq.~(\ref{Hcoll}) as 
\begin{equation}
H_{\rm coll}=\frac{1}{2\alpha^2}\left[
\left(1 - \chi\left(k^{\prime}\right)\right)\mbox{\boldmath$I$}^2 + 
 \chi\left(k^{\prime}\right)\left(\mbox{\boldmath$j$} - 
\mbox{\boldmath$r$}\right)^2
+ \chi\left(k^{\prime}\right)\left(\chi\left(k^{\prime}\right) 
- 1 \right)
\mbox{\boldmath$K^{\prime}$}^2 \right] \ . 
\label{Hcoll 2}
\end{equation}
The mass splittings within each given 
$k^{\prime}$ multiplet due to $H_{\rm coll}$ 
are displayed in Table~\ref{tab:1}.
\begin{table}[htbp]
\begin{center}
\small
\begin{tabular}{c|c|c|c|c|c}
 $I$ & $k^{\prime}$ 
 & $\left\vert
      \mbox{\boldmath$K$}^\prime+\mbox{\boldmath$J$}^{\rm sol}
   \right\vert$
 & $V$ & $\alpha^2 H_{\rm coll}$ 
&{Candidates for $r=0$} \\
 $=J^{\rm sol}$ &~ &~ &~ &~ &  {missing states}~~~~~ \\
\hline
 & 0 & 0 & $-\frac{3}{2} d_{\rm S}\, F^{\prime}(0)$ & 0 
& $\Lambda_{Q}\left(\frac{1}{2}^{-}\right)$\\
~{\Large 0}~ & 1 & 1  & $\lambda$ & $\chi^2$ 
&  $\left\{\Lambda_{Q}\left(\frac{1}{2}^{-}\right), 
\Lambda_{Q}\left(\frac{3}{2}^{-}\right)\right\}$\\
 & 2 & 2  & $-\frac{1}{2}d_{\rm T}\, F^{\prime}(0) $ & $\frac{3}{16}$ 
&  $\left\{\Lambda_{Q}\left(\frac{3}{2}^{-}\right), 
\Lambda_{Q}\left(\frac{5}{2}^{-}\right)\right\}$\\
\hline
 & 0  &  1  & $-\frac{3}{2}d_{\rm S}\, F^{\prime}(0) $ & 1  
&  $\left\{\Sigma_{Q}\left(\frac{1}{2}^{-}\right), 
\Sigma^{\prime}_{Q}\left(\frac{3}{2}^{-}\right)\right\}_{1}$\\
 & 1  &  0  & $\lambda $ & $\left(\chi - 1\right)^2$ & \\
 & 1  &  1  & $''$& $\left(\chi - 1\right)^2 + \chi$
&  $\left\{\Sigma_{Q}\left(\frac{1}{2}^{-}\right), 
\Sigma_{Q}\left(\frac{3}{2}^{-}\right)\right\}_{2}$\\
~{\Large 1}~ & 1  &  2  & $''$& $\left(\chi - 1\right)^2 + 3\,\chi$ & \\
 & 2  &  1  & $-\frac{1}{2}d_{\rm T}\,F^{\prime}(0)$ & $\frac{7}{16}$
&  $\left\{\Sigma_{Q}\left(\frac{1}{2}^{-}\right), 
\Sigma_{Q}\left(\frac{3}{2}^{-}\right)\right\}_{3}$\\
 & 2  &  2  & $''$ & $\frac{15}{16}$ &  \\
 & 2  &  3  & $''$ & $\frac{27}{16}$ &  \\
\end{tabular}
\end{center}
\caption{
Contributions to energies of new predicted $\ell=1$ states.
Here, 
$\lambda=\frac{1}{4}F^{\prime}(0) 
\left[\left(d_{\rm S} + \frac{5}{3} d_{\rm T} \right) - 
\sqrt{\left(d_{\rm S} - \frac{5}{3} d_{\rm T} \right)^2 +
\frac{32}{3} |f_{\rm ST}|^2} \right]$ is the presumed negative binding 
potential 
in the $k^{\prime}=1$ channel. Furthermore $\chi=\chi(1)$ in
Eq.~(\ref{def: chi});  
it satisfies $-\frac{1}{4}\leq \chi \leq \frac{1}{2}$.
} 
\label{tab:1}
\end{table}
\noindent
This table also shows the splitting of the $k^{\prime}$ multiplets from 
each other due to the classical potential in 
Eqs.~(\ref{kprime=0})--(\ref{kprime=1}). Note that the slope of the 
Skyrme profile function $F^{\prime}(0)$ is of order $1$ GeV. The 
coupling constants $d_{\rm S}, d_{\rm T}, f_{\rm ST}$, 
based on $d\simeq 0.5$ for the 
ground state heavy meson, are expected to be of the order unity.
Hence the binding potentials $V$ are expected to 
be of the rough order of $500$ MeV. The inverse moment of inertia 
 $1/\alpha^2$ is of the order of $200$ MeV which 
(together with $-\frac{1}{4}\leq \chi \leq \frac{1}{2}$) sets the scale 
for the ``$1/N_{c}$'' corrections due to $H_{\rm coll}$. 
As mentioned before, 
if the coupling constants satisfy the inequalities 
(\ref{dst constraints}) and (\ref{fst constraints}), all the 
$\Lambda_{Q}$ multiplets shown will be bound. At first  
glance we might expect all the $\Sigma_{Q}$ states listed 
also to be bound. However the $H_{\rm coll}$ corrections increase as
$I$ 
increases, which is a possible indication that many of the 
$\Sigma_{Q}$'s might be only weakly bound. In a more complete 
model they may become unbound. Hence it is interesting to ask which 
of the three displayed candidates for the single missing 
$\Sigma_{Q}$ multiplet is mostly tightly bound in the present 
model. Neglecting the effect of $V$ we can see that $H_{\rm coll}$ 
raises the  energy of candidate 3 less than those of candidates 1  
and 2. Furthermore, for the large range of $\chi$, 
$-\frac{1}{4}\leq \chi \leq 1 - \frac{\sqrt{7}}{4}$, candidate 3 suffers 
the least unbinding due to $H_{\rm coll}$ of any of the $I=1$ heavy
baryons  listed. 
The $\Lambda_{Q}$ states suffer still less unbinding
due  
to $H_{\rm coll}$.

\section{Extension to the Higher Orbital Excitations}
\label{sec:5}

We have already explicitly seen that the ``missing'' first orbitally
excited heavy baryon states in the bound state picture might be
generated if the model is extended to also include binding the first 
orbitally excited heavy mesons in the background field of a Skyrme 
soliton. From the correspondence (\ref{correspondence:2}) and 
associated discussion we expect that any of the higher excited heavy 
baryons of the CQM might be similarly generated by binding the 
appropriately excited heavy mesons. In this section we will show 
in detail how this result can be achieved in the general case.
An extra complication, which was neglected for simplicity in the last 
section, is the possibility of baryon states constructed by binding
heavy mesons of different $\ell$, mixing with each other.
For example $\{r=1\,,\,\ell=0\}$ 
type states can mix with $\{r=1\,,\,\ell=2\}$ 
type states, other quantum numbers being the same.
Since $\mbox{\boldmath$r$}+\mbox{\boldmath$\ell$}$ must add to $1$, 
this channel could not mix with $\{r=1\,,\,\ell=4\}$.
An identical type of mixing -- between $\{L_E=1\,,\,L_I=0\}$ and
$\{L_E=1\,,\,L_I=2\}$ -- may also exist in the CQM. The present model, 
however, provides a simple way to study this kind of mixing as a 
perturbation.

\begin{table}[htbp]
\begin{center}
\begin{tabular}{cccc}
field & $\ell$ & $j_l$ & $J^P$ \\
\hline
$H$ & 0 & $1/2$ & $0^-$, $1^-$ \\
& & & \\
${\cal H}$ & 1 & $1/2$ & $0^+$, $1^+$ \\
${\cal H}_\mu$ & 1 & $3/2$ & $1^+$, $2^+$ \\
& & & \\
$H_\mu$ & 2 & $3/2$ & $1^-$, $2^-$ \\
$H_{\mu\nu}$ & 2 & $5/2$ & $2^-$, $3^-$ \\
& & & \\
\vdots & & & \\
& & & \\
$H_{\mu_1\cdots\mu_{\ell-1}}$ & $\ell=\mbox{even}$
 & $\ell-1/2$ & $(\ell-1)^-$, $\ell^-$ \\
$H_{\mu_1\cdots\mu_\ell}$ & $\ell=\mbox{even}$
 & $\ell+1/2$ & $\ell^-$, $(\ell+1)^-$ \\
& & & \\
${\cal H}_{\mu_1\cdots\mu_{\ell-1}}$ & $\ell=\mbox{odd}$
 & $\ell-1/2$ & $(\ell-1)^+$, $\ell^+$ \\
${\cal H}_{\mu_1\cdots\mu_\ell}$ & $\ell=\mbox{odd}$
 & $\ell+1/2$ & $\ell^+$, $(\ell+1)^+$ \\
& & & \\
\vdots & & & \\
\end{tabular}
\end{center}
\caption{%
Notation for the heavy meson multiplets.
$j_l$ is the angular momentum of the ``light cloud'' surrounding the
heavy quark while $J^P$ is the spin parity of each heavy meson in the
multiplet.
}
\label{table:3}
\end{table}
To start the analysis it may be helpful to refer to
Table~\ref{table:3}, which shows our notations for the excited heavy
meson multiplet  ``fluctuation'' fields. 
The straight $H$'s contain negative parity mesons and 
the curly ${\cal H}$'s contain positive parity mesons. 
Further details are given 
in Ref.~\cite{Falk-Luke}. Note that each field is symmetric in 
all Lorentz indices and obeys the constraints
\begin{equation}
V^{\mu_1} H_{\mu_1\cdots\mu_n} = 
H_{\mu_1\cdots\mu_n} \gamma^{\mu_1} = 0 \ ,
\end{equation}
as well as for ${\cal H}_{\mu_1\cdots\mu_n}$.
The general chiral invariant interaction with the lowest number of 
derivatives is
\begin{equation}
{\cal L}_{\rm d} + {\cal L}_{\rm f} + {\cal L}_{\rm g} \ ,
\end{equation}
where
\begin{eqnarray}
{\cal L}_{\rm d} &=& - M \sum_{n=0} d_{{\rm P}n} 
\mbox{\rm Tr}\, 
\left[
  H^{\mu_1\cdots\mu_n} p^\mu \gamma_\mu \gamma_5 
  \bar{H}_{\mu_1\cdots\mu_n}
\right]
\nonumber\\
&& {}
- M \sum_{n=0} d_{{\rm S}n} 
\mbox{\rm Tr}\, 
\left[
  {\cal H}^{\mu_1\cdots\mu_n} p^\mu \gamma_\mu \gamma_5 
  \bar{\cal H}_{\mu_1\cdots\mu_n}
\right]
\ ,
\nonumber\\
{\cal L}_{\rm f} &=&  M \sum_{n=0} f_{{\rm P}n} 
\mbox{\rm Tr}\, 
\left[
  H^{\mu_1\cdots\mu_n} p^\mu \gamma_5 \bar{H}_{\mu_1\cdots\mu_n\mu}
\right]
+ \mbox{h.c.}
\nonumber\\
&& {}
+ M \sum_{n=0} f_{{\rm S}n} \mbox{\rm Tr}\, 
\left[
  {\cal H}^{\mu_1\cdots\mu_n} p^\mu \gamma_5 
  \bar{\cal H}_{\mu_1\cdots\mu_n\mu}
\right]
+ \mbox{h.c.}
\ .
\label{general d f term}
\end{eqnarray}
The final piece,
\begin{equation}
{\cal L}_{\rm g} =  M \sum_{n=0} g_{n} 
\mbox{\rm Tr}\, 
\left[
  {\cal H}^{\mu_1\cdots\mu_n} p^\mu \gamma_\mu \gamma_5 
  \bar{H}_{\mu_1\cdots\mu_n}
\right]
+ \mbox{h.c.}
\label{general g term}
\end{equation}
exists in general, but does not contribute for our {\it ansatz}.
Terms of the form
\begin{equation}
\mbox{\rm Tr}\, 
\left[
  H^{\mu_1\cdots\mu_n\mu} p_\mu \gamma_5 
  \bar{\cal H}_{\mu_1\cdots\mu_n} \right]
\ , \qquad
\mbox{\rm Tr}\, 
\left[
  {\cal H}^{\mu_1\cdots\mu_n\mu} p_\mu \gamma_5 
  \bar{H}_{\mu_1\cdots\mu_n}
\right]
\end{equation}
can be shown to vanish by the heavy spin symmetry.
In the notation of Eq.~(\ref{Int for curlH}),
$d_{\rm S}=d_{{\rm S}0}$, $d_{\rm T}=d_{{\rm S}1}$ and 
$f_{\rm ST}=f_{{\rm S}0}$.
A new type of coupling present in Eq.~(\ref{general d f term})
also connects multiplets to others differing by $\Delta \ell=\pm2$.
These are the terms with odd (even) $n$ for $H$ (${\cal H}$)--type
fields. The interactions in Eq.~(\ref{general g term}) 
connecting multiplets
differing by $\Delta\ell=\pm1$ turn out not to contribute in our 
model. In the interest of simplicity we will consider all heavy 
mesons to have the same mass. This is clearly an approximation which 
may be improved in the future.

The rest frame {\it ans\"atze} for the bound state wave functions which
generalize Eq.~(\ref{curlH: matrix}) are (note $j_l=n+1/2$):
\begin{equation}
\left(\bar{H}_{i_1\cdots i_n}\right)_{\rm c} \rightarrow
\left\{
\begin{array}{l}
\displaystyle
\bar{h}^a_{i_1\cdots i_n,lh} \otimes
\left(
\begin{array}{cc}
0 & 0 \\ 1 & 0
\end{array}
\right)
\ , \qquad j_l=\ell+\frac{1}{2} \ ,
\\
\displaystyle
\bar{h}^a_{i_1\cdots i_n,lh} \otimes
\left(
\begin{array}{cc}
1 & 0 \\ 0 & 0
\end{array}
\right)
\ , \qquad j_l=\ell-\frac{1}{2}  \ ,
\end{array}
\right.
\label{general ansatz}
\end{equation}
with identical structures for $\bar{H} \rightarrow \bar{\cal H}$.
Note that again $a$, $l$, $h$ represent respectively the isospin, 
light spin and heavy spin bivalent indices.
Extracting a factor of 
$\hat{\mbox{\boldmath$x$}}\cdot\mbox{\boldmath$\tau$}$ as we did
before in Eqs.~(\ref{h: classical}) and (\ref{f: classical}) leads to
\begin{equation}
\bar{h}^a_{i_1\cdots i_n,lh} = 
\frac{u\left(\left\vert\mbox{\boldmath$x$}\right\vert\right)}%
{\sqrt{M}}
\left(\hat{\mbox{\boldmath$x$}}\cdot\mbox{\boldmath$\tau$}\right)_{ad}
\psi_{i_1\cdots i_n,dl}\left(k^\prime,k_3^\prime,r\right)\,\chi_h
\end{equation}
with similar notations.
The relevant wave--functions are the 
$\psi_{i_1\cdots i_n,dl}\left(k^\prime,k_3^\prime,r\right)$.
$k^\prime$ was defined in Eq.~(\ref{def: k prime});
we will see that it remains a good quantum number.
Since the terms which connect the positive parity ($H$ type) and
negative parity (${\cal H}$ type) heavy mesons 
(Eq.~(\ref{general g term})) vanish when the {\it ans\"atze}
(\ref{general ansatz}) are substituted, the baryon states associated 
with each type do not mix with each other in our model.
We thus list separately the potentials for each type.
For the $\ell=\mbox{even}$ baryons (associated with $H$ mesons),
\begin{eqnarray}
V\left[k'=0\right] &=&
- \frac{3}{2} d_{{\rm P}0} \, F'(0) \ ,
\nonumber\\
V\left[k'\neq0\right] &=& F'(0) \,
\left[ \begin{array}{cc}
\displaystyle
- \left(-1\right)^{k'} \frac{d_{{\rm P}(k'-1)}}{2} &
\displaystyle
\sqrt{\frac{2}{3}} f_{{\rm P}(k'-1)} \\
\displaystyle
 \sqrt{\frac{2}{3}} f^{\ast}_{{\rm P}(k'-1)} &
\displaystyle
- \left(-1\right)^{k'} \frac{2k'+3}{2k'+1} \frac{d_{{\rm P}k'}}{2} 
\end{array} \right]
\ ,
\label{pot: gen p}
\end{eqnarray}
while for the $\ell=\mbox{odd}$ baryons (associated with ${\cal H}$
mesons), 
\begin{eqnarray}
V\left[k'=0\right] &=&
- \frac{3}{2} d_{{\rm S}0} \, F'(0) \ ,
\nonumber\\
V\left[k'\neq0\right] &=& F'(0) \,
\left[ \begin{array}{cc}
\displaystyle
- \left(-1\right)^{k'} \frac{d_{{\rm S}(k'-1)}}{2} &
\displaystyle
\sqrt{\frac{2}{3}} f_{{\rm S}(k'-1)} \\
\displaystyle
 \sqrt{\frac{2}{3}} f^{\ast}_{{\rm S}(k'-1)} &
\displaystyle
- \left(-1\right)^{k'} \frac{2k'+3}{2k'+1} \frac{d_{{\rm S}k'}}{2} 
\end{array} \right]
\ .
\label{pot: gen s}
\end{eqnarray}
Details of the derivations of Eqs.~(\ref{pot: gen p}) and 
(\ref{pot: gen s}) are given in Appendix~\ref{app:a}.
The ordering of matrix elements in Eqs.~(\ref{pot: gen p}) and
(\ref{pot: gen s}), for a given $k'$, is such that the first heavy
meson has a light spin, $j_l = k' - \frac{1}{2}$ while the 
second has $j_l=k'+\frac{1}{2}$.  The $H$ type (${\cal H}$ type) 
channels with $k'=\mbox{even}$ (odd) involve two mesons with the 
same $\ell=k'$.  The $H$ type (${\cal H}$ type) channels with 
$k'=\mbox{odd}$ (even) involve two mesons differing by $\Delta
\ell=2$, {\it i.e.}, $\ell=k'-1$ and $\ell=k'+1$. This pattern is, for
convenience, illustrated in Table~\ref{table:4}.
\begin{table}[htbp]
\begin{center}
\begin{tabular}{cc|cc|cc}
& & \multicolumn{2}{c|}{$H$ mesons} & 
 \multicolumn{2}{c}{${\cal H}$ mesons} \\
$k'$ & $j_l$ & $\ell$ & \# & $\ell$ & \# \\
\hline\hline
$0$ & $1/2$ & $0$ & $1$ & $1$ & $1$ \\
\hline
$1$ 
 & \multicolumn{1}{c|}{$\begin{array}{c} $1/2$ \\ $3/2$ \end{array}$}
 & \multicolumn{1}{c}{$\begin{array}{c} $0$ \\ $2$ \end{array}$}
 & $0$ 
 & \multicolumn{1}{c}{$\begin{array}{c} $1$ \\ $1$ \end{array}$}
 & 1 \\
\hline
$2$ 
 & \multicolumn{1}{c|}{$\begin{array}{c} $3/2$ \\ $5/2$ \end{array}$}
 & \multicolumn{1}{c}{$\begin{array}{c} $2$ \\ $2$ \end{array}$}
 & $1$ 
 & \multicolumn{1}{c}{$\begin{array}{c} $1$ \\ $3$ \end{array}$}
 & 2 \\
\hline
$3$ 
 & \multicolumn{1}{c|}{$\begin{array}{c} $5/2$ \\ $7/2$ \end{array}$}
 & \multicolumn{1}{c}{$\begin{array}{c} $2$ \\ $4$ \end{array}$}
 & $0$ 
 & \multicolumn{1}{c}{$\begin{array}{c} $3$ \\ $3$ \end{array}$}
 & 1 \\
\end{tabular}
\end{center}
\caption{
Pattern of states for Eqs.~(\ref{pot: gen p}) and (\ref{pot: gen s}).
Note that $j_l=n+\frac{1}{2}$ is the light cloud spin of the heavy
meson.
The columns marked \# stand for the number of channels which are
expected to be bound, for that particular $k'$, according to the CQM.
}
\label{table:4}
\end{table}
Also shown, for each $k'$, are the number of channels which are
expected to be bound according to the CQM.

It is important to note that Table~\ref{table:4} holds for any value
of the angular momentum $r$, which is a good quantum number in our
model.  
For the reader's orientation, we now locate the previously considered
cases in Table~\ref{table:4}.  
The standard ``ground state'' heavy baryons discussed in 
section~\ref{sec:preliminaries} are made from the $H$ meson with
$\ell=0$ and $j_l=1/2$.
They have $r=0$ and $k'=0$.
The seven negative parity heavy baryons discussed in 
section~\ref{sec:preliminaries} also are made from the $H$ meson with
$\ell=0$ and $j_l=1/2$.
They still have $k'=0$, but now $r=1$.
The seven ``missing'' first excited heavy baryons discussed in 
section~\ref{sec:4} have $r=0$ and are made from the $\ell=1$, 
${\cal H}$ and ${\cal H}_\mu$ mesons with $j_l=1/2$ and 
$j_l=3/2$.
There should appear one bound state for $k'=0$, one bound state for
$k'=1$ and one bound state for $k'=2$ in the ``${\cal H}$--meson''
section of Table~\ref{table:4}.
Note that the number of states expected in the CQM model for $k'=2$ 
is listed in Table~\ref{table:4} as two, rather than one. In the 
absence of $\Delta \ell=2$ terms connecting ${\cal H}_\mu$ and
${\cal H}_{\mu\nu}$ (see the last term in 
Eq.~(\ref{general d f term})) $\ell$ would be conserved for our model 
and only the $\ell=1$ state would be relevant. This was the
approximation we made, for simplicity, in section~\ref{sec:4}. The
other entry would have $\ell=3$ and would decouple.  
When the $\Delta\ell=2$ mixing terms are turned on, the $\ell=1$ and
$\ell=3$, $k'=2$ 
channels will mix. One diagonal linear combination should be counted
against the $L_I=1$ CQM states and one against the $L_I=3$ CQM states.

To summarize: for the $H$--type mesons, the even $k'$ channels should
each have one bound state, while the odd $k'$ channels should have
none.  The situation is very different for the ${\cal H}$--type 
mesons; then the even $k'\neq0$ channels should contain two bound
states while the odd $k'$ channels should contain one bound state. The
$k'=0$ channel should have one bound state.

For the $H$--type meson case, the pattern of bound states mentioned
above would be achieved dynamically if the coupling constants 
satisfied:
\begin{eqnarray}
&& d_{{\rm P}0} > 0 \ ,
\nonumber\\
&& 
\left(-1\right)^{k'}
\left[
  d_{{\rm P}(k'-1)} d_{{\rm P}k'} \left( \frac{2k'+3}{2k'+1} \right)
  - \frac{8}{3} \left\vert f_{{\rm P}(k'-1)} \right\vert^2
\right]
< 0 \ , \quad (k'>0)
\nonumber\\
&&
d_{{\rm P}(k'-1)} + \left(\frac{2k'+3}{2k'+1} \right)
d_{{\rm P}k'} > 0
\ , \quad (k'=\mbox{odd}) \ .
\label{constraint: p}
\end{eqnarray}
These follow from requiring only one negative eigenvalue of 
Eq.~(\ref{pot: gen p}) for $k'=\mbox{even}$ and none for
$k'=\mbox{odd}$. Similarly requiring for the ${\cal H}$--type 
meson case in Eq.~(\ref{pot: gen s}), a negative eigenvalue for 
$k'=0$, one negative eigenvalue for $k'=\mbox{odd}$ and two 
negative eigenvalues for $k'>0$ and even leads to the criteria,
\begin{eqnarray}
&& d_{{\rm S}0} > 0 \ ,
\nonumber\\
&& 
\left(-1\right)^{k'}
\left[
  d_{{\rm S}(k'-1)} d_{{\rm S}k'} \left( \frac{2k'+3}{2k'+1} \right)
  - \frac{8}{3} \left\vert f_{{\rm S}(k'-1)} \right\vert^2
\right]
> 0 \ , \quad (k'>0)
\nonumber\\
&&
d_{{\rm S}(k'-1)} + \left(\frac{2k'+3}{2k'+1} \right)
d_{{\rm S}k'} > 0
\ , \quad (k'=\mbox{even}\neq0) \ .
\label{constraint: s}
\end{eqnarray}
{}From Eqs.~(\ref{constraint: p}) and (\ref{constraint: s}) it can be
seen that all the $d$'s are required to be positive.
Furthermore these equations imply that the 
$\left\vert f \right\vert$'s which connect heavy mesons with 
$\Delta \ell=2$ are relatively small (compared to the $d$'s) while the
$\left\vert f \right\vert$'s which connect heavy mesons with 
$\Delta \ell=0$ are relatively large. In detail this means that
$\left\vert f_{{\rm P}(k'-1)} \right\vert$ should be small for odd
$k'$ and large for even $k'$ with just the reverse for 
$\left\vert f_{{\rm S}(k'-1)} \right\vert$.
This result seems physically reasonable.

As in the example in the preceding section we should introduce the
collective variable $A(t)$ in order to define states of good isospin
and angular momentum. This again yields some splitting of the different 
$\left\vert \mbox{\boldmath$K$}^\prime + 
\mbox{\boldmath$J$}^{\rm sol} \right\vert$
members of each $k'$ bound state. Now, each $k'$ channel (except for 
$k'=0$) is described by a $2\times2$ matrix. Thus there will be an 
appropriate mixing angle $\theta$, analogous to the one introduced in 
Eq.~(\ref{rotation}), for each $k'$ and parity choice ({\it i.e.}, 
$H$--type or ${\cal H}$--type field). The collective Lagrangian is 
still given by Eq.~(\ref{def: Lcoll}) but, in the general case,
\begin{equation}
\chi_{\pm}(k') = \frac{1}{2k'(k'+1)}
\left[ 
  \frac{1}{2} \pm \left( k' + \frac{1}{2} \right) \cos 2\theta
\right]
\ .
\end{equation}
In this formula the different signs correspond to the two possible
eigenvalues,
\begin{equation}
\lambda_{\pm} = 
\left[
  \frac{\left(-1\right)^{k'-1}}{4}
  \left( d_{(k'-1)} + \frac{2k'+3}{2k'+1} d_{k'} \right)
  \pm \frac{1}{4}
  \sqrt{ \left( d_{(k'-1)} - \frac{2k'+3}{2k'+1} d_{k'} \right)^2
    + \frac{32}{3} \left\vert f_{(k'-1)} \right\vert^2 
  }
\right]
F'(0)
\end{equation}
of the potential matrix.
For example, referring to Table~\ref{table:4}, we would expect the 
$k'=2$,
${\cal H}$--type meson case to provide two distinct bound states and
hence both $\chi_+(2,{\cal H})$ and $\chi_-(2,{\cal H})$ would be
non-zero. On the other hand, we would expect no bound states in 
the $k'=3$, $H$--type meson case so $\chi_{\pm}(3,H)$ should be 
interpreted as zero.

It is convenient to summarize the energies of the predicted states in
tabular form, generalizing the example presented in Table~\ref{tab:1}.
The situation for baryons with $\mbox{parity}=- (-1)^r$ 
(${\cal H}$--type mesons) is presented in Table~\ref{table:5}.
For definiteness we have made the assumption that the constraints
(\ref{constraint: s}) above are satisfied.
\begin{table}[htbp]
\begin{center}
\small
\begin{tabular}{c|c|c|c|c|c}
$I$ & $k'$ 
 & $\left\vert
      \mbox{\boldmath$K$}^\prime+\mbox{\boldmath$J$}^{\rm sol}
   \right\vert$
 & $V$ & $\alpha^2 \times H_{\rm coll}$
 & Candidates for $r=0$ 
\\
$=J^{\rm sol}$ & & & & & missing states ~~~~~~
\\
\hline
 & $2n-1$ & $2n-1$ & $\lambda_+$ & $n(2n-1)\chi_-^2$
 & $\left\{
     \Lambda\left((2n-3/2)^-\right)\,,\,
     \Lambda\left((2n-1/2)^-\right)
   \right\}$
\\
\cline{2-6}
{\large$0$} & $2n$ & $2n$ & $\lambda_+$ & $n(2n+1)\chi_+^2$
 & $\left\{
     \Lambda\left((2n-1/2)^-\right)\,,\,
     \Lambda\left((2n+1/2)^-\right)
   \right\}$
\\
 & & & $\lambda_-$ & $n(2n+1)\chi_-^2$ & $''$ 
\\
\hline
 & $2n-1$ & $2n-2$ & & $n(2n-1)\chi_+^2+1-2n\chi_+$ & 
\\
 & & $2n-1$ & $\lambda_+$ 
 & $n(2n-1)\chi_+^2+1-\chi_+$ 
 & $\left\{
     \Sigma\left((2n-3/2)^-\right)\,,\,
     \Sigma\left((2n-1/2)^-\right)
   \right\}_1$
\\
 & & $2n$ & & $n(2n-1)\chi_+^2+1+(2n-1)\chi_+$ & 
\\
\cline{2-6}
 & $2n$ & $2n-1$ & & $n(2n+1)\chi_+^2+1-(2n+1)\chi_+$
 & $\left\{
     \Sigma\left((2n-3/2)^-\right)\,,\,
     \Sigma\left((2n-1/2)^-\right)
   \right\}_2$
\\
 {\large$1$} & & $2n$ & $\lambda_+$ & $n(2n+1)\chi_+^2+1-\chi_+$ & 
\\
 & & $2n+1$ & & $n(2n+1)\chi_+^2+1+2n\chi_+$
 & $\left\{
     \Sigma\left((2n+1/2)^-\right)\,,\,
     \Sigma\left((2n+3/2)^-\right)
   \right\}_3$
\\
\cline{3-6}
 & & $2n-1$ & & $n(2n+1)\chi_-^2+1-(2n+1)\chi_-$
 & $\left\{
     \Sigma\left((2n-3/2)^-\right)\,,\,
     \Sigma\left((2n-1/2)^-\right)
   \right\}_4$
\\
 & & $2n$ & $\lambda_-$ & $n(2n+1)\chi_-^2+1-\chi_-$ & 
\\
 & & $2n+1$ & & $n(2n+1)\chi_-^2+1+2n\chi_-$
 & $\left\{
     \Sigma\left((2n+1/2)^-\right)\,,\,
     \Sigma\left((2n+3/2)^-\right)
   \right\}_5$
\\
\end{tabular}
\end{center}
\caption{
Contributions to energies of the new predicted states made from 
${\cal H}$--type heavy mesons.
Note that $n$ is a positive integer.
The $n=0$ case is given in Table~\ref{tab:1}.
The $\lambda_+$ entries in the $V$ column are more tightly bound than
the $\lambda_-$ entries.
$\left\vert \mbox{\boldmath$K$}^\prime+\mbox{\boldmath$J$}^{\rm sol}
\right\vert$ is the light part of the heavy baryon angular momentum
for $r=0$ (See Eq.~(\ref{baryon jlight}).).
}
\label{table:5}
\end{table}
In order to explain Table~\ref{table:5} let us ask which states
correspond to the ($L_I=3$, $L_E=0$) states in the CQM.
Reference to Table~\ref{table:1} shows that three negative parity
$\Lambda$--type heavy multiplets and one negative parity
$\Sigma$--type heavy multiplet should be present.
The correspondence in Eq.~(\ref{correspondence:2}) instructs us to set
$r=0$ and, noting Eq.~(\ref{def: k prime}) , to identify
\begin{equation}
\mbox{\boldmath$K$}^\prime + \mbox{\boldmath$J$}^{\rm sol}
\leftrightarrow 
\mbox{\boldmath$L$}_I + \mbox{\boldmath$S$}
\ .
\end{equation}
The $\Lambda$--type particles are of type c) in 
Eq.~(\ref{condition:CQM}) so we must take $S=1$.
Hence, since $J^{\rm sol}=0$ for $\Lambda$--type particles, we learn
that $k'$ can take on the values $2$, $3$ and $4$.
For $k'=2$, the second line of the $k'$ column yields two possible
multiplets (energies $\lambda_+$ and $\lambda_-$) with $n=1$ and
structure $\left\{ \Lambda\left(\frac{3}{2}^-\right)\,,\,
\Lambda\left(\frac{5}{2}^-\right)  \right\}$.
We should choose one of these to be associated with ($L_I=3$, $L_E=0$)
and the other with ($L_I=1$, $L_E=0$) in the CQM.
We remind the reader that $\ell$ is not a good quantum number so that 
the correspondence $\mbox{\boldmath$\ell$} \leftrightarrow
\mbox{\boldmath$L$}_I$ in Eq.~(\ref{correspondence:2}) only holds when
the $\Delta \ell=2$ mixing terms are neglected.
For $k'=3$, the first line of the $k'$ column correctly yields one
multiplet with $n=2$ and structure 
$\left\{ \Lambda\left(\frac{5}{2}^-\right)\,,\,
\Lambda\left(\frac{7}{2}^-\right)  \right\}$.
For $k'=4$, the second line of the $k'$ column yields two multiplets
with $n=2$ and structure
$\left\{ \Lambda\left(\frac{7}{2}^-\right)\,,\,
\Lambda\left(\frac{9}{2}^-\right)  \right\}$.
One of these is to be associated with ($L_I=3$, $L_E=0$) and the other
with ($L_I=5$, $L_E=0$) in the CQM.
Now let us go on to the $\Sigma$--type heavy multiplets.
These are of type d) in Eq.~(\ref{condition:CQM}) and yield $S=0$.
Hence $\mbox{\boldmath$K$}^\prime + \mbox{\boldmath$J$}^{\rm sol}
\leftrightarrow \mbox{\boldmath$L$}_I$ and 
$\left\vert\mbox{\boldmath$K$}^\prime + 
\mbox{\boldmath$J$}^{\rm sol} \right\vert=3$.
Five candidates for this 
$\left\{ \Sigma\left(\frac{5}{2}^-\right)\,,\,
\Sigma\left(\frac{7}{2}^-\right)  \right\}$
multiplet are shown in the last column of Table~\ref{table:5}.
These consecutively correspond to the choices $n=2$, $2$, $1$, $2$,
$1$ in the $\left\vert\mbox{\boldmath$K$}^\prime + 
\mbox{\boldmath$J$}^{\rm sol} \right\vert$ column.
As before it is necessary for an exact correspondence with the CQM
that one of these should be dynamically favored (much more tightly
bound) over the others.
Again, note that the choice
$\left\vert\mbox{\boldmath$K$}^\prime + 
\mbox{\boldmath$J$}^{\rm sol} \right\vert=3$ does not uniquely
constrain the value of $\ell$.

Next, the situation for baryons with
$\mbox{parity}=\left(-1\right)^{r}$ ($H$--type baryons) is presented
in Table~\ref{table:6}..
\begin{table}[htbp]
\begin{center}
\small
\begin{tabular}{c|c|c|c|c|c}
$I$ & $k'$ 
 & $\left\vert
      \mbox{\boldmath$K$}^\prime+\mbox{\boldmath$J$}^{\rm sol}
   \right\vert$
 & $V$ & $\alpha^2 \times H_{\rm coll}$
 & Candidates for $r=0$ 
\\
$=J^{\rm sol}$ & & & & & missing states ~~~~~~
\\
\hline
$0$ & $2n$ & $2n$ & $\lambda_+$ & $n(2n+1)\chi_+^2$
 & $\left\{
     \Lambda\left((2n-1/2)^+\right)\,,\,
     \Lambda\left((2n+1/2)^+\right)
   \right\}$
\\
\hline
 & & $2n-1$ & & $n(2n-1)\chi_+^2+1-(2n+1)\chi_+$ 
 & $\left\{
     \Sigma\left((2n-3/2)^+\right)\,,\,
     \Sigma\left((2n-1/2)^+\right)
   \right\}_1$
\\
 1 & $2n$ & $2n$ & $\lambda_+$ & $n(2n+1)\chi_+^2+1-\chi_+$
 & $\left\{
     \Sigma\left((2n-1/2)^+\right)\,,\,
     \Sigma\left((2n+1/2)^+\right)
   \right\}_2$
\\
 & & $2n+1$ & & $n(2n+1)\chi_+^2+1+2n\chi_+$
 & $\left\{
     \Sigma\left((2n+1/2)^+\right)\,,\,
     \Sigma\left((2n+3/2)^+\right)
   \right\}_3$
\\
\end{tabular}
\end{center}
\caption{
Contributions to energies of the new predicted states made from 
$H$--type heavy mesons.
Other details as for Table~\ref{table:5}.
}
\label{table:6}
\end{table}
For definiteness we have made the assumption that the constraints
(\ref{constraint: p}) above are satisfied.
This eliminates the odd $k'$ states and agrees with the CQM counting.
For example, we ask which states correspond to the ($L_I=2$, $L_E=0$)
states in the CQM.
Reference to Table~\ref{table:1} shows that one positive parity
$\Lambda$--type heavy multiplet and three positive parity
$\Sigma$--type heavy multiplets should be present.
For $r=0$ we have the correspondence $\mbox{\boldmath$K$}^\prime +
\mbox{\boldmath$J$}^{\rm sol} \leftrightarrow \mbox{\boldmath$L$}_I +
\mbox{\boldmath$S$}$.
The $\Lambda$--type particles are of type a) in
Eq.~(\ref{condition:CQM}) so we must set $k'=2$.
The first line in Table~\ref{table:6} then yields, with $n=1$ the
desired $\left\{\Lambda\left(\frac{3}{2}^+\right) \,,\,
\Lambda\left(\frac{5}{2}^+\right)\right\}$ heavy multiplet.
The $\Sigma$ particles are of type b) in Eq.~(\ref{condition:CQM}) so
that $\left\vert \mbox{\boldmath$K$}^\prime + 
\mbox{\boldmath$J$}^{\rm sol} \right\vert$ can take on the values $1$,
$2$ and $3$.
The last three lines in Table~\ref{table:6}, with $n=1$, give the
desired multiplets:
$\left\{\Sigma\left(\frac{1}{2}^+\right) \,,\,
\Sigma\left(\frac{3}{2}^+\right)\right\}$,
$\left\{\Sigma\left(\frac{3}{2}^+\right) \,,\,
\Sigma\left(\frac{5}{2}^+\right)\right\}$ and 
$\left\{\Sigma\left(\frac{5}{2}^+\right) \,,\,
\Sigma\left(\frac{7}{2}^+\right)\right\}$.
In this case all the states should be bound so that the splittings 
due to $H_{\rm coll}$ are desired to be relatively small.
The present structure is simpler than the one shown in
Table~\ref{table:5} for the ${\cal H}$--type cases.

\section{Conclusions for the Generalized Heavy Baryon Model}
\label{sec:6}

In this chapter \cite{HSSW:planets} we have pointed out 
the problem of getting, in the
framework of a bound state picture, the excited states which are
expected on geometrical grounds from the constituent quark model.
We treated the heavy baryons and made use of the Isgur--Wise heavy
spin symmetry.
The approach may also provide some insight into the understanding of
light excited baryons.
The key problem to be solved is the introduction of an additional
``source'' of angular momentum in the model.
It was noted that this might be achieved in a simple way by
postulating that excited heavy mesons, which have ``locked--in" 
angular momentum, are bound in the background Skyrmion field.
The model was seen to naturally have the correct kinematical structure
in order to provide the excited states which were missing in earlier
models.

An important aspect of this work is the investigation of 
which states in the model are actually bound.
This is a complicated issue since there are many interaction terms
present with {\it a priori} unknown coupling constants.
Hence, for the purpose of our initial investigation we included only
terms with the minimal interactions of the light pseudoscalar mesons.
The large $M$ limit was also assumed and nucleon recoil as well as
mass splittings among the heavy excited meson multiplets were
neglected. We expect, based on previous work, that the most important 
improvement of the present calculation would be to include the 
interactions of the light vector mesons. It is natural to expect 
that possible interactions of the light higher spin mesons 
also play a role.  In the calculation of the ground state heavy baryons 
the light vectors were actually slightly more important than the light 
pseudoscalars and reinforced the binding due to the latter.
Another complicating factor is the presence, expected from
phenomenology, of radially excited mesons along with orbitally excited
ones.

It is interesting to estimate which of the first excited states,
discussed in section~\ref{sec:4}, are bound.
The criteria for actually obtaining the missing states in the model
with only light pseudoscalars present are given in 
Eqs.~(\ref{dst constraints}) and (\ref{fst constraints}).
Based on the use of chiral symmetry for relating the coupling constants
to axial matrix elements and using a quark model argument to estimate
the axial matrix elements, Falk and Luke~\cite{Falk-Luke}
presented the estimates (their Eqs.~(2.23) and (2.24))
$d_{\rm T} = 3 d_{\rm S} = d$ and 
$\left\vert f_{\rm ST} \right\vert = \frac{2}{\sqrt{3}} d$.
With these estimates 
Eqs.~(\ref{dst constraints}) and (\ref{fst constraints}) are
satisfied.
Note that $d>0$ provides binding for the ground state heavy baryons.
However we have checked this and find that, although we are
in agreement for $\left\vert f_{\rm ST} \right\vert$ we obtain instead 
$d_{\rm T} = 3 d_{\rm S} = - d$.
Assuming that this is the case then it is easy to see that the 
only bound multiplet will have $k'=1$.
This leads to the desired $\Sigma$--type multiplet and one of the
three desired $\Lambda$--type multiplets being bound, but not the
$k'=0$ and $2$, $\Lambda$--type multiplets.
Clearly, it is important to make a more detailed calculation of the
light meson--excited heavy meson coupling constants.
We also plan to investigate the effects of including light vector
mesons in the present model.
It is hoped that the study of these questions will lead to a better
understanding of the dynamics of the excited heavy particles.

Finally we would like to add a few remarks on studies 
of the excited ``light'' hyperons within 
the bound state approach to the SU(3) Skyrme model. 
In that model the heavy spin symmetry is not maintained since 
the vector counterpart of the kaon, the $K^\ast$, is omitted;
while the kaons themselves couple to the pions as prescribed by 
chiral symmetry. On the other hand the higher 
orbital angular momentum channels ({\it i.e.} $r\ge2$) have 
been extensively studied. The first study 
was performed by the SLAC group \cite{Kr86}.
However, they were mostly interested in the amplitudes for 
kaon--nucleon scattering and for simplicity omitted 
flavor symmetry breaking terms in the effective Lagrangian.
Hence they did not find any bound states, except for 
zero modes. These symmetry breaking terms were, however, 
included in the scattering analysis of all higher orbital angular 
momentum channels by Scoccola \cite{Sc90}. The only bound states 
he observed were those for P-- and S--waves. After collective 
quantization these are associated with the ordinary hyperons 
and the $\Lambda(1405)$. As a matter of fact these states 
were already found in the original study by Callan and Klebanov 
\cite{Callan-Klebanov}. 
It is clear that the orbital excitations found in the bound 
state approach to the Skyrme model should be identified as the 
$\ell=0$ states. Furthermore when the dynamical coupling of 
the collective coordinates ($A,\mbox{\boldmath $\Omega$}$) 
is included in the scattering analysis \cite{Sch92} the only 
resonances which are observed obey the selection rule 
 $|J-1/2|\le r \le |J+1/2|$, where $r$ denotes the kaon 
orbital angular momentum. This rule is consistent 
with $\ell=0$ in our model. In order to find states with 
$\ell\ne0$ in this model one would also have to include
pion fluctuations besides the kaon fluctuations for the 
projectile--state. As indicated in the previous sections,
 these fluctuating 
fields should be coupled to carry the good quantum number $\ell$.
The full calculation would not only require this complicated 
coupling but also an expansion of the Lagrangian up to 
fourth order in the meson fluctuations off the background 
soliton. Such a calculation seems impractical, 
indicating that something like our present approximation, 
which treats these 
coupled states as elementary particles, is needed.

%% file: appendixtot.tex
\chapter{Part I Appendix}
\section{Scattering kinematics}
The general partial wave scattering matrix for the multichannel case 
can be written as:

\begin{equation}
S_{ab}=\delta_{ab}+2iT_{ab}\ .
\label{scatt}
\end{equation}
For simplicity, the diagonal isospin and angular momentum labels have
not been indicated. 

By requiring the unitarity condition $S^{\dagger}S=1$ one deduces for
the two--channel case the
following relations:
\begin{eqnarray}
Im ({T}_{11}) &=&|{T}_{11}|^2 + |{T}_{21}|^2 \nonumber \ ,\\
Im ({T}_{22}) &=&|{T}_{22}|^2 + |{T}_{12}|^2 \ , \\
Im ({T}_{12}) &=&{T}_{11}^*~{T}_{12} + 
{T}_{12}^*{T}_{22}\ , \nonumber 
\label{constr}
\end{eqnarray}
where ${T}_{12}={T}_{21}$. In the present case we will
identify 1 as the $\pi\pi$ channel and 2 as the $K\overline{K}$ channel.
In order to get the relations between the relative phase shifts and the
amplitude we need to consider the following parameterization of the
scattering amplitude:
\begin{equation}
S= \pmatrix
    {&\eta~e^{2i\delta_{\pi}}&\pm i\sqrt{1-\eta^2}~e^{i\delta_{\pi K}} \cr
     &\pm i\sqrt{1-\eta^2}~ e^{i\delta_{\pi K}} & \eta ~e^{2i\delta_K}
}\ ,
\label{param}
\end{equation} 
where $\delta_{\pi K}=\delta_\pi +\delta_K$ and $0< \eta < 1$ is the elasticity parameter.  By comparing 
eq.~(\ref{param}) and eq.~(\ref{scatt}) one can easily deduce:
\begin{equation}
\eta^2=1 - 4|{T}_{12}|^2\ .
\label{eta2} 
\end{equation}
Analogously, for ${T}_{aa}$ we have:   
\begin{equation}
{T^I_{aa;l}}(s)=\frac{(\eta^I_l (s)~e^{2i\delta^I_{a;l}(s)}-1)}{2i}\ ,
\end{equation}
where $l$ and $I$ label the angular momentum and isospin respectively.
Extracting the real
and imaginary parts via
\begin{eqnarray}
R^I_{aa;l}&=&\frac{\eta^I_l ~\sin(2\delta^I_{a;l})}{2}\ , \nonumber\\
I^I_{aa;l}&=&\frac{1-\eta^I_l ~\cos(2\delta^I_{a;l})}{2}
\label{real-imaginary}
\end{eqnarray}
\noindent
leads to the very important bounds
\begin{equation}
\big|R^I_{aa;l}\big{|}\leq\frac{1}{2}\ ,~~~~~~~0\leq I^I_{aa;l}\leq 1\ .
\label{eq:bound}
\end{equation}
Unitarity also requires $|T^I_{12;l}|< 1/2$\ .

Now we relate these partial wave amplitudes to the invariant amplitudes.
The invariant
amplitude for $ \pi_i(p_1) + \pi_j(p_2) \rightarrow  \pi_k(p_3) + \pi_l(p_4)
$ is 
decomposed as:
\begin{equation}
 \delta_{ij}\delta_{kl} A(s,t,u) + \delta_{ik}\delta_{jl} A(t,s,u)
+ \delta_{il}\delta_{jk} A(u,t,s)\ ,
\label{eq:def}
\end{equation}
where $s$, $t$ and $u$ are the usual Mandelstam variables.
Note that the phase of eq.~(\ref{eq:def}) corresponds to
simply  taking
the matrix element of the Lagrangian density of a four--point contact
interaction.
   Projecting out amplitudes of definite isospin yields:
\begin{eqnarray}
T_{11}^0(s,t,u) &=& 3A(s,t,u)+A(t,s,u)+A(u,t,s)\ ,\nonumber\\
T_{11}^1(s,t,u) &=& A(t,s,u)-A(u,t,s)\ ,\nonumber\\
T_{11}^2(s,t,u) &=& A(t,s,u)+A(u,t,s)\ .
\label{eq:isospin}
\end{eqnarray}

\noindent
The needed $I=0$ $\pi\pi \rightarrow K\overline{K}$ amplitude can be 
obtained as:

\begin{equation}
T_{12}^0(s,t,u) = -\sqrt{6}A(\pi^0(p_1)\pi^0(p_2),K^+(p_3)K^-(p_4))\ .
\end{equation}
\noindent
We then
define
the partial wave isospin amplitudes  according to the following formula:
\begin{equation}
T_{ab;l}^{I}(s)\equiv \frac{1}{2} \sqrt{\rho_{a}\rho_{b}}~\int^{1}_{-1}d\cos
\theta
P_l(\cos \theta) T_{ab}^I(s,t,u)\ ,
\label{eq:wave}
\end{equation} 
where $\theta$ is the scattering angle and 
\begin{equation} 
\rho_a=\frac{1}{S~16\pi}\sqrt{\frac{s-4 m_{\pi}^2}{s}}~\theta(s-4 m_a^2)\ .
\label{kfact}
\end{equation}
$S$ is a symmetry factor which is 2 for identical particles
($\pi\pi$ case) and $1$ for distinguishable particles ($K\overline{K}$
case). 
\section{Unregularized amplitudes}
\subsection{Amplitudes for the $\pi\pi \rightarrow \pi \pi$ channel}
The current algebra contribution to $A(s,t,u)$ is
\begin{equation}
A_{ca}(s,t,u)=2\frac{(s-m_{\pi}^2)}{F_{\pi}^2}\ .
\label{current}
\end{equation}
The amplitude for the vectors can be expressed in the following form
\begin{equation}
A_{\rho}(s,t,u)=-\frac{g^2_{\rho\pi\pi}}{2m^2_{\rho}}\left[
\frac{t(u-s)}{m^2_{\rho}-t}+\frac{u(t-s)}{m^2_{\rho}-u}\right]\ ,
\label{vector}
\end{equation}
where $g_{\rho\pi\pi}$ is the coupling of the vector to two pions.

For the scalar particle we deduce 
\begin{equation}
A_{f_0}(s,t,u)=\frac{\gamma_0^2}{2}\frac{\left(s-2m_{\pi}^2\right)^2}
{m_{f_0}^2-s}\ .
\label{scalar}
\end{equation}

To calculate the tensor exchange diagram we need the spin 2 propagator 
\cite{tensor}
\begin{equation}
\frac{i}{q^2 - m^2_{f_2}}\left[
\frac{1}{2}\left(
\theta_{\mu_1\nu_1} \theta_{\mu_2\nu_2}+
\theta_{\mu_1\nu_2}\theta_{\mu_2\nu_1}\right)-
\frac{1}{3}\theta_{\mu_1\mu_2}\theta_{\nu_1\nu_2}\right]\ ,
\label{eq:tensorpropag}
\end{equation}
\noindent
where
\begin{equation}
\theta_{\mu\nu}=-g_{\mu\nu}+\frac{q_\mu q_\nu}{m^2_{f_2}}\ .
\end{equation}
\noindent
A straightforward computation then yields the $f_2$ contribution to the
$\pi\pi$ scattering amplitude:
\begin{eqnarray}
A_{f_2}(s,t,u)&=&\frac{\gamma^2_2}{2(m^2_{f_2}-s)}
\left(
-\frac{16}{3}m_{\pi}^4
+\frac{10}{3}m_{\pi}^2 s
-\frac{1}{3}s^2
+\frac{1}{2}(t^2+u^2)\right.\nonumber\\
&~&\left.-\frac{2}{3}\frac{m_{\pi}^2s^2}{m^2_{f_2}}
-\frac{s^3}{6m^2_{f_2}}
+\frac{s^4}{6m^4_{f_2}}
\right)\ .
\label{eq:tensorampl}
\end{eqnarray}

\subsection{Amplitudes for $\pi^0\pi^0 \rightarrow K^+ K^-$ }

Current algebra amplitude:
\begin{eqnarray}
A_{ca}(\pi^0\pi^0,K^+K^-)&=&\frac{s}{2F_{\pi}^2}\ .
\label{12_current}
\end{eqnarray}
\noindent
Vector meson contribution:
\begin{eqnarray}
A_{Vector}(\pi^0\pi^0,K^+K^-)&=&\frac{g^2_{K^* K \pi}}{8 m^2_{K^*}}
\left[\frac{t(s-u)}{m^2_{K^*}-t} +\frac{u(s-t)}{m^2_{K^*}-u}
\right. \nonumber \\
&+&\left.
(m_k^2-m_{\pi}^2)^2\left(\frac{1}{m^2_{K^*}-t}+\frac{1}{m^2_{K^*}-u}\right)
\right]\ .
\end{eqnarray}
\noindent
Direct--channel contribution for the scalar:
\begin{eqnarray}
A_{f_0}(\pi^0\pi^0,K^+K^-)&=&\frac{1}{4}{\gamma_{f_0\pi\pi}\gamma_{f_0K\overline{K}}}
\frac{(s-2m_{\pi}^2)(s-2m_k^2)}{m_{f_0}^2-s}\ .
\end{eqnarray}
\noindent
Cross--channel contribution for the scalar:
\begin{eqnarray}
A_{K^*_0}(\pi^0\pi^0,K^+K^-)&=&\frac{\gamma_{K^*_0 K\pi}^2}{8}
\left[\frac{(m_K^2+m_{\pi}^2-t)^2}{m^2_{K^*_0}-t}+
\frac{(m_K^2+m_{\pi}^2-u)^2}{m^2_{K^*_0}-u}\right]\ .
\end{eqnarray}
Direct channel tensor contribution:
\begin{eqnarray}
A_{f_2}(\pi^0\pi^0,K^+K^-)&=&\frac{\gamma_{2\pi\pi}\gamma_{2K\overline{K}}}
{2(m_{f_2}^2 -s) }\left[ \left( \frac{s^2}{4m_{f_2}^2}+\frac{t}{2}-
\frac{(m_{\pi}^2+m_{K}^2)}{2}\right)^2\right.
\nonumber\\
&+&\left.
\left(\frac{s^2}{4m_{f_2}^2}+\frac{u}{2}-\frac{(m_{\pi}^2+m_{K}^2)}{2}
\right)^2\right.
\nonumber \\
&-&\left.
\frac{2}{3}\left(\frac{s^2}{4m_{f_2}^2}-\frac{s}{2}+m_{\pi}^2
\right)\left(\frac{s^2}{4m_{f_2}^2}-\frac{s}{2}+m_{K}^2
\right)\right]\ .
\end{eqnarray}
\noindent
Cross--channel tensor contribution:
\begin{eqnarray}
A_{K^*_2}(\pi^0\pi^0,K^+K^-)&=&
\frac{\gamma_{2K\pi}^2}
{16(m_{K^*_2}^2 -t) }
\left\{\left[(2m_{\pi}^2-s)-\frac{1}{2m_{K^*_2}^2}(m_{\pi}^2-m_{K}^2+t)^2
\right]  \right. \nonumber \\
&\times& \left. 
\left[(2m_{K}^2-s)-\frac{1}{2m_{K^*_2}^2}(m_{K}^2-m_{\pi}^2+t)^2
\right] \right. \nonumber \\
&+&\left. \left[(u-m_{\pi}^2-m_{K}^2) + \frac{1}{2m_{K^*_2}^2}(t^2-
(m_{K}^2 - m_{\pi}^2)^2) \right]^2 \right. \nonumber \\
&-&\left. \frac{2}{3}
\left[(t-m_{\pi}^2-m_{K}^2) - \frac{1}{2m_{K^*_2}^2}(t^2-
(m_{K}^2 - m_{\pi}^2)^2) \right]^2 \right\} \nonumber \\
&+&( t \longleftrightarrow u )\ .
\end{eqnarray}

\chapter{Part II Appendix}

\section{Classical Potential}
\label{app:a}
Here we will show how to compute the relevant matrix 
elements associated with the classical potential.

For any fixed value of  $k^{\prime}\ne 0$ the heavy meson 
light cloud spin (${\mbox{\boldmath$J$}}_{\rm light}$) 
takes the values $j_{l}=k^{\prime}\mp \frac{1}{2}$ 
since 
${\mbox{\boldmath$K$}}^{\prime}= {\mbox{\boldmath$J$}}_{\rm light} + 
{\mbox{\boldmath$I$}}_{\rm light}$
, where 
$\mbox{\boldmath$I$}_{\rm light}$ is the heavy meson isospin. Hence
the classical potential will be, in general, a $2\times 2$ matrix
schematically  
represented as 
\begin{equation}
V\left(k^{\prime}\ne 0\right)=
\left( \begin{array}{cc}
\langle 
 H_{\mu_1\cdots\mu_{k^\prime-1}}|V|H_{\mu_1\cdots\mu_{k^\prime-1}}
\rangle 
& 
\langle 
 H_{\mu_1\cdots\mu_{k^\prime-1}} |V|H_{\mu_1\cdots\mu_{k^\prime}}
\rangle
\\ 
\langle 
 H_{\mu_1\cdots\mu_{k^\prime~~}} |V| H_{\mu_1\cdots\mu_{k^\prime-1}} 
\rangle  
&  
\langle
 H_ {\mu_1\cdots\mu_{k^{\prime~~}}}|V| {H}_{\mu_1\cdots\mu_{k^\prime}}
\rangle
\end{array}\right)
 \ .
\label{kprime matrix}
\end{equation}
Here $|H_{\mu_1\cdots\mu_{k^{\prime}-1}}\rangle$ corresponds to the  
$j_{l}=k^{\prime}-\frac{1}{2}$  state 
while 
$|H_{\mu_1\cdots\mu_{k^{\prime}}}\rangle$ corresponds to 
 $j_{l}=k^{\prime}+\frac{1}{2}$.
 In order to compute the potential 
there is no need to distinguish even parity heavy mesons ${\cal H}$ 
from odd parity ones $H$. 
The diagonal matrix elements are 
obtained by substituting the appropriate rest frame {\it ansatz}
(\ref{general ansatz}) into the general potential term as:
\begin{eqnarray}
&+&  M\, d_{n}\,  \int d^3x \, 
\mbox{Tr}\, \left[ H^{\mu_1\cdots\mu_{n}}\gamma^{\alpha}
\gamma_5 p_{\alpha}\bar{H}_{\mu_1\cdots\mu_{n}}\right] \nonumber \\ 
&=& d_n\, \frac{F^{\prime}(0)}{2}\,(-1)^n  \int d\Omega \, 
\psi^{\ast}_{i_{1}\cdots i_{n},dl}
\left(k^{\prime},k^{\prime}_3,r\right) 
{\mbox{\boldmath$\sigma$}}_{ll^{\prime}}\cdot
{\mbox{\boldmath$\tau$}}_{dd^{\prime}}
\psi_{i_{1}\cdots i_{n},d^{\prime}l^{\prime}}
\left(k^{\prime},k^{\prime}_3,r\right) \ ,
\label{diagonal element}
\end{eqnarray}
where $j_{l}=n + \frac{1}{2}$ and $n=k^{\prime}\mp 1$ for the two 
diagonal matrix elements. 
The operator which mesures the total 
light cloud  spin $j_{l}$  is 
\begin{eqnarray}
 \left(J^a_{\rm light}\right)_{i_1 j_1,\cdots,i_n j_n;ll^{\prime}} 
&=&\frac{\sigma^a_{ll^{\prime}}}{2}
\otimes\delta_{i_1j_1}\otimes\cdots\otimes\delta_{i_nj_n}+
\delta_{ll^{\prime}}\otimes\left(-i\epsilon_{ai_1j_1}\right)\otimes
\delta_{i_2j_2}\otimes\cdots\otimes\delta_{i_nj_n}  \nonumber \\
&+&
\cdots +\delta_{ll^{\prime}}\otimes\delta_{i_1j_1}\otimes\cdots\otimes
\delta_{i_{n-1}j_{n-1}}\otimes\left(-i\epsilon_{ai_n j_n}\right) \ .
\label{jlight}
\end{eqnarray}
where $\epsilon_{aij}$ is the totally antisymmetric 
tensor. 
The isospin operator is
\begin{equation}
\mbox{\boldmath$I$}_{\rm light} = 
\frac{\mbox{\boldmath$\tau$}}{2} \ .
\end{equation}
We can write Eq.~(\ref{jlight}) compactly in the following way 
\begin{equation}
{\mbox{\boldmath$J$}}_{\rm light}
=\mbox{\boldmath$s$}+\hat{\mbox{\boldmath$l$}} \ ,
\end{equation}
where $\mbox{\boldmath$s$}\equiv\frac{\mbox{\boldmath$\sigma$}}{2}$.
Due to the 
total symmetrization of the vectorial indices we have $\hat{l}=n$.
We want to stress that $\mbox{\boldmath$s$}$ and 
$\hat{\mbox{\boldmath$l$}}$ do not necessarily agree with 
$\mbox{\boldmath$S$}_{\rm light}$ and $\mbox{\boldmath$\ell$}$.
Indeed for $\Phi_{ld}$ associated with ${\cal H}$ in 
Eq.~(\ref{f: classical}), $\hat{l}=0$ and 
$\mbox{\boldmath$J$}_{\rm light}=\mbox{\boldmath$s$}=
\mbox{\boldmath$S$}_{light} + \mbox{\boldmath$\ell$}$ while
for associated $\Phi_{i,ld}$ with ${\cal H}_\mu$, $\hat{l}=1$.
Now we have,
for fixed $n=j_l-\frac{1}{2}$, the following 
useful result:
\begin{equation}
\int d\Omega \, \psi^{\ast}\mbox{\boldmath$s$}\psi =  
\frac{\int d\Omega \psi^{\ast}\,
  \left(\mbox{\boldmath$s$}\cdot\mbox{\boldmath$J$}_{\rm light}\right)
\psi}{j_l (j_l+1)} \int d\Omega \, 
\psi^{\ast}{\mbox{\boldmath$J$}}_{\rm light}\psi 
=
\frac{1}{2\, j_{l}} \int d\Omega \, 
\psi^{\ast}{\mbox{\boldmath$J$}}_{\rm light}\psi \ .
\label{wigner}
\end{equation}
By using Eq.~(\ref{wigner}) we can 
write Eq.~(\ref{diagonal element}) as
\begin{eqnarray}
&~&(-1)^n\,  d_n \frac{F^{\prime}(0)}{j_{l}} 
\int d\Omega \, 
\psi^{\ast}\left(k^{\prime},k^{\prime}_3,r\right)
{\mbox{\boldmath$J$}}_{\rm light}\cdot 
{\mbox{\boldmath$I$}}_{\rm light} \psi 
\left(k^{\prime},k^{\prime}_3,r\right)\nonumber \\  
&=&
(-1)^n\, d_n \frac{F^{\prime}(0)}{2 j_{l}} 
\left[k^{\prime}(k^{\prime}+1) - j_{l}(j_{l}+1) - \frac{3}{4} \right] 
\ .
\label{second step}
\end{eqnarray}
For $j_l=k^{\prime}\mp \frac{1}{2}$ we get the diagonal matrix
elements for both, the $H$ type as well as the ${\cal H}$ type fields
\begin{equation}
(-1)^{k^\prime-1}\frac{F^\prime(0)}{2}\cdot
\left\{\begin{array}{ll}
\displaystyle
d_{k^\prime-1} \ , & j_l=k^\prime-\frac{1}{2} \ ,\\
\displaystyle
d_{k^\prime}\, \left(\frac{2\,k^{\prime}+3}{2\, k^\prime+1}\right) \ ,
& j_l=k^{\prime}+\frac{1}{2} \ ,
\end{array}\right.
\label{def: final diag elements}
\end{equation}
where we used $n=j_l-1/2$.

{}For the non--diagonal matrix elements we consider the contribution 
to the potential due to the following 
$f$ type term:
\begin{eqnarray}
&& -  M\, f_{n}\, \int d^3x \, 
\mbox{Tr}\, \left[ H^{\mu_1\cdots\mu_{n}}p^{\mu} 
\gamma_5 \bar{H}_{\mu_1\cdots\mu_{n}\mu}\right] \nonumber \\ 
&=& i f_n\, \frac{F^{\prime}(0)}{2}\,  \int d\Omega \, 
\psi^{\ast}_{i_{1}\cdots i_{n},dl}
\left(k^{\prime},k^{\prime}_3,r\right) 
\tau^i_{dd^{\prime}}
\psi_{i_{1}\cdots i_{n}i,d^{\prime}l}
\left(k^{\prime},k^{\prime}_3,r\right) \ .
\label{non diagonal element}
\end{eqnarray}
This corresponds to the transition between $j_l=n+\frac{1}{2}$ and 
$j_l=n+\frac{3}{2}$ states.
Now we notice that by construction any wave function $\psi$ must 
satisfy the condition
\begin{equation}
\left(P^{3/2}\right)_{ii_{1};ll^{\prime}}
\psi_{i_1i_2\cdots i_n,dl^{\prime}}=
\psi_{ii_2\cdots i_n,dl} \ ,
\label{rarita2}
\end{equation}
where $P^{3/2}$ is the spin $3/2$ projection operator 
\begin{equation}
\left(P^{3/2}\right)_{ik;ll^{\prime}}=\frac{2}{3} 
\left(\delta_{ik}\delta_{ll^{\prime}}- \frac{i}{2}\epsilon_{jik} 
\sigma^j_{ll^{\prime}}\right) \ .
\end{equation}
The condition (\ref{rarita2}) yields the following identity
\begin{eqnarray}
 &\int& d\Omega \, 
\psi^{\ast}_{i_{1}\cdots i_{n},dl}
\left(k^{\prime},k^{\prime}_3,r\right) 
\tau^i_{dd^{\prime}}
\psi_{i_{1}\cdots i_{n}i,d^{\prime}l}
\left(k^{\prime},k^{\prime}_3,r\right)= \nonumber \\
 &\int& d\Omega \, 
\psi^{\ast}_{i_{1}\cdots i_{n},dl}
\left(k^{\prime},k^{\prime}_3,r\right) 
\tau^j_{dd^{\prime}}\left(P^{3/2}\right)_{jk;ll^{\prime}}
\psi_{i_{1}\cdots i_{n}k,d^{\prime}l^{\prime}}
\left(k^{\prime},k^{\prime}_3,r\right) \ . 
\end{eqnarray}
Using the fact that $P^{3/2}\mbox{\boldmath$\tau$}$ commutes
with $\mbox{\boldmath$K$}^\prime$, we get
\begin{equation}
\left(P^{3/2}\right)_{jk;ll^{\prime}}\tau^{k}_{dd^{\prime}}
\psi_{i_{1}\cdots i_{n},d^{\prime}l^{\prime}}
\left(k^{\prime},k^{\prime}_3,r\right) 
= N \psi_{i_{1}\cdots i_{n}j,dl}
\left(k^{\prime},k^{\prime}_3,r\right) \ , 
\end{equation}
where $N$ is a normalization constant.
It is evaluated as
\begin{equation}
|N|^2=\int d\,\Omega\, 
\psi^{\ast}_{i_{1}\cdots i_{n},dl}
\left(k^{\prime},k^{\prime}_3,r\right) 
\tau^{c}_{dd^{\prime}}
\left(P^{3/2}\right)_{ck;ll^{\prime}}
\tau^{k}_{d^{\prime}d^{\prime\prime}}
\psi_{i_{1}\cdots i_{n},d^{\prime\prime}l^{\prime}}
\left(k^{\prime},k^{\prime}_3,r\right)=\frac{8}{3} \ .
\end{equation}
The non--diagonal matrix element is, up to a phase factor  
 in Eq.~(\ref{non diagonal element})  
\begin{equation}
i f_n \, F^{\prime}(0) \sqrt{\frac{2}{3}} \ , \quad ~~~\forall 
\, k^{\prime}\neq 0\ .
\end{equation}

{}For $k^{\prime}=0$ we have only one diagonal element with 
$j_{l}=\frac{1}{2}$. The second line of 
Eq.~(\ref{def: final diag elements})  
provides
\begin{equation}
V(k^{\prime}=0)=-\frac{3}{2}F^{\prime}(0)d_0\ .
\end{equation}

\section{Collective Lagrangian}
\label{app:b}

Here the relevant 
matrix elements associated with the collective coordinate Lagrangian 
are computed. 
We will restrict $k'$ to be nonzero since
there is no contribution 
for $k^{\prime}=0$ 
to the collective Lagrangian.

The kinetic Lagrangian for $H$ type and ${\cal H}$ type fields is:
\begin{equation}
{\cal L}_{\rm kin}=+i\,M V^{\mu}\sum_{n}  \mbox{Tr}\, 
\left[ H^{\mu_1\cdots\mu_n} D_\mu \bar{H}_{\mu_1\cdots\mu_n} \right]
-i\,M V^{\mu}\sum_n 
\mbox{Tr}\, 
\left[ {\cal H}^{\mu_1\cdots\mu_n} D_\mu 
\bar{{\cal H}}_{\mu_1\cdots\mu_n} \right]
 \ .
\end{equation}
In the following we will not distinguish between the 
$H$ and ${\cal H}$ types of field.  
We need to consider the collective coordinate Lagrangian  for  a given 
 $k^{\prime}$ classical bound channel 
in the heavy meson rest frame. 
{}For $k^{\prime}\neq 0 $  the bound state wave--function can 
schematically be represented as
\begin{equation}
|Bound~State; k^{\prime} \rangle
= \alpha\, |H_{\mu_1\cdots\mu_{k^{\prime}-1}} \rangle
+ \beta\, |H_{\mu_1\cdots\mu_{k^{\prime}}} \rangle \ ,
\end{equation}
where $|\alpha|^2 + |\beta|^2=1$.

The collective coordinate Lagrangian 
($\delta L_{\rm coll}$), induced by the heavy meson kinetic term, 
is obtained by   
generalizing 
Eqs.~(\ref{def: collective}) and (\ref{def: curl collective}) 
to the higher excited heavy  
meson fields, introducing the collective coordinate $A(t)$ rotation
via  
\begin{equation}
\bar{H}_{i_1\cdots i_n}(\mbox{\boldmath$x$},t)=
A(t)\bar{H}_{i_1\cdots i_n {\rm c}}
(\mbox{\boldmath$x$})\ ,
\end{equation}
where the $\bar{H}_{i_1\cdots i_n {\rm c}}
(\mbox{\boldmath$x$})$ classical 
ansatz is given in Eq.~(\ref{general ansatz}). 
The contribution for fixed $k^{\prime}\neq 0$ is:
\begin{eqnarray}
\delta L_{\rm coll}&=& -\Omega^{q} \, 
\left[|\alpha|^2 \int d\,\Omega\, 
\psi^{\ast}_{i_{1}\cdots i_{k^{\prime}-1},dl}
\left(k^{\prime},k^{\prime}_3,r\right) 
\frac{\tau^q_{dd^{\prime}}}{2}
\psi_{i_{1}\cdots i_{k^{\prime}-1},d^{\prime}l} 
\left(k^{\prime},k^{\prime}_3,r\right) \right.\nonumber \\ 
&~&+\left.|\beta|^2 \int d\,\Omega \, 
\psi^{\ast}_{i_{1}\cdots i_{k^{\prime}},dl}
\left(k^{\prime},k^{\prime}_3,r\right) 
\frac{\tau^q_{dd^{\prime}}}{2}
\psi_{i_{1}\cdots i_{k^{\prime}},d^{\prime}l}
\left(k^{\prime},k^{\prime}_3,r\right) \right]\nonumber \\
&\equiv&-|\alpha|^2 
\int d\Omega\,
\psi^{\ast}
\left(k^{\prime},k^{\prime}_3,j_l=k^{\prime}-1/2\right)
\mbox{\boldmath$\Omega$}\cdot\mbox{\boldmath$I$}_{\rm light}
\psi
\left(k^{\prime},k^{\prime}_3,j_l=k^{\prime}-1/2\right)
\nonumber \\
&~&-|\beta|^2 
\int d\Omega
\psi^{\ast}
\left(k^{\prime},k^{\prime}_3,j_l=k^{\prime}+1/2\right)
\mbox{\boldmath$\Omega$}\cdot\mbox{\boldmath$I$}_{\rm light}
\psi
\left(k^{\prime},k^{\prime}_3,j_l=k^{\prime}+1/2\right)
\ , \nonumber\\ 
~&~&~
\label{collective kinetic term}
\end{eqnarray}
where the over all minus sign in Eq.~(\ref{collective kinetic term})
is required, as explained in section \ref{sec:4}.
According to the Wigner-Eckart theorem:
\begin{equation}
\int d \Omega \, \psi^{\ast} 
\mbox{\boldmath$I$}_{\rm light}
\psi = 
\frac{\left[k^{\prime}(k^{\prime}+1)-j_l(j_l+1) + \frac {3}{4}\right]}
{2\,k^{\prime}(k^{\prime}+1)} 
\int d\Omega \, \psi^{\ast}
\mbox{\boldmath$K$}^{\prime}
\psi
\ ,
\label{collective  matrix element}
\end{equation}
we thus obtain the following heavy meson contribution to the
collective coordinate Lagrangian for 
$k^{\prime}\neq 0$
\begin{equation}
\delta L_{\rm coll}
= 
-\chi(k^{\prime})\,
\mbox{\boldmath$\Omega$}\cdot\mbox{\boldmath$K$}^{\prime}
\ .
\end{equation}
The quantity $\chi(k')$ is given by
\begin{equation}
 \chi(k^{\prime})=\frac{1}{2\,k^{\prime}(k^{\prime}+1)}
\left[ \frac{1}{2} \pm \left(k^{\prime} + \frac{1}{2}\right) 
\cos 2 \theta\right] \ ,
\label{collective chi}
\end{equation}
where $|\alpha|^2 - |\beta|^2=\pm \cos 2 \theta$ was used.
In Eq.~(\ref{collective chi}) 
the $\pm$ sign corresponds to the two possible eigenvalues in the 
potential matrix for given $k^{\prime}\neq 0$.